\documentclass[a4paper]{aa}
\usepackage{graphics,epsfig,txfonts}
\usepackage[utf8]{inputenc}
\usepackage[section]{placeins}
\usepackage{amsmath}
\usepackage{color}
\usepackage{xcolor}
\usepackage{graphicx}
\usepackage{xparse,xcoffins}
\usepackage{natbib}
\bibpunct{(}{)}{;}{a}{}{,}

\ExplSyntaxOn
\NewCoffin\imagecoffin
\NewCoffin\labelcoffin

\keys_define:nn { miguel/label }
 {
  label   .tl_set:N = \l_miguel_label_tl,
  labelbox .bool_set:N = \l_miguel_label_box_bool,
  labelbox .default:n = true,
  fontsize .tl_set:N = \l_miguel_label_size_tl,
  fontsize .initial:n = \footnotesize,
  pos .choice:,
  pos/nw .code:n = \tl_set:Nn \l_miguel_label_pos_tl { left,up },
  pos/ne .code:n = \tl_set:Nn \l_miguel_label_pos_tl { right,up },
  pos/sw .code:n = \tl_set:Nn \l_miguel_label_pos_tl { left,down },
  pos/se .code:n = \tl_set:Nn \l_miguel_label_pos_tl { right,down },
  pos/n .code:n = \tl_set:Nn \l_miguel_label_pos_tl { hc,up },
  pos/w .code:n = \tl_set:Nn \l_miguel_label_pos_tl { left,vc },
  pos/s .code:n = \tl_set:Nn \l_miguel_label_pos_tl { hc,down },
  pos/e .code:n = \tl_set:Nn \l_miguel_label_pos_tl { right,vc },
  pos .initial:n = nw,
  unknown .code:n   = \clist_put_right:Nx \l_miguel_label_clist
                       { \l_keys_key_tl = \exp_not:n { #1 } }
 }
\clist_new:N \l_miguel_label_clist
\box_new:N \l_miguel_label_box
\box_new:N \l_miguel_label_image_box

\NewDocumentCommand{\xincludegraphics}{O{}m}
 {
  \group_begin:
  \tl_clear:N \l_miguel_label_tl
  \clist_clear:N \l_miguel_label_clist
  \keys_set:nn { miguel/label } { #1 }
  \tl_if_empty:NTF \l_miguel_label_tl
   {
    \miguel_includegraphics:Vn \l_miguel_label_clist { #2 }
   }
   {
    \SetHorizontalCoffin\imagecoffin
     {
      \miguel_includegraphics:Vn \l_miguel_label_clist { #2 }
     }
    \SetHorizontalCoffin\labelcoffin
     {
      \raisebox{\depth}
       {
        \bool_if:NTF \l_miguel_label_box_bool
         { \fcolorbox{white}{white}{\l_miguel_label_size_tl\l_miguel_label_tl} }
         { \l_miguel_label_size_tl\l_miguel_label_tl }
       }
     }
    \SetVerticalPole\imagecoffin{left}{3pt+\CoffinWidth\labelcoffin/2}
    \SetVerticalPole\imagecoffin{right}{\Width-3pt-\CoffinWidth\labelcoffin/2}
    \SetHorizontalPole\imagecoffin{up}{\Height-3pt-\CoffinHeight\labelcoffin/2}
    \SetHorizontalPole\imagecoffin{down}{3pt+\CoffinHeight\labelcoffin/2}
    \use:x{\JoinCoffins\imagecoffin[\l_miguel_label_pos_tl]\labelcoffin[vc,hc]} 
    \TypesetCoffin\imagecoffin
   }
   \group_end:
 }
\NewDocumentCommand{\setlabel}{m}
 {
  \keys_set:nn { miguel/label } { #1 }
 }

\cs_new_protected:Nn \miguel_includegraphics:nn
 {
  \includegraphics[#1]{#2}
 }
\cs_generate_variant:Nn \miguel_includegraphics:nn { V }

\ExplSyntaxOff

\def \MSUN{\ifmmode{\mathrm{M_{\odot}}} \else {$\mathrm{M_{\odot}}$}\fi}
\def\vmeas {\ifmmode{\mathrm{v}} \else {$\mathrm{v}$}\fi} 
\def\vsys {\ifmmode{V_\mathrm{sys}} \else {$V_\mathrm{sys}$}\fi} 
\def\vrot {\ifmmode{V_\mathrm{rot}} \else {$V_\mathrm{rot}$}\fi} 
\def\PAkin {\ifmmode{PA_\mathrm{{kin}}} \else {$PA_\mathrm{{kin}}$}\fi}  
\def\PAphot {\ifmmode{PA_\mathrm{{phot}}} \else {$PA_\mathrm{{phot}}$}\fi}  
\def\re {\ifmmode{R_\mathrm{{e}}} \else {$R_\mathrm{{e}}$}\fi} 
\def\fex {\ifmmode{{f_\mathrm{exsitu}}} \else {$f_\mathrm{exsitu}$}\fi}
\def \rsoft {\ifmmode{r_\mathrm{{soft}}} \else {$r_\mathrm{{soft}}$}\fi} 
\defcitealias{2020A&A...641A..60P}{Paper~I}
\title{The stellar halos of ETGs in the IllustrisTNG simulations: \\ II. Accretion, merger history, and dark halo connection}
\author{C. Pulsoni\inst{1,2}
     \and {O. Gerhard\inst{1}}
     \and {M. Arnaboldi\inst{3}}
     \and {A. Pillepich\inst{4}}
     \and {V. Rodriguez-Gomez\inst{5}}
     \and {D. Nelson\inst{6,7}}
     \and \\ {L. Hernquist\inst{8}}
     \and {V. Springel\inst{6}}}

\authorrunning{C. Pulsoni et al.}
\titlerunning{ETG stellar halos in IllustrisTNG: Accretion, merger history, and dark halo connection}

\institute{Max-Planck-Institut f\"ur extraterrestrische Physik, Giessenbachstra{\ss}e, 85748 Garching, Germany
	   \and Excellence Cluster Universe, Boltzmannstra{\ss}e 2, 85748 Garching, Germany 
	   \and European Southern Observatory, Karl-Schwarzschild-Stra{\ss}e 2, 85748 Garching, Germany
	   \and Max-Planck-Institut für Astronomie, K\"onigstuhl 17, 69117 Heidelberg, Germany
	   \and Instituto de Radioastronom\'ia y Astrof\'isica, Universidad Nacional Aut\'onoma de M\'exico, A.P. 72-3, 58089 Morelia, Mexico
	   \and Max-Planck-Institut f\"ur Astrophysik, Karl-Schwarzschild-Str. 1, 85748 Garching, Germany
	   \and Universit\"{a}t Heidelberg, Zentrum f\"{u}r Astronomie, Institut f\"{u}r theoretische Astrophysik, Albert-Ueberle-Str. 2, 69120 Heidelberg, Germany
	   \and Harvard-Smithsonian Center for Astrophysics, 60 Garden Street, Cambridge, MA 02138, USA
	   }

\date{16 January 2021}

\abstract{
    Stellar halos in early-type galaxies (ETGs) are shaped by their accretion and merger histories.
    We use a sample of 1114 ETGs in the TNG100 simulation of the IllustrisTNG suite with stellar masses $10^{10.3}\leq M_{*}/\MSUN{}\leq 10^{12}$, selected at $z=0$ within the range of $g-r$ colour and $\lambda$-ellipticity diagram populated by observed ETGs. We study how the rotational support and intrinsic shapes of the stellar halos depend on the fraction of stars accreted, overall and separately by major, minor, and mini mergers.
    Accretion histories in TNG100 ETGs as well as the final radial distributions of ex-situ stars $\fex{}(R)$ relative to in-situ ('accretion classes') strongly correlate with stellar mass. 
    Low-mass galaxies have characteristic peaked rotation profiles and near-oblate shapes with rounder halos that are completely driven by the in-situ stars.
    At high \fex{} major mergers decrease the in-situ peak in rotation velocity, flatten the $V_{*}/\sigma_{*}(R)$ profiles, and increase the triaxiality of the stellar halos. Kinematic transition radii do not trace the transition between in-situ and ex-situ dominated regions, but for systems with $M_{*}>10^{10.6}\MSUN{}$ the local rotational support of the stellar halos decreases with the local ex-situ fraction $\fex{}(R)$ at fixed $M_{*}$, and their triaxiality increases with $\fex{}(R)$.
    These correlations between rotational support, intrinsic shapes and local $\fex{}$ are followed by fast and slow rotators alike with a continuous and overlapping sequence of properties, but slow rotators are concentrated at the high $\fex{}$ end dominated by dry major mergers.
    We find that in $\sim20\%$ of high-mass ETGs the central regions are dominated by stars from a high-redshift compact progenitor.
    Merger events dynamically couple stars and dark matter: in high mass galaxies and at large radii where $\fex{}\gtrsim0.5$, both components tend to have similar intrinsic shapes and rotational support, and nearly aligned principal axes and spin directions.
    Based on these results we suggest that extended photometry and kinematics of massive ETGs ($M_{*}>10^{10.6}\MSUN{}$) can be used to estimate the local fraction of ex-situ stars, and to approximate the intrinsic shapes and rotational support of the co-spatial dark matter component. 
}

\keywords{Galaxies: elliptical and lenticular, cD - Galaxies: evolution - Galaxies: halos - Galaxies: kinematics and dynamics - Galaxies: photometry - Galaxies: structure - 
} 

\begin{document}

\maketitle

\section{Introduction}

The $\Lambda$CDM cosmology predicts that structures form hierarchically, so that more massive systems form through the accretion of less massive objects \citep{1978MNRAS.183..341W, 1980lssu.book.....P}. In this model, the formation of massive early type galaxies (ETGs) is believed to have occurred in two phases \citep[e.g.][]{2010ApJ...725.2312O}. After an initial assembly stage, where ETGs form stars in a brief intense burst which is quickly quenched \citep[e.g.][]{2005ApJ...621..673T, 2014ApJ...780...33C, 2010ApJ...721..193P}, at $z\lesssim1$ galaxies grow efficiently in size through a series of merger episodes, mainly dry minor mergers \citep[e.g][]{2009ApJ...699L.178N}, which enrich the galaxies with accreted (ex-situ) stars.
More massive galaxies can have accreted fractions larger than 80\%, while lower mass systems are mostly made of in-situ stars and their accreted component is mainly deposited in the outskirts \citep[e.g][]{2013MNRAS.434.3348C,2016MNRAS.458.2371R, 2018MNRAS.475..648P, 2019MNRAS.487.5416T, 2020MNRAS.tmp.1947D}. 

Elliptical galaxies essentially divide into two classes with distinct physical properties \citep[e.g.][]{2009ApJS..182..216K, 2016ARA&A..54..597C}: those with low to intermediate masses and coreless luminosity profiles and those, which are frequently among the most massive galaxies, with cored profiles. This dichotomy in light distribution roughly corresponds to different kinematic properties, with coreless disky objects being rotationally supported, and cored boxy galaxies having low rotation \citep{1987MitAG..70..226B}. In the last decade, the advent of integral field kinematics has lead to a kinematics-based division between fast rotators (FRs) and slow rotators (SRs, \citealt{2011MNRAS.414..888E, 2017ApJ...835..104V, 2017MNRAS.471.1428V, 2018MNRAS.477.4711G}): low mass, coreless, FR ellipticals share similar properties with lenticular galaxies, which are also included in the FR family, while massive cored ellipticals are typically SRs.

State-of-the-art cosmological simulations agree in that the progenitors of FR and SR at high redshifts ($z\sim1$) are indistinguishable \citep[][with Illustris, Eagle, and Magneticum, respectively]{2017MNRAS.468.3883P, 2017MNRAS.464.3850L, 2018MNRAS.480.4636S}. Present-day SR/FR (i.e. the core/coreless) classes result from different formation pathways characterized by different numbers of mergers, merger mass ratio, timing, and gas fractions \citep[][see also the discussion in \citealt{2009ApJS..182..216K}]{2014MNRAS.444.3357N, 2017MNRAS.468.3883P}, although the details still depend on the star formation and AGN feedback models adopted by the numerical models \citep{2017ARA&A..55...59N}. In general, the result of a formation history dominated by gas dissipation is most likely a coreless FR, while dry major mergers often result in SRs. 

A consequence of the two-phase formation is that ETG central regions are the remnants of the stars formed in-situ while the external stellar halos are principally made of accreted material \citep{2005ApJ...635..931B, 2010MNRAS.406..744C}, even though the details strongly depend on stellar mass \citep{2018MNRAS.475..648P}. 
Because of the different origin of central regions and stellar halos, galaxies are expected to show significant variation of physical properties with radius, such as shapes of the light profiles \citep{2013ApJ...768L..28H, 2014MNRAS.443.1433D,2017A&A...603A..38S, 2017MNRAS.466.4888B}, stellar populations \citep{2010MNRAS.407L..26C, 2014MNRAS.442.1003P, 2020MNRAS.491.3562Z}, and kinematics \citep{2009MNRAS.394.1249C, 2012ApJS..203...17R,2014ApJ...791...80A,2016MNRAS.457..147F}.

\citet{2018A&A...618A..94P} found evidence for a kinematic transition between the central regions and the outskirts in the majority of the ETGs from the ePN.S survey \citep{2017IAUS..323..279A}, consisting in variations of the rotational support or changes in the direction of rotation at large radii \citep[see also][]{2009MNRAS.394.1249C, 2016MNRAS.457..147F}. As a result, ePN.S ETG halos display a variety of kinematic behaviors despite the FR/SR dichotomy of their central regions, with most of the stellar halos displaying similar rotational support $V/\sigma$ across classes (where $V$ is the rotation velocity and $\sigma$ the velocity dispersion). 
These findings suggest the idea that at large radii the dynamical structure of FRs and SRs could be much more similar than in their centers: if halos are mainly formed from accreted material, their common origin would explain their similarities. The measured kinematic transition radii and their dependence on the galaxy stellar mass seem to support such an interpretation. 
Recently, \citet{2020MNRAS.493.3778S} using the Magneticum Pathfinder simulations found that the kinematic transition radius, estimated as the position of the peak in rotation for a subset of FRs with decreasing $V/\sigma$ profiles, is a good estimator of radius of the transition between in-situ and ex-situ dominated regions, especially in galaxies that did not undergo major mergers in their evolution.

In \citet[][hereafter \citetalias{2020A&A...641A..60P}]{2020A&A...641A..60P} we investigated the kinematic and photometric properties of simulated ETGs from the IllustrisTNG simulations, TNG50 and TNG100. We found that simulations reproduce the diversity of kinematic properties observed in ETG halos: FRs divide in one third having flat rotation profiles and high halo rotational support, a third with gently decreasing profiles, and another third with low halo rotation; SRs tend to increase rotational support in the outskirts and half of them exceed angular momentum parameter $\lambda=0.2$. Simulated stellar halos are also characterized by a large variety of intrinsic shapes which are strongly related to their rotational support: high rotation is associated with flattened near-oblate shapes; decreasing rotational support with radius is accompanied by a change towards more spheroidal intrinsic shape, with a wide range of triaxiality.
These variations in the intrinsic structure of ETGs results into a blurring of the FR/SR bimodality at large radii, with the two families showing a gradual transition in stellar halo properties.

This current work builds upon the results of \citetalias{2020A&A...641A..60P}. The goal is to relate the kinematic and photometric properties of stellar halos to the galaxy accretion history, parametrized by the total fraction of accreted stellar mass, the mass fraction contributed by different mass ratio mergers, and the fraction of recently accreted cold gas. Then we extend the investigation to the properties of the dark matter halos and their relation with the stellar component depending on accretion. 

Dark matter halo and galaxy properties are known to be related. Primordially, large scale tidal fields generate tidal stretching \citep{2001MNRAS.320L...7C} and tidal torquing \citep[][for a review]{2009IJMPD..18..173S} on both baryons and dark matter inducing shape and spin alignments. In the subsequent evolution, baryonic physics modifies the shapes of inner dark matter halos from triaxial towards rounder and more oblate shapes 
\citep[e.g.,][]{2004ApJ...611L..73K, 2013MNRAS.429.3316B, 2019MNRAS.484..476C}, while galaxy central regions and outer dark matter halos are more uncorrelated \citep[e.g.,][]{2005ApJ...627L..17B, 2011MNRAS.415.2607D, 2015MNRAS.453..469T}. This is due to the fact that both centers and outskirts evolve away from the initial shape and angular momentum correlations predicted from the tidal torque theory. On one hand the outer dark matter halos continues to accumulate accreted material, on the other the central regions gain angular momentum through non linear tidal torques and dissipation exerted by the gas flows entering the dark matter halos, and are affected by disk instabilities and feedback processes that makes the galaxy spins (and shapes) deviate from the halo spins \citep[][]{2017MNRAS.466.1625Z, 2019MNRAS.487.1607G}. Since mergers deposit together stars and dark matter in the galaxy outskirts, which both follow a collisionless dynamics, then we would expect that the (accreted) stellar and dark matter halos have a similar structure at large radii. This is investigated in the last section of this paper.

The paper is organized as follows. Section~\ref{sec:SelectionSample} summarizes the main characteristics of the IllustrisTNG simulations and describes the selection of the analyzed ETG sample. Section~\ref{sec:measuring_quantities} details how physical quantities are measured from the simulated galaxies. Section~\ref{sec:history} gives an overview of the accretion parameters used in the main analysis to parametrize galaxy accretion history and their dependence on stellar mass, and studies the different radial distribution of accreted stars in TNG ETGs. Sections~\ref{sec:Vsigma_profiles} and \ref{sec:Intrinsic_shapes} analyse the dependence of the rotational support $V/\sigma(R)$, of the axis ratio $q(r)$, and of the triaxiality $T(r)$ profile shapes on the fraction of in-situ stars. We study how mergers shape the galaxy structure and set up local correlations between physical parameters and accreted fraction. Section~\ref{sec:dmVSstars} investigates the connection between stellar and dark matter halos. Section~\ref{sec:discussion_RNcores} considers the fraction of TNG ETGs with cores made of stars from early compact progenitors, comparing their evolution with respect to ETGs of similar masses. Section~\ref{sec:discussion} discusses some of the main results of the paper. Section~\ref{sec:summary} summarizes the work and Sect.~\ref{sec:conclusions} lists our conclusions.

\section{Selection of the sample of ETGs in the IllustrisTNG simulations} \label{sec:SelectionSample}

The IllustrisTNG simulations are a new generation of cosmological magnetohydrodynamical simulations that model the formation and evolution of galaxies within the $\Lambda$CDM paradigm \citep{2018MNRAS.475..676S, 2018MNRAS.475..624N, 2018MNRAS.475..648P,2018MNRAS.477.1206N, 2018MNRAS.480.5113M}. 
The model for galaxy formation, described in \citet{2017MNRAS.465.3291W} and \citet{2018MNRAS.473.4077P}, includes prescriptions for star formation and evolution, chemical enrichment of the
ISM, gas cooling and heating, black hole and supernova feedback. It builds and improves upon the Illustris simulation \citep{2014MNRAS.445..175G, 2014MNRAS.444.1518V} including improvements in the models for chemical enrichment and feedback, and introduces new physics such as the growth and amplification of seed magnetic fields.
Overall the TNG model has been demonstrated to agree satisfactorily with many observational constraints \citep[e.g.][]{2018MNRAS.474.3976G, 2018MNRAS.475..624N} and to return a reasonable mix of morphological galaxy types \citep{2019MNRAS.483.4140R}.

In this study we consider the highest resolution realization of the intermediate $110.7^3$ Mpc$^3$ cosmological volume, TNG100 \citep[which is now publicly available,][]{2019ComAC...6....2N}, in order to exploit the statistically significant number of simulated objects and, at the same time, to be able to well resolve their inner structure. The simulation follows the evolution of $2\times1820^3$ initial resolution elements. The mean mass of the stellar particles is $1.4\times10^6\MSUN{}$, while the dark matter component is sampled by $7.5 \times 10^6\MSUN{}$ mass particles. The Plummer equivalent gravitational softening length for the collisionless component at redshift $z=0$ is $\rsoft{} = 0.74$ kpc.

The purpose of this paper is to study the properties of ETG halos and their relation to the accretion history. Here we refer to stellar halos as the outer regions of the galaxies beyond a few effective radii ($R_e$), where the physical properties are often markedly different from those of the central regions within $\sim 1 R_e$.

We consider a volume- and stellar mass-limited sample of simulated ETGs, selected as in \citetalias{2020A&A...641A..60P}. There we used a selection in color and intrinsic shape to extract a sample of ETGs with properties consistent with observations. In particular we considered the Atlas3D sample properties to validate our selection criteria, as this survey was especially targeted to study a volume-limited sample of ETGs \citep{2011MNRAS.413..813C}. 

In brief, from the $(g-r)$ color-stellar mass diagram we isolated galaxies in the red sequence by imposing
\begin{equation}
    (g-r) \geq 0.05\log_{10}(M_{*}/\MSUN{}) + 0.1 \mathrm{mag}.
    \label{eq:color_mass_sel}
\end{equation}
We restricted the stellar mass range to $10^{10.3} \leq M_{*} \leq 10^{12} \MSUN{}$. This limit in stellar mass assures that the galaxies are resolved by at least $2\times10^4$ stellar particles. In addition we imposed that the galaxies' effective radius $\re{} \geq 2\rsoft{}$ to guarantee that the regions at $r=\re{}$ of all simulated galaxies are well resolved. This excludes 42 galaxies at the low mass end.

The final sample was obtained by further selecting galaxies in the $\lambda_e - \varepsilon$ diagram, where $\varepsilon$ is the ellipticity. There we excluded a fraction of bar-like objects, whose $\lambda_e - \varepsilon$ properties are not consistent with those of observed ETGs. In  \citetalias{2020A&A...641A..60P} we found that this latter selection does not affect our results on the halo properties of the simulated galaxies. This final criterion restricts the sample to 1114 ETGs, including both centrals (61\% of the sample) and satellites, of which 855 are FRs and 259 are SRs. The classification in FRs and SRs uses the threshold $\lambda_e = 0.31\sqrt{\varepsilon}$ defined by \cite{2011MNRAS.414..888E}, applied to the edge-on projected galaxies to minimize inclination effects.
In  \citetalias{2020A&A...641A..60P} we showed that the mass function of the selected sample is in good agreement with the Atlas3D ETGs. In addition the comparison of the ellipticity distribution at $1\re{}$ of the simulated galaxies with the Atlas3D sample showed that overall the selected sample of ETGs contains a balance between disks-dominated and spheroid-dominated galaxies consistent with the observations of a volume and magnitude-limited sample of ETGs.

\section{Methods and definitions}
\label{sec:measuring_quantities}

In this section we describe how physical quantities are evaluated from the simulated galaxies. 
We define the total stellar mass $M_{*}$ as the total bound stellar mass of the galaxy, $M_{*}= \sum_n m_n$, where the index $n$ runs over the stellar particles of mass $m_n$. The effective radius $\re{}$, used throughout the paper to normalize radial distances, is the edge-on projected half-mass radius. This is the semi-major axis of the elliptical aperture that contains half of the total bound stellar mass. The ellipticity of the aperture is given by the projected flattening of the galaxy at $1\re{}$, measured using the 2D inertia tensor as in  \citetalias{2020A&A...641A..60P}.

\subsection{Characterizing the
galaxy merger and accretion histories}\label{sec:measuring_accretion}

Stellar particles in the simulated galaxies are tagged as accreted (or ex-situ) or as in-situ following the definition in \citet{2016MNRAS.458.2371R}. In-situ stars form within the "main progenitor branch" of the galaxy merger tree, independently of the origin of the star-forming gas. Ex-situ stars form outside the main progenitor branch and are subsequently accreted onto the host galaxy through mergers or stripping events \citep[the fraction of smoothly accreted stars, which were not bound to other subhalos at formation, is negligible,][]{2016MNRAS.458.2371R}.
Since mergers can contribute to a significant fraction of gas in the host, which can lead to in-situ star formation, the total ex-situ stellar mass fraction does not directly quantify the merger history but measures the relative importance of dry merging with respect to dissipative processes that increase the fraction of in-situ stars \citep[e.g.][]{2010ApJ...725.2312O}. 

By definition, the total bound stellar mass of a galaxy is given by the sum of the mass in the in-situ and the ex-situ components $M_{*} = M_{\rm insitu} + M_{\rm exsitu}$. The total in-situ or ex-situ mass fraction is then $M_{i}/M_{*}$, with $i\in\{{\rm insitu, exsitu}\}$. Throughout the paper we consider also the local in-situ and ex-situ mass fractions, which are the stellar mass fractions measured within a given radial bin. These are indicated by $f_{\rm insitu}$ and $\fex{}$, respectively.

The ex-situ stars can be distinguished according to whether they were stripped from a surviving galaxy, for example in a flyby or in a merger that is still on-going at $z=0$, or whether they originated from completed mergers. In case of completed mergers, we classify the stars according to the merger stellar mass ratio. We distinguish between major mergers (i.e., with stellar mass ratio $\mu>1/4$), minor mergers ($1/10<\mu\leq1/4$), and mini mergers ($\mu\leq1/10$). 

To describe the galaxy accretion histories, we consider the following parameters:
\begin{itemize}
    \item the total in-situ mass fraction;
    \item the mean accreted "cold" (star forming) gas fraction of the accreted galaxies at all epochs, weighted by their stellar mass. The gas fraction and stellar mass of the accreted galaxy are measured at the time when its stellar mass is maximum \citep{2017MNRAS.467.3083R};
    \item the redshift of its last major merger $z_{last}$;
    \item the fraction of in-situ stars that are produced after a certain redshift $\bar{z}$, $\Delta_{\rm insitu, z\leq\bar{z}}$:
    \begin{equation}
        \Delta_{\rm insitu, z\leq\bar{z}} = [M_{\rm insitu}(z=0) - M_{\rm insitu} (\bar{z})] / M_{\rm *}(z=0). 
    \end{equation}
\end{itemize}

\subsection{Intrinsic shapes}
\label{sec:measuring_IntrinsicShape}

The intrinsic shapes of the galaxies are evaluated using the moment of inertia tensor.
For each galaxy we define a coordinate system $(x,y,z)$ centered on the position of the most bound particle, and aligned with the principal axes of the stellar component, such that $x$ is along the intrinsic major axis of the galaxy and $z$ is along the intrinsic minor axis. Since the direction of the principal axes of a galaxy might change with the distance from the center, we choose $(x,y,z)$ to be the directions of the principal axes derived for the 50\% most bound particles using the inertia tensor $I_{ij}$,
\begin{equation}
    I_{ij} = \frac{\sum_n m_{n} x_{n,i} x_{n,j}}{\sum_n m_{n}},
\label{eq:inertia_tensor}
\end{equation}
where the sum is performed over the particles concerned, and $x_{n,i}$ are their coordinates in a system of reference centered on the galaxy and with axes oriented along the sides of the simulation box. Then the edge-on projection of each galaxy is obtained by choosing the $y$-axis as the line-of-sight. 
In this work we indicate with the lower-case letters $(x,y,z)$, $(v_x,v_y,v_z)$, and $r$ the 3D coordinates, velocities, and radii. We reserve capital letters for the edge-on projected quantities. The coordinate $r$ indicates the three-dimensional radius, in case of quantities calculated in spherical bins, or the intrinsic semi-major axis distance, in case of quantities calculated in ellipsoidal bins.  The coordinate $R$ indicates the edge-on projected semi-major axis distance. 

The intrinsic shapes of the galaxies are evaluated by diagonalizing the inertia tensor $I_{ij}$ in Eq.~\eqref{eq:inertia_tensor} summed over particles in ellipsoidal shells.  We follow the same iterative procedure outlined in  \citetalias{2020A&A...641A..60P} for the stellar and dark matter components separately. 
Each component is divided into spherical shells between galactocentric distances $r$ and $r+\Delta r$. In each shell the tensor $I_{i,j}$ is derived: the square root of the ratio of its eigenvalues gives the axis ratios $p = b/a$ and $q = c/a$ of the principal axes of the ellipsoid (with $a\geq b \geq c$), the eigenvectors $\hat{e}_j$ (with $j=a,b,c$) their directions. The spherical shell is then iteratively rotated and deformed to a homeoid of semi-axes $a=r$, $b = p a$ and $c = q a$ until convergence in the values $p$ and $q$ is reached, that is, the fractional difference between two iteration steps in both axis ratios is smaller than 1\%.

The triaxiality parameter is defined as
\begin{equation}
    T(r)=\frac{1 - p(r)^2}{1-q(r)^2},
\end{equation} 
where $r$ is the intrinsic semi-major axis distance. We will consider shapes with $T\leq0.3$ as near-oblate and shapes with $T>0.7$ as near-prolate. Intermediate values of the triaxiality parameter define triaxial shapes.

In  \citetalias{2020A&A...641A..60P} we verified that the intrinsic shape measurements are affected by the resolution of the gravitational potential and by the number of particles only for the lowest mass systems, for which we estimated an absolute error of 0.1 on both $p$ and $q$ measured at $1\re{}$. More massive galaxies $M_{*}>10^{11}\MSUN{}$ are much better resolved at $r\sim1\re{}$. At $r\geq9\rsoft{}$, where $\rsoft{}$ is the softening length of the simulation, (i.e., at $\gtrsim 3.5\re{}$ for $M_{*}\sim10^{10.3}\MSUN{}$ and at $\gtrsim 1\re{}$ for $M_{*}\sim10^{11}\MSUN{}$) the resolution effects are negligible compared to the errors from limited particle numbers, which we quantified being of the order of $0.02$. This implies an uncertainty on the triaxiality parameter of $\Delta T\sim0.2$, for typical FR axis ratios (i.e. $q\sim0.5$ and $p=0.9$). At $r<9\rsoft{}$, $\Delta T$ increases and then intrinsic shapes are better quantified by $p$ and $q$. 
The error on the direction of the principal axes is generally very small, but it increases with the axis ratios. For axis ratio $p$ or $q/p = 0.5$ it is of the order of $1^\circ$, and reaches $\sim6^\circ$ for axis ratio of $0.9$. At higher values of $p$ and of $q/p$ the uncertainties on the direction of the major axis and minor axis, respectively, grow exponentially. 
These considerations hold for both stars and dark matter, because the gravitational softening length is identical for both components; the above limitations are derived from the results of \citet{2019MNRAS.484..476C} for the Illustris dark matter only simulation. Because the full physics simulation TNG100 has very similar resolution and a larger number of particles in the central regions compared to this dark matter only simulation, it is reasonable to expect similar, if not better, convergence in TNG100.

In the paper we consider as reliable shape measurements from shells with at least 1000 particles \citep{2011ApJS..197...30Z}. This means that we are able to measure intrinsic shapes for the stellar component out to $8\re{}$ for 96\% of the selected galaxies. Beyond that limit the lower mass objects begin to lack sufficient numbers of stellar particles (see \citetalias{2020A&A...641A..60P}).

\subsection{Rotational support}
\label{sec:measuring_rotational_support}

Ordered rotation in a galaxy is quantified by the ratio $V/\sigma$, where $V$ is the mean velocity and $\sigma$ the velocity dispersion.
Summing the components of the angular momentum vectors of all particles in a shell $S$, weighted by one over the product of the particle mass $m_n$ times radius $r_{n}=|\overrightarrow{r_n}|$, we define a mean rotation velocity vector in the shell,
\begin{equation}
        u_i(r) \equiv \left[\frac{\left(\sum_{n} m_n \overrightarrow{r_n} \times \overrightarrow{v_n}\right)_{i}}{ \sum_{n} |\overrightarrow{r_n}| m_n}\right]_{n|\overrightarrow{r_n} \in S[r, r+dr]},
\label{eq:specific_j_over_r}
\end{equation}
from which, together with twice the kinetic energy per unit mass $k$ in the shell,
\begin{equation}
    k(r) = \left[\frac{\sum_{n} m_n (v_{x,n}^2 + v_{y,n}^2 + v_{z,n}^2)}{\sum_{n} m_n}\right]_{n|\overrightarrow{r_n} \in S[r, r+dr]},
\label{eq:specific_k}
\end{equation} 
we define a mean 3D velocity 
\begin{equation}
    V(r) = \sqrt{u_x^2(r) + u_y^2(r) + u_z^2(r)}
\label{eq:V_3D}
\end{equation}
and velocity dispersion
\begin{equation}
    \sigma(r) = \sqrt{k(r) - V(r)^2}.
\label{eq:sigma_3D}
\end{equation}
The index $n$ runs over the particles within the chosen ellipsoidal or spherical shell $S[r, r+dr]$.
In Sect.~\ref{sec:dmVSstars}, where we compare the rotational support of the stellar component to that of the dark matter, we derive $V/\sigma$ for the two components separately, summing over stellar or dark matter particles within spherical shells of radii $r$ and $r+dr$.

In Sect.~\ref{sec:Vsigma_profiles} we study the rotational support of galaxies using their edge-on projections, in order to minimize inclination effects but at the same time to use quantities as close as possible to observables. In this case we calculate quantities within elliptical radial bins of semi-major axis $R$ and $R+dR$ and consider only the particle velocities along the line-of-sight $y$, $V_{Y,n} = v_{y,n}$, so that
\begin{equation}
V(R) = \sqrt{U_{X}^2(R) + U_{Z}^2(R)}
\label{eq:V_2D} 
\end{equation} 
and
\begin{equation}
\sigma(R) = \sqrt{k(R) - V(R)^2}.
\label{eq:sigma_2D}
\end{equation}
In Eq.~\eqref{eq:V_2D} $R$ is the semi-major axis of the elliptical annulus and $U_X$ and $U_Z$ are defined analogously to Eq.~\eqref{eq:specific_j_over_r}, where the particles are weighted by one over their mass times projected circular radius. The ratio between Eqs.~\eqref{eq:V_2D} and \eqref{eq:sigma_2D} gives the edge-on projected rotational support which, for the stellar component, is denoted by $V_{*}/\sigma_{*}(R)$. 

\begin{figure}
    \centering
    \includegraphics[width=\linewidth]{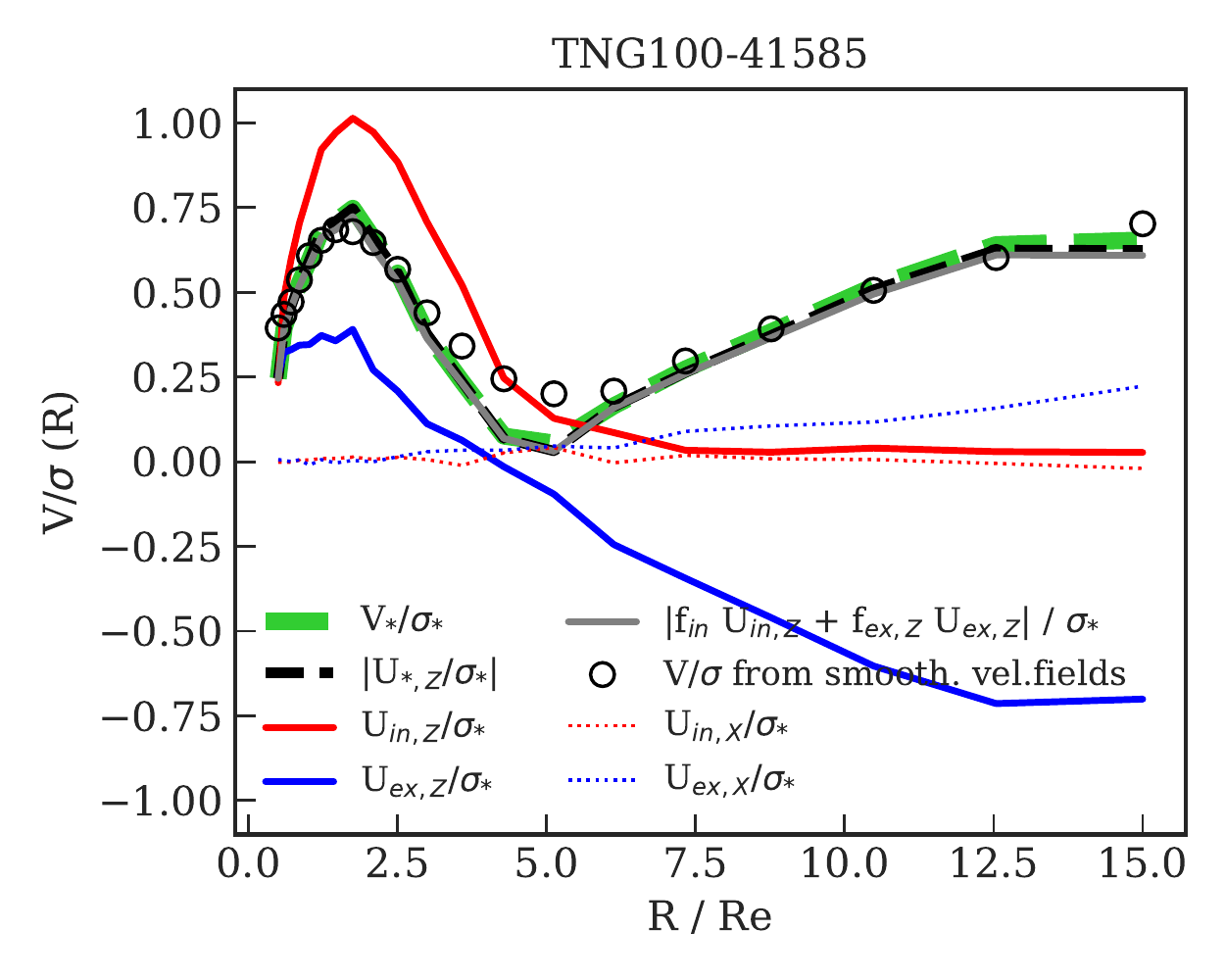}
    \caption{$V/\sigma$ profiles for an example FR galaxy with $M_{*}=10^{11.4}\MSUN{}$ and counter-rotating ex-situ component in the halo. The $V_{*}/\sigma_{*}(R)$ profile for the stars (Eqs.~\ref{eq:V_2D} and \ref{eq:sigma_2D}, green dashed line) is compared with the $V/\sigma(R)$ profile derivable from the smoothed velocity fields as in Eq.~\ref{eq:particlesVSvelfields} (open circles), with the $|U_{*,Z}(R)|/[\sigma_{*}(R)]$ profile (dashed black line), and with the weighted sum of the in-situ and ex-situ contributions (as in Eq.~\ref{eq:Vsigma_decomp_inex}, dashed gray curve). The individual un-weighted profiles of the in-situ and ex-situ stars are shown with red and blue curves: the solid lines show the contribution to the mean velocity from the major axis rotation $U_{\rm i,Z}(R) / [\sigma_{*}(R)]$, the dotted lines show the contribution from minor axis rotation $U_{\rm i,X}(R) / [\sigma_{*}(R)]$. }
    \label{fig:Method_Vsigma_example}
\end{figure}

The $V_{*}/\sigma_{*}(R)$ ratio defined in this way differs from the quantities presented in \citetalias{2020A&A...641A..60P} where, instead of the mean velocity, we used the rotational velocity derived from 2D mean velocity fields which maximizes the $V/\sigma$ ratio. 
The $V_{*}/\sigma_{*}(R)$ measured here directly from the star particles is roughly equal to the ratio of the mass weighted average velocity and the mass weighted average velocity dispersion obtained from the 2D mean velocity fields as built in \citetalias{2020A&A...641A..60P}:
\begin{equation}
V_{*}/\sigma_{*}(R) \sim \left[\frac{\sum_n m_{n} \left < V_{n} \right >}{\sum_n m_{n} \left < \sigma_{n} \right > } \right]_{n|\overrightarrow{R_n}\in S[R, R+dR]}.
\label{eq:particlesVSvelfields}
\end{equation}
Here the $n$ index runs over the Voronoi bins of total mass $m_n$, mean velocity $<V_{n}>$, and mean velocity dispersion $<\sigma_{n}>$ in the regions where the velocity fields are Voronoi binned \citep{2003MNRAS.342..345C}, and it runs over the stellar particles of mass $m_n$, mean velocity $<V_{n}>$, and mean velocity dispersion $<\sigma_{n}>$ in the regions where the velocity fields are smoothed with the adaptive kernel technique \citep{2009MNRAS.394.1249C}.
Figure~\ref{fig:Method_Vsigma_example} shows the similarity between the $V_{*}/\sigma_{*}(R)$ profile (dashed green line) with the rotational support derived from the smoothed velocity fields (Eq.~\ref{eq:particlesVSvelfields}, open circles) for one example galaxy.

\subsection{Rotational support of the in-situ and ex-situ components}
\label{sec:measuring_rotational_support_insit_exsitu}

We derive the rotational support for the in-situ ($V_{\rm insitu}(R)$) and the ex-situ stars ($V_{\rm exsitu}(R)$)  using Eq.~\eqref{eq:V_2D} for the two components separately. We then normalize the mean velocities by the velocity dispersion of the total stellar component $\sigma_{*}(R)$, although we note that the results of Sect.~\ref{sec:Vsigma_profiles} are unchanged if we normalized by the respective $\sigma_{\rm insitu}(R)$ and $\sigma_{\rm exsitu}(R)$. For all three components we use elliptical radial bins with identical edges in each galaxy. 

In the same radial bins we derive the surface mass density of all the stellar particles $\Sigma_{*}(R)$, of the in-situ stars $\Sigma_{\rm insitu}(R)$, and of the ex-situ stars $\Sigma_{\rm exsitu}(R)$. From these, we define the local in-situ and ex-situ fractions as $f_i(R)=\Sigma_i(R)/\Sigma_{*}(R)$ with $i\in\{{\rm insitu, exsitu}\}$. Since each star is either classified as in-situ or ex-situ, $M_{*} = M_{\rm insitu} + M_{\rm exsitu}$ and $f_{\rm insitu}(R)+f_{\rm exsitu}(R) =1$.

By approximating $V_{*}/\sigma_{*}(R) \sim \big{|}U_{*,Z}(R)/\sigma_{*}(R)\big{|}$, we find that the $V/\sigma(R)$ profile of the galaxies are almost exactly approximated by the sum of the rotational support of the in-situ and ex-situ components, weighted by the their local mass fraction $f_{i}(R)$ profiles

\begin{equation}
\begin{aligned}
   & V_{*}/\sigma_{*}(R) \sim \big{|} U_{*,Z}(R)\big{|} /[\sigma_{*}(R)] \sim \\
   &\sim \big{|} \left[f_{\rm insitu}(R)\, U_{{\rm insitu},Z}(R) + f_{\rm exsitu}(R)\, U_{{\rm exsitu},Z}(R)\right] \big{|} / \sigma_{*}(R).
\label{eq:Vsigma_decomp_inex}
\end{aligned}
\end{equation} 
The $U_{i,Z}$ components need to be summed with their sign, as it is not uncommon that the in-situ and the ex-situ stars counter-rotate ($\sim20\%$ of the cases for our sample). We assume a positive sign for the component rotating with the same sign as the total $U_{*,Z}$ at its maximum rotation.

The approximation $V_{*}\sim \big{|} U_{*,Z}\big{|}$ holds for the majority of the selected TNG galaxies, for which the $\mathrm{median}(\big{|}U_{*,X}/U_{*,Z}\big{|})= 0.04$. For the in-situ and ex-situ components separately the $U_{i,X}$ component is also negligible in most of the cases (${\rm median}\left(\big{|}U_{{\rm insitu},X}/U_{{\rm insitu},Z}\big{|}\right)=0.04$ and ${\rm median}\left(\big{|}U_{{\rm exsitu},X}/U_{{\rm exsitu},Z}\big{|}\right)=0.14$) but not at large radii and for high mass systems (e.g., at $R=8\re{}$ for $M_{*}>10^{11.5}\MSUN{}$, ${\rm median}\left(\big{|}U_{*,X}/U_{*,Z}\big{|}\right)=0.45$).

Figure~\ref{fig:Method_Vsigma_example} shows as an example a massive FR galaxy with $M_{*}=10^{11.4}\MSUN{}$. Its $V_{*}/\sigma_{*}(R)$ profile, derived from the ratio of Eqs.~\ref{eq:V_2D} and \ref{eq:sigma_2D} and shown with a green dashed curve, has a peak at $R\sim2\re{}$, it decreases to 0 at $\sim5\re{}$, and it increases again at larger radii. The dashed black curve shows that, for this galaxy, the approximation of the mean velocity $V_{*}(R)$ with the mean major axis rotation $|U_{*,Z}(R)|$ holds at all radii. The gray curve shows the weighted sum of the major axis mean rotation of the in-situ and ex-situ stars divided by the total velocity dispersion, as in Eq.~\eqref{eq:Vsigma_decomp_inex}: solid green, dashed black, and solid gray curves agree almost perfectly along the whole radial range considered. The rotational support of the in-situ and ex-situ stars is shown by separating the major ($U_{i,Z}(R)$) and minor axis ($U_{i,Y}(R)$) rotation contributions. For this galaxy the contribution from minor axis rotation is negligible at all radii for the in-situ component and out to $\sim8\re{}$ for the ex-situ component. 
The in-situ stars have a "peaked-and-outwardly-decreasing" rotation profile (solid red curve), while the ex-situ stars have a co-rotating central component and a counter-rotating component at large radii contributing to almost all the galaxy rotational support at $R>5\re{}$.

\section{The accretion histories of ETGs in TNG100}\label{sec:history}

We begin by studying the dependence of the main accretion parameters on stellar mass for the selected sample of ETGs. 
For each galaxy we consider the total fraction of in-situ stars, the mean accreted "cold" (star forming) gas fraction weighted by the mass of the merged galaxies at all epochs, the fraction of in-situ stars produced since $z=1$, and the redshift of its last major merger (i.e. with mass ratio $\mu>1/4$);
see Sect.~\ref{sec:measuring_accretion}.

\begin{figure}[ht]
    \centering
    \includegraphics[width=1.\linewidth]{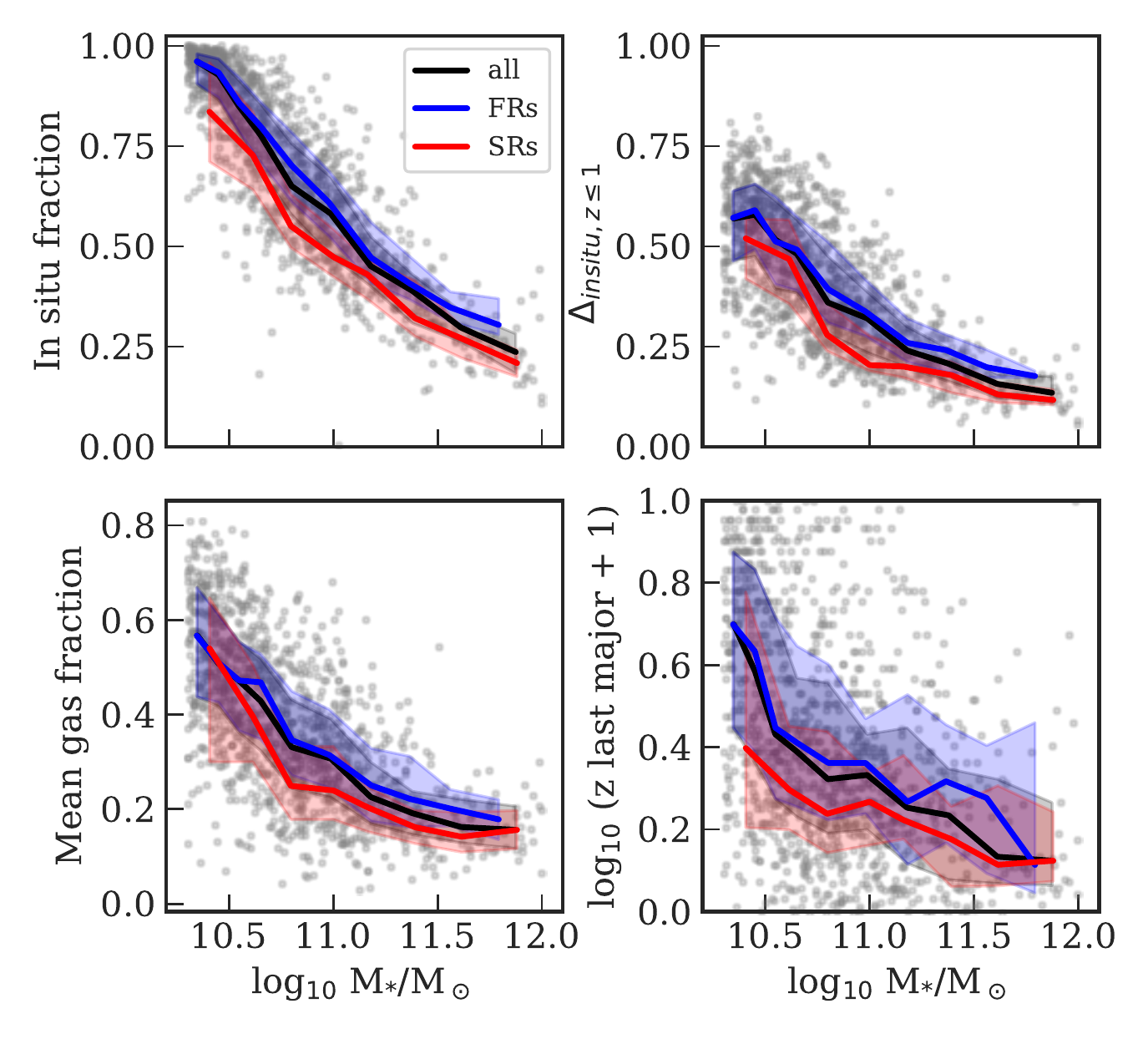}
    \caption{Accretion parameters as a function of stellar mass for ETGs in TNG100, and separately for FRs and SRs: total in-situ mass fraction (\textbf{top left}), fraction of new in-situ stars since $z=1$ (\textbf{top right}), mean accreted gas fraction (\textbf{bottom left}), and redshift of the last major merger (\textbf{bottom right)}. The solid lines show the median profiles for the whole sample (black) and for FRs (blue) and SRs (red). The shaded regions show the quartiles of the distributions. FRs and SRs follow similar trends with $M_{*}$ but, at fixed $M_{*}$, SRs have on average more accreted stars, they accreted less gas, and they had more recent major mergers.}
    \label{fig:trendswithmass}
\end{figure}

Figure \ref{fig:trendswithmass} shows the distribution of these parameters for all the sample galaxies and for the FRs and SRs separately.
We find a strong dependence with stellar mass in that more massive galaxies have lower in-situ fractions, had recent major mergers, and on average the accreted satellites were poorer in gas (dry mergers). 
The total ex-situ mass fractions for the IllustrisTNG were already presented by \citet{2018MNRAS.475..648P}. Their strong variation with stellar mass was observed there and in different sets of simulations (e.g \citealt{2016MNRAS.458.2371R} in the original Illustris, \citealt{2020MNRAS.tmp.1947D} in EAGLE), although with some differences due to the different galaxy formation models. 

FRs and SRs follow similar trends of the accretion parameters versus stellar mass $M_{*}$ in Figure \ref{fig:trendswithmass} but, at fixed $M_{*}$, the SRs have lower in-situ fractions and had more recent and drier mergers. This is consistent with previous studies on the relation between galaxy angular momentum, gas accretion, and morphology \citep[e.g.][]{2014MNRAS.444.3357N, 2017MNRAS.467.3083R, 2018MNRAS.473.4956L,  2019MNRAS.487.5416T}. 
 
Each relation comes with a fair amount of scatter since, at each stellar mass, galaxies are characterized by an individual history parametrized by a combination of parameters including the fraction of accreted stellar mass, when this mass has been accreted, and how much gas was involved in the accretion. 
This is shown by the comparison of the relations between the total in-situ mass fraction and the fraction of new in-situ stars since $z=1$ ($\Delta_{\rm insitu, z\leq1}$) with total stellar mass. Similar to the total in-situ fraction,
$\Delta_{\rm insitu, z\leq1}$ is a steep function of stellar mass. The most massive galaxies have little in-situ star formation overall and also at $z\leq1$, while galaxies at the low mass end with total in-situ fractions larger than $90\%$ and $\Delta_{\rm insitu, z\leq1}\sim0.6$ exhibit a large variety of recent in-situ star formation histories (the scatter in $\Delta_{\rm insitu, z\leq1}$ is $0.2$ while the scatter in in-situ mass fraction is $0.07$), also influenced by the different timing of their gas accretion and star formation.

In addition to the considered accretion parameters, the orbital parameters of the mergers also play a role in determine the properties of the remnant \citep{2018MNRAS.473.4956L}. This additional investigation is outside of the scope of this study.

\begin{figure}
    \centering
    \includegraphics[width=1.\linewidth]{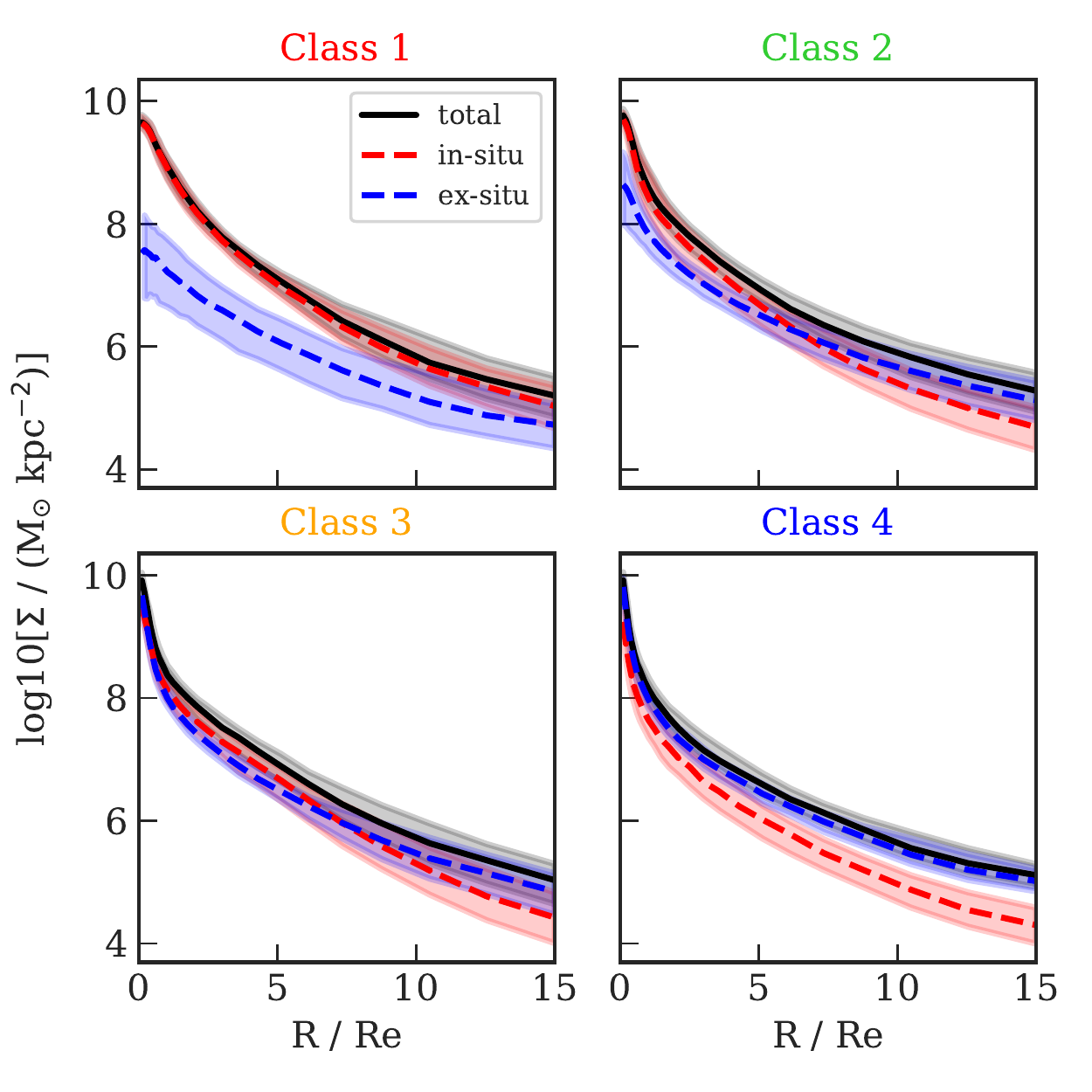}
    \includegraphics[width=\linewidth]{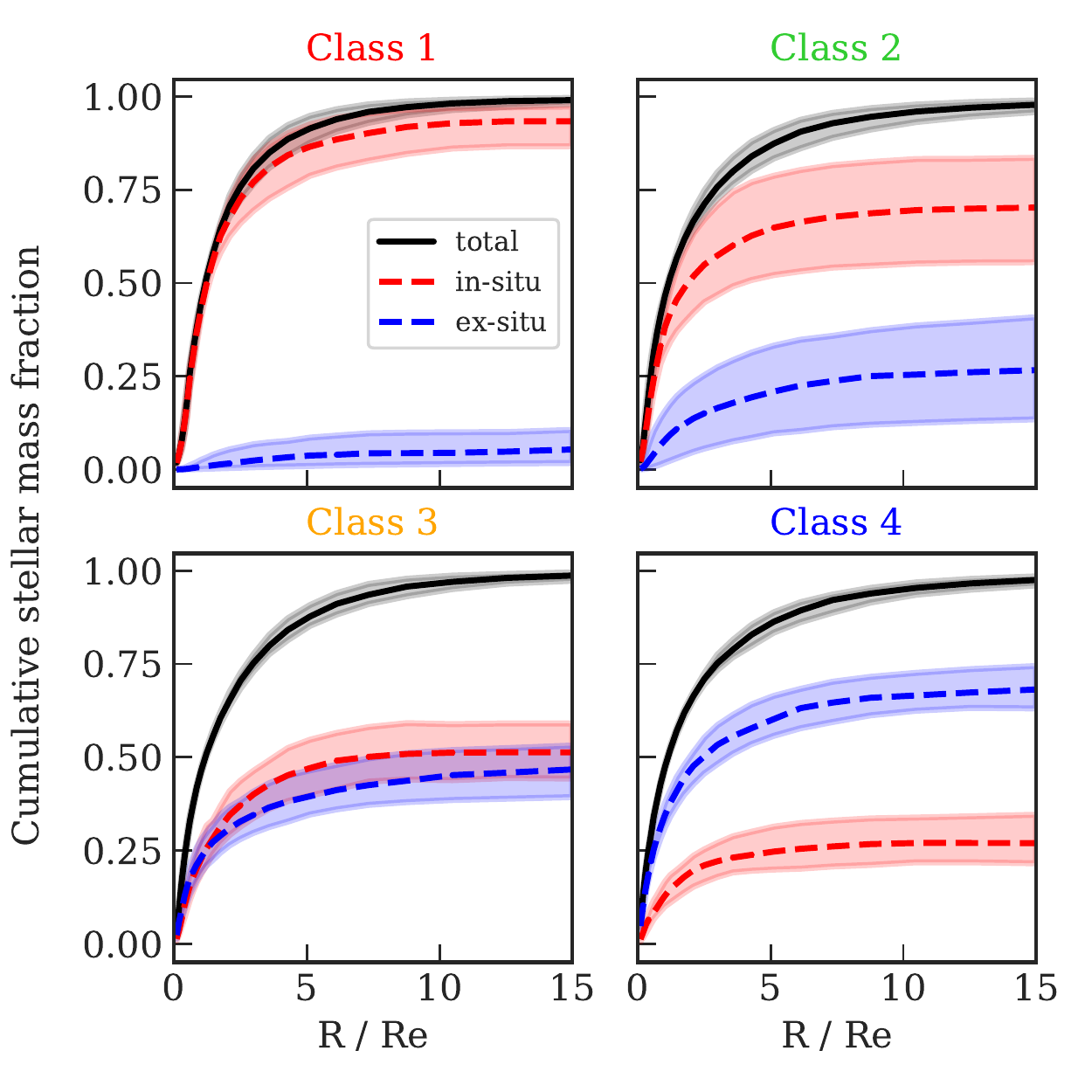}
    \caption{Accretion classes for the simulated sample of ETGs. \textbf{Top}: Median stellar mass density profiles for each of the four classes defined in Sect.  \ref{sec:accretion_classes}. The black lines show the median total density profiles, the red and blue lines show the median density profiles of the in-situ and the ex-situ stars respectively. The shaded regions show the quartiles of the distributions. 
    \textbf{Bottom}: Median cumulative stellar mass fraction profiles for the in-situ and ex-situ components in the four accretion classes. The shaded regions show the quartiles of the distributions.
    }
    \label{fig:definition_classes}
\end{figure}

\begin{figure}
    \centering
    \includegraphics[width=\linewidth]{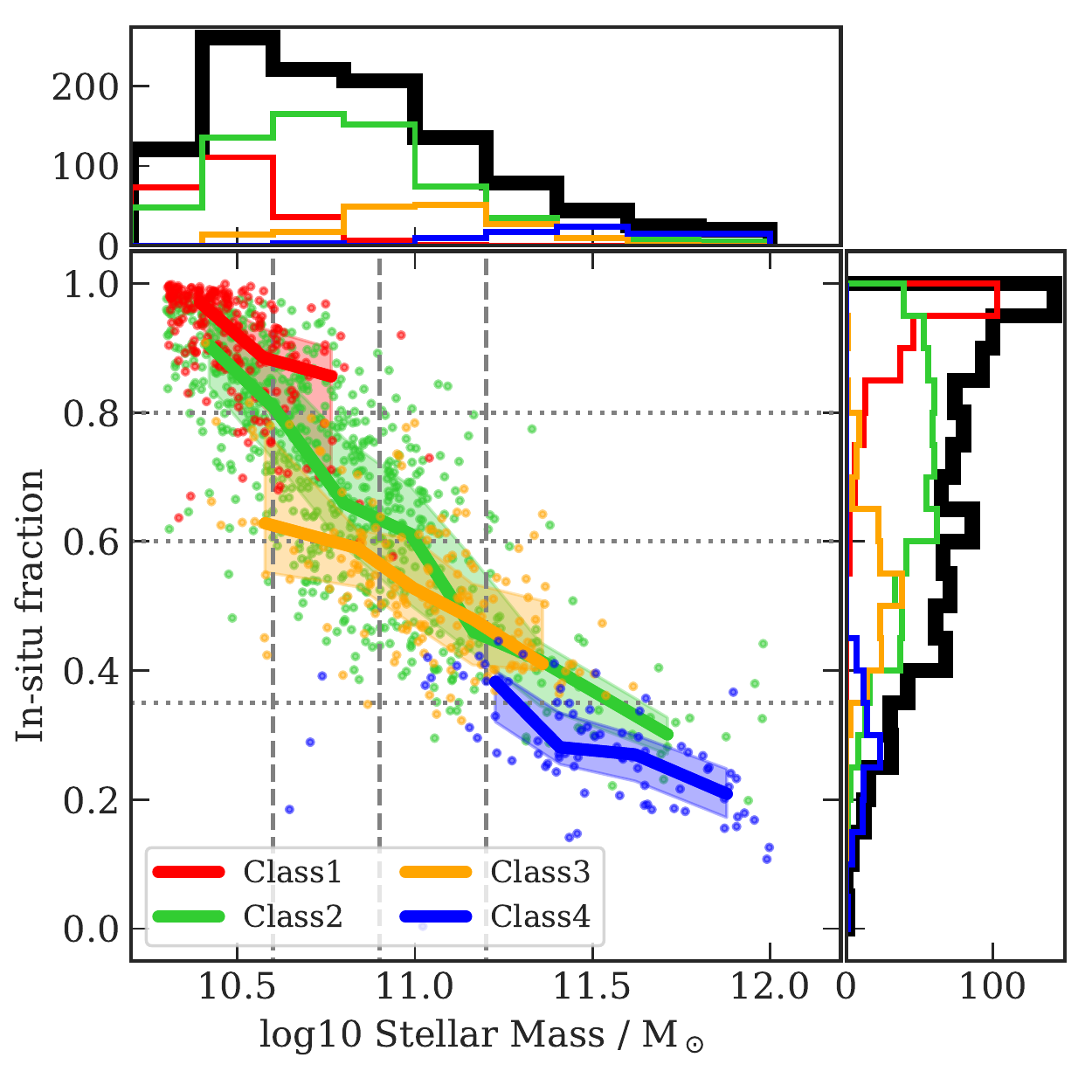}
    \caption{Total in-situ mass fraction as a function of the stellar mass for the four accretion classes. Solid lines show the median profiles for each class and the shaded regions the 25th - 75th percentiles.
    Dashed vertical lines show the stellar mass bins considered in the text. The dotted horizontal lines show the bins in in-situ mass fraction. Each accretion class dominates at a different stellar masses and is characterized by different total in-situ mass fractions.
    }
    \label{fig:insituVSmass_classes}
\end{figure}

\subsection{Accretion classes}
\label{sec:accretion_classes}
        
For each simulated galaxy we derive edge-on projected stellar mass density profiles in elliptical radial bins with flattening given by the ellipticity profile of the galaxy. In each bin we obtain the stellar mass density of all the stars $\Sigma_{*}(R)$ and of the in-situ and ex-situ stars ($\Sigma_{\rm insitu}(R)$ and $\Sigma_{\rm exsitu}(R)$). 
Within the selected TNG galaxies we can identify four groups, or accretion classes, according to their relative radial distribution of ex-situ stars with respect to the in-situ component. The assignment of each galaxy to an accretion class is done by analysing the sign of the  function $\Sigma_{\rm insitu}(R) - \Sigma_{\rm exsitu}(R)$ and the positions of its zeros. The top panel of Fig.~\ref{fig:definition_classes} displays the median stellar mass density profiles of all the galaxies in each accretion class. The median cumulative stellar mass fraction profiles are also shown in the bottom panels. The median stellar mass density and cumulative stellar mass fraction profiles are available in tabulated form in Appendix~\ref{appendix:tables}. Figure~\ref{fig:insituVSmass_classes} shows that each of the defined accretion classes dominates in a different interval of stellar mass and is characterized by different accreted fractions.

\textbf{Class 1} - Galaxies in class1 are in-situ dominated at all radii, and represent 20\% of the selected galaxies. More than 53\% of the low mass galaxies ($M \leq10^{10.5}\MSUN{}$) belong to this group.

\textbf{Class 2} - Galaxies in this class are dominated by the in-situ stars in the central regions, and by the ex-situ stars in the outskirts. For these galaxies it is possible to define a transition radius $R_{\rm exsitu}$, as the galactocentric distance at which the ex-situ stars dominate over the in-situ \citep{2013MNRAS.434.3348C, 2016MNRAS.458.2371R}. These galaxies represent 57\% of the selected TNG100 galaxies, most of them with intermediate stellar masses. 

\textbf{Class 3} - Class 3 contains all the galaxies where the in-situ and ex-situ stars interchange dominance at different radii ($\sim14\%$ of the total sample), and galaxies with ex-situ core and in-situ outskirts ($\sim1\%$ of the total sample). 

\textbf{Class 4} - Galaxies in this group are dominated by ex-situ stars at all radii. This group constitutes circa $8\%$ of the sample, although at stellar masses $M_{*}\geq10^{11.5}\MSUN{}$ nearly $64\%$ of the galaxies are in class 4. 

Since different accretion classes dominate at different stellar masses and the in-situ mass fraction tightly correlates with the stellar mass (Fig. \ref{fig:insituVSmass_classes}), measuring the stellar mass of an ETG gives a good first prediction for its accretion properties.

We note that in the original Illustris simulation the great majority (85\%) of the galaxies with $M_{*}\geq10^{10.3}\MSUN{}$ belong to class 2, that is, they have segregated in-situ dominated central regions and ex-situ dominated outskirts \citep[see also][]{2016MNRAS.458.2371R}. In TNG100 60\% (57\%) of all (ETG) galaxies with $M_{*}\geq10^{10.3}\MSUN{}$ are in class 2. These differences between the distributions of in-situ and ex-situ stars in galaxies from the two simulations are likely due to the changes implemented in the feedback model for TNG100, as discussed by \citet{2019MNRAS.487.5416T}.

\subsection{Local in-situ and ex-situ density fractions and the contributions from different mass-ratio mergers}
\label{sec:accretion_classes_fprofiles}

\begin{figure}
    \centering
    \includegraphics[width=1.\linewidth]{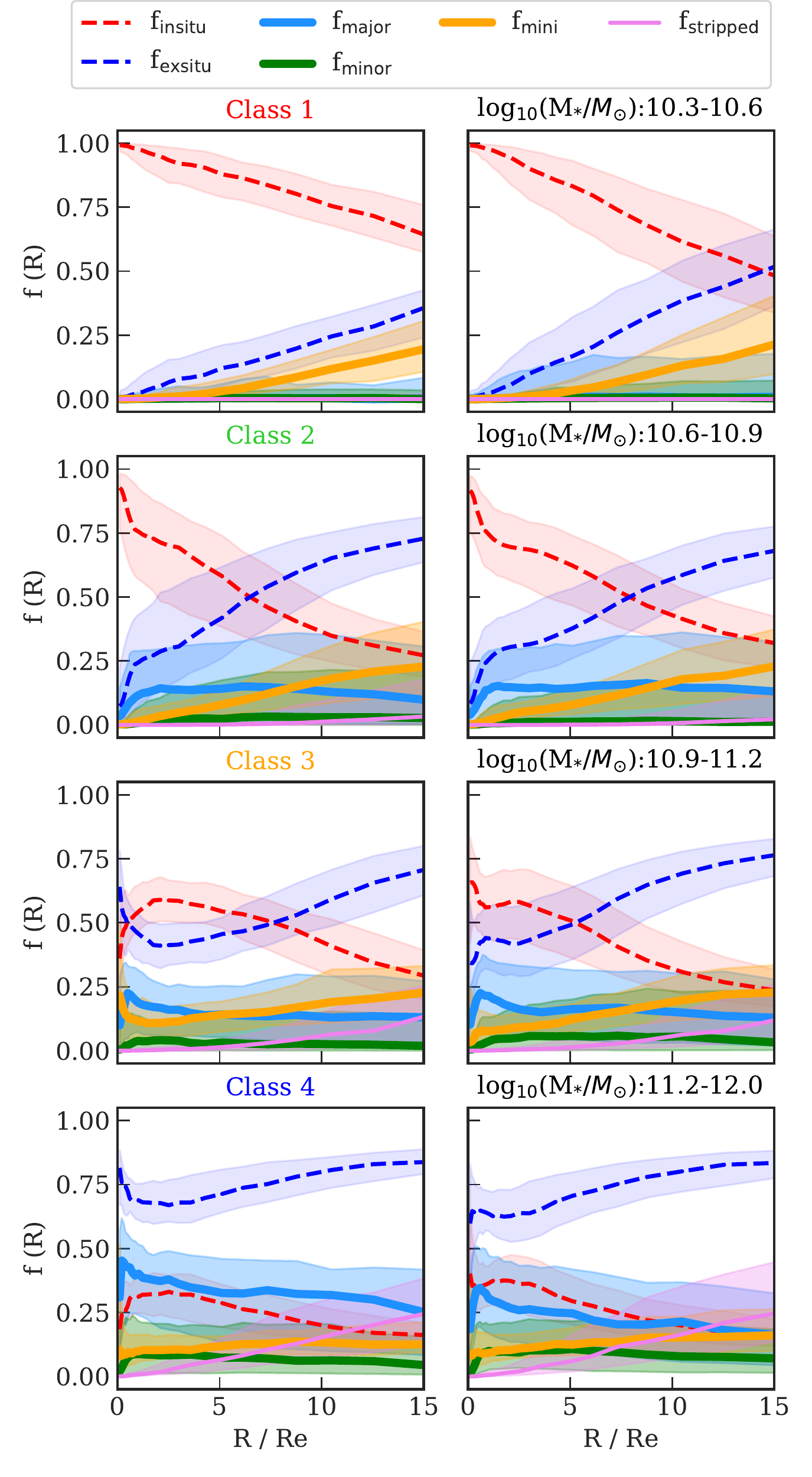}
    \caption{Local median mass fraction $f(R)$ of the in-situ and ex-situ stars (dashed lines). The ex-situ component is then split according to its origin from major, minor, mini mergers, or from stripped galaxies (solid lines, see legend). Galaxies are shown divided in accretion classes (left column) and in stellar mass bins (right column). The spatial distributions of the in-situ and ex-situ components, as well as of the accreted stars contributed by different kinds of mergers, vary for different accretion classes and stellar mass bins.}
    \label{fig:f_profiles}
\end{figure}

From the surface density profiles of the in-situ and ex-situ stellar components we can define the local in-situ and ex-situ fractions $f_i = \Sigma_{i}(R) / \Sigma_{*}(R)$ with $i\in \{ {\rm insitu, \,exsitu} \}$.
We tagged the ex-situ stars according to their origin (i.e. whether they where stripped or accreted through mergers of different $\mu$, see Sect.~\ref{sec:measuring_accretion}) and derived the local fractions of the ex-situ stars accreted from the different merger classes from the corresponding edge-on projected density profiles $f_{j} (R)= \Sigma_{j}(R) / \Sigma_{*}(R)$ with $j\in \{ {\rm major, \, minor, \, mini, \, stripped} \}$, such that $\sum_{j} f_{j} = \fex{}$. 
Figure \ref{fig:f_profiles} shows radial profiles of the median local fraction of in-situ stars $f_{\rm insitu}(R)$, ex-situ stars $\fex{}(R)$, and of the accreted stars contributed by different mass ratio mergers. Galaxies are divided according to their accretion class (left column) and in stellar mass bins that best "isolate" the accretion classes (as shown in Fig.~\ref{fig:insituVSmass_classes} with vertical dashed lines). 

Fig.~\ref{fig:f_profiles} shows how the spatial distributions of these stellar components differ between the accretion classes and stellar mass bins. For lower mass galaxies the accreted stars represent less than $10\%$ of the total $M_{*}$. These come mainly from mini mergers and are distributed at large radii, determining growing $\fex{}(R)$ profiles. At progressively higher stellar masses the accreted mass fraction increases. The local fraction from major mergers $f_{\rm major}(R)$ is overall higher at all radii and dominates in the central regions over the other components. The $f_{\rm mini}(R)$ profiles instead tend to become flatter at increasing $M_{*}$. 
The median $f_{\rm minor}(R)$ profiles show that, on average, the fraction of stellar mass contributed by minor mergers is much lower than that from mini and major mergers. This is driven by the fact that in each mass bin there is a significant fraction of galaxies in which the accreted satellites with mass ratio $1/10<\mu<1/4$ contribute less than 10\% of the total accreted mass. 
This can also be seen in the right panels of  Fig.~\ref{fig:accretion_classes_merger} where we show the median mass fraction from different components as a function of the total stellar mass.

On the left column of Fig.~\ref{fig:f_profiles} we can see again, in a different form as Fig.~\ref{fig:definition_classes}, how the different accretion classes are defined in terms of their $f_{\rm insitu}(R)$ and $\fex{}(R)$ profiles.  The median $f_{j}(R)$ (with $j\in \{ {\rm major, \,minor, \,mini, \, stripped} \}$) profiles follow roughly the same behavior as the corresponding stellar mass bin, considering that class 2 galaxies span a wide range of stellar masses (from $10^{10.4}\MSUN{}$ to above $10^{11.2}\MSUN{}$).
We observe that most of the galaxies in class 3 have centrally peaked $f_{\rm mini}(R)$ profiles. These could be explained
by the accretion of small compact satellites that are able to reach the central regions of the galaxies before disrupting \citep[e.g.,][]{2017MNRAS.464.2882A}. 

The overall trends of the median $f_{\rm insitu}(R)$ and $\fex{}(R)$ profiles with stellar mass are consistent with the results of \citet{2016MNRAS.458.2371R} for the original Illustris galaxies, although with some quantitative differences as already noted. The median accreted mass fraction from all mergers is larger in IllustrisTNG than in Illustris for all galaxies with $M_{*}\geq10^{10.3}\MSUN{}$ \citep[Fig.~10 in ][]{2019MNRAS.487.5416T}. However, in both simulations the median fraction of stellar mass contributed by minor mergers ($\Delta M_{\rm *,minor}/M_{*}$) is consistently lower than those from major ($\Delta M_{\rm *,major}/M_{*}$) or mini mergers ($\Delta M_{\rm *,mini}/M_{*}$) \citep[see Fig.~\ref{fig:accretion_classes_merger} above and Fig.~4 in][]{2016MNRAS.458.2371R}.

\subsection{Accretion classes and merger histories}
\label{sec:accretion_classes_mergerhistory}

\begin{figure}
    \centering
    \includegraphics[width=1\linewidth]{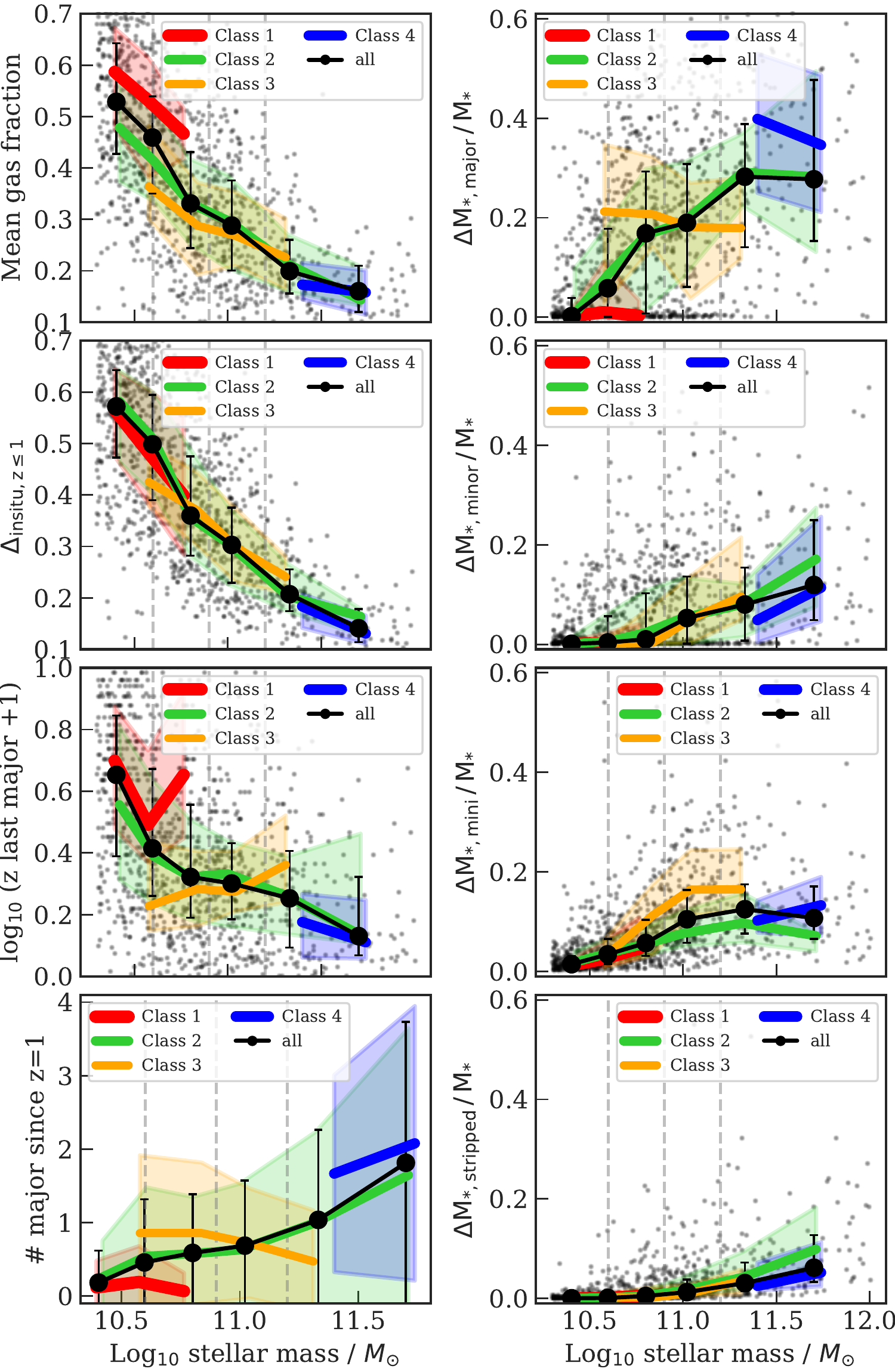}
    \caption{Merger history in different accretion classes. \textbf{Left:} Median accretion history parameters as a function of the stellar mass. \textbf{Right:} Median mass fractions from different mass ratio mergers and from stripped galaxies as a function of the stellar mass. Shaded regions or error-bars show the 25th-75th percentiles of the distributions. In the bottom left panel (number of major mergers since $z=1$) we show mean profiles and standard deviations around the mean. Galaxies are divided in the accretion classes as shown in the legend. The accretion classes defined in Sect.~\ref{sec:accretion_classes} are the results of different merger histories.}
    \label{fig:accretion_classes_merger}
\end{figure}

In Sect.~\ref{sec:accretion_classes_fprofiles} we observed that galaxies of different stellar mass or, almost equivalently, different accretion classes distribute their in-situ and accreted stars differently. In this section we show that these are the result of different merger histories. 

The left column of Fig.~\ref{fig:accretion_classes_merger} shows the different accretion history parameters, defined in Sect.~\ref{sec:measuring_accretion} and specified by in the headings of the plot, as a function of the stellar mass. On the right column is the stellar mass fraction of the stars accreted from different mass ratio mergers and from stripping events as a function of the total stellar mass.  

Galaxies of \textbf{class 1} had few major mergers at early times (high $z_{last}$). These contributed to few ex-situ stars, compared to the total stellar mass at $z=0$, but brought in a large fraction of gas which contributed to the in-situ star formation. Their recent history (after $z\sim2$) is relatively more quiet compared to galaxies of class 2 with similar $M_{*}$: they accreted very little and most of their ex-situ stars comes from mini mergers which are deposited in the outskirts (Fig.~\ref{fig:f_profiles}). 

\textbf{Class 2} contain a large variety of objects. The least massive are similar to class 1 galaxies, with early gas rich major mergers and outskirts enriched with ex-situ stars from low mass (i.e. minor and mini) mergers, although by comparison their last major mergers happened more recently ($z_{\rm last}\sim2$ versus $z_{\rm last}\sim4$) and with less amount of gas (45\% versus 55\%). This implies a larger fraction of ex-situ stars from major mergers, as shown in the top right panel of Fig.~\ref{fig:accretion_classes_merger}. 
The more massive galaxies of class 2 have larger ex-situ fractions: these are contributed by both major and low mass mergers in similar amount at  $10^{10.5}<M_{*}/\MSUN{}<10^{11}$ ($\sim 15\%$ and $\sim 15\%$), but at higher stellar masses the contribution from major mergers dominates. Some of the most massive systems ($M_{*}\gtrsim10^{11.2}\MSUN{}$) had multiple recent ($z<1$) major mergers.

The ex-situ dominated \textbf{class 4} contains massive galaxies whose evolution is dominated by dry mergers (lowest mean gas fraction). Almost all of class 4 galaxies had recent ($z\leq1$) dry major mergers, half of them more than once. 
Compared to the most massive systems of class 2, the class 4 galaxies had a more gas-poor history, on average more recent and more numerous major mergers. 

\begin{figure}
    \centering
    \includegraphics[width=\linewidth]{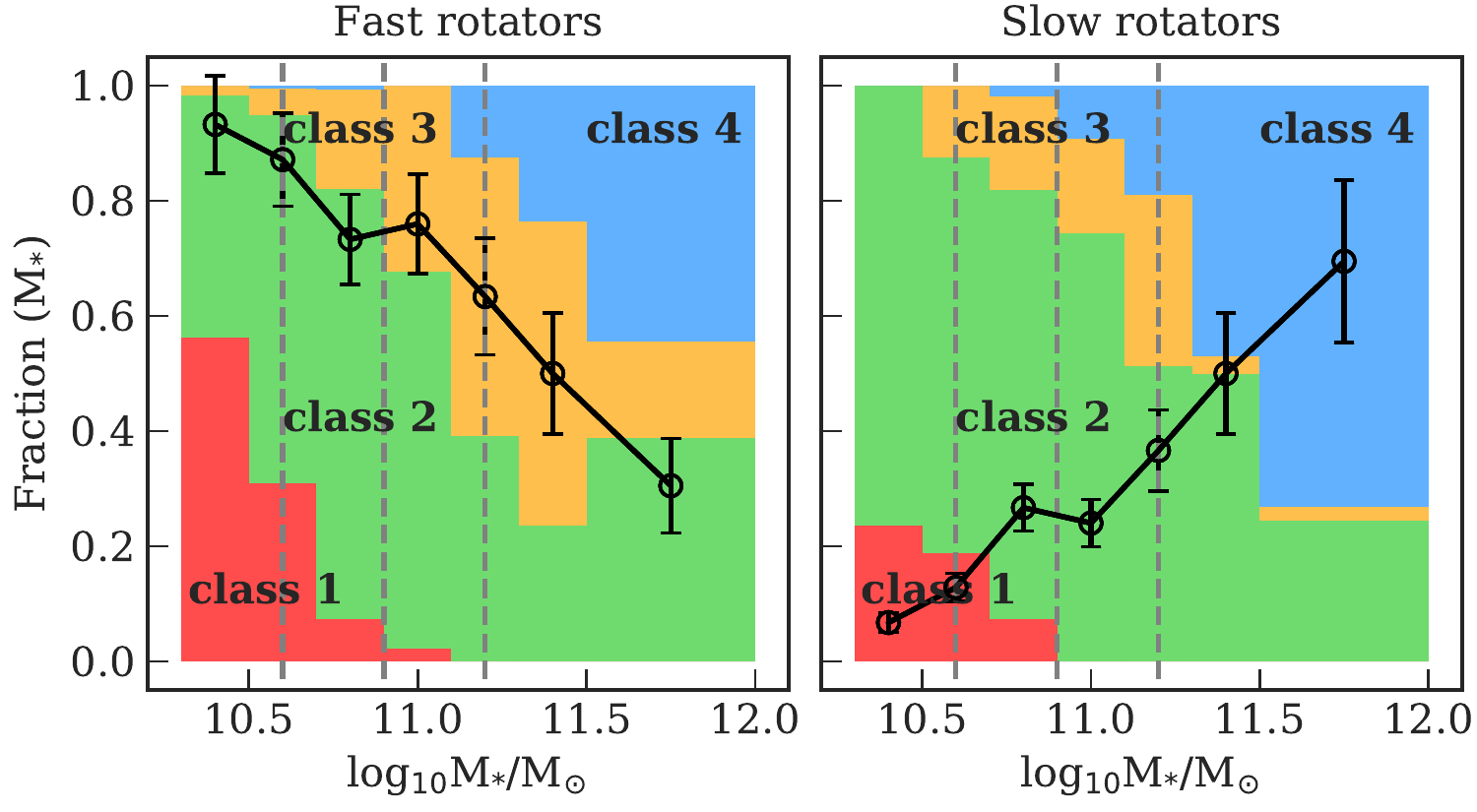}
    \caption{Fraction of FRs (\textbf{left}) and SRs (\textbf{right}) that populate each accretion class as a function of the stellar mass, represented with a cumulative bar plot. The open circles show the fraction of fast or slow rotators in each mass bin. The vertical dashed lines represent the stellar mass bins adopted in the paper. FRs and SRs populate the four accretion classes differently, highlighting the different accretion histories of the two classes, as well as the fact that different formation pathways can lead to the formation of a FR or a SR galaxy.}
    \label{fig:fraction_FRandSR_accretion_classes}
\end{figure}

\textbf{Class 3} collects objects with accreted fractions $\sim 0.5$, but with stellar masses ranging from $10^{10.5}$ to  $10^{11.5} \MSUN{}$, and very different accretion histories. The galaxies with ($M_{*}\lesssim10^{11.2} \MSUN{}$) are recent (at $z<1$) major mergers. These accretion events are more gas-poor compared to galaxies of class 2 with similar masses.
The high mass end of class 3 are galaxies which did not have recent major mergers. They had on average larger mean accreted gas fraction and higher fractions of newly formed in-situ stars compared to class 2 and 4 galaxies of similar stellar masses. We also note that, overall, class 3 galaxies have the highest fraction of stars accreted by mini mergers. 81\% of class 3 galaxies are fast rotators.

As shown in Fig.~\ref{fig:fraction_FRandSR_accretion_classes}, FRs and SRs populate all four accretion classes but with very different relative fractions: only 3\% of all the FRs populate class 4 and only 5\% of the SRs are in class 1. FRs with high $M_{*}>10^{11.2}$ divide between 33\% belonging to class 2,  41\% to class 3, and 25\% to class 4. Massive SRs belong almost exclusively to class 4 (57\%) and class 2 (37\%).
This underlines both the importance of (dry) mergers in the formation of SRs, as well as the fact that different formation pathways can result in a fast or a slow rotating galaxy in agreement with previous studies \citep[e.g.][]{2014MNRAS.444.3357N, 2017MNRAS.468.3883P}.

\section{$\mathrm{V/\sigma}(R)$ profiles and accretion history} \label{sec:Vsigma_profiles}

\begin{figure*}
    \centering
    \includegraphics[width=1.\linewidth]{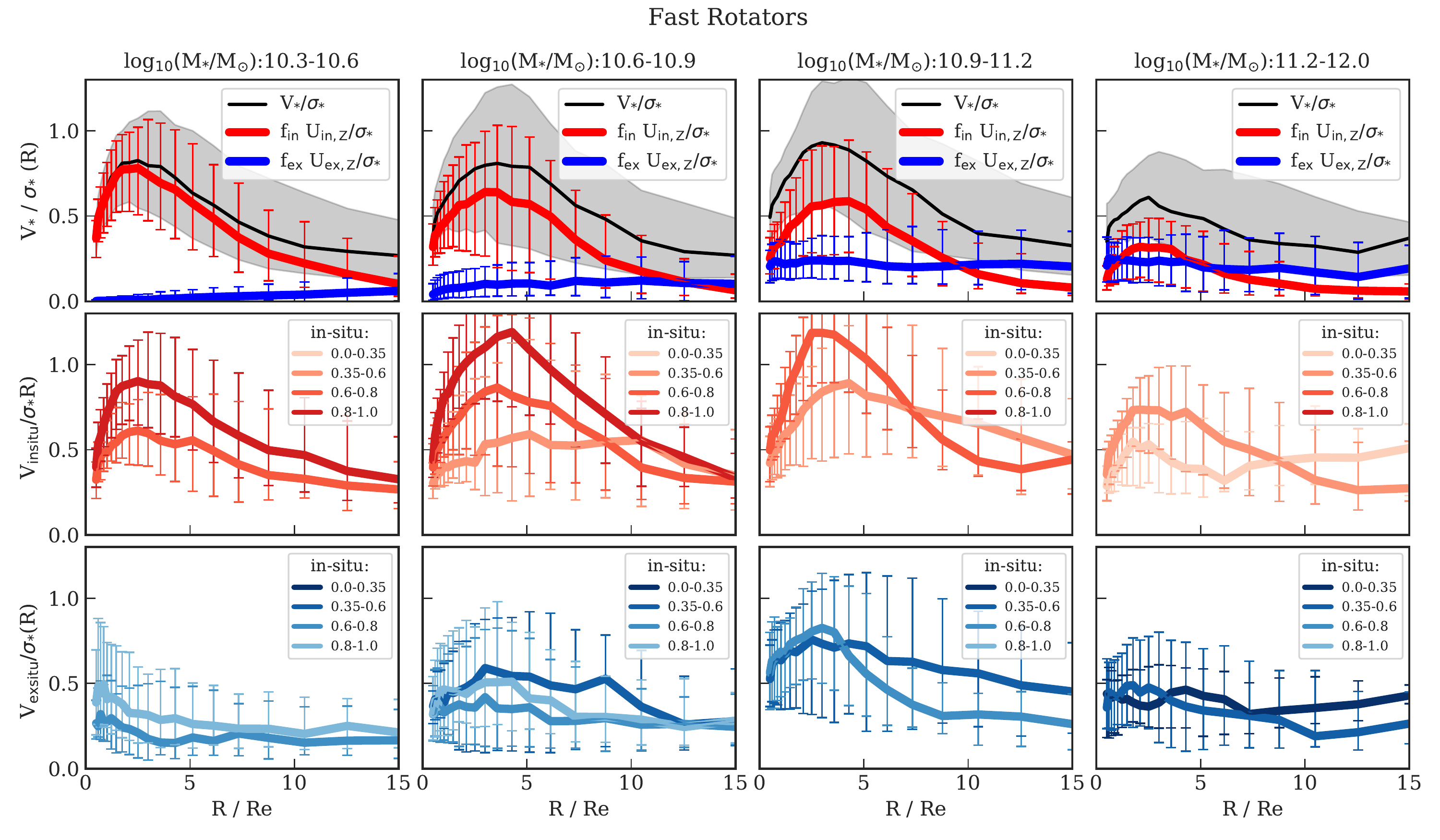}
    \includegraphics[width=1.\linewidth]{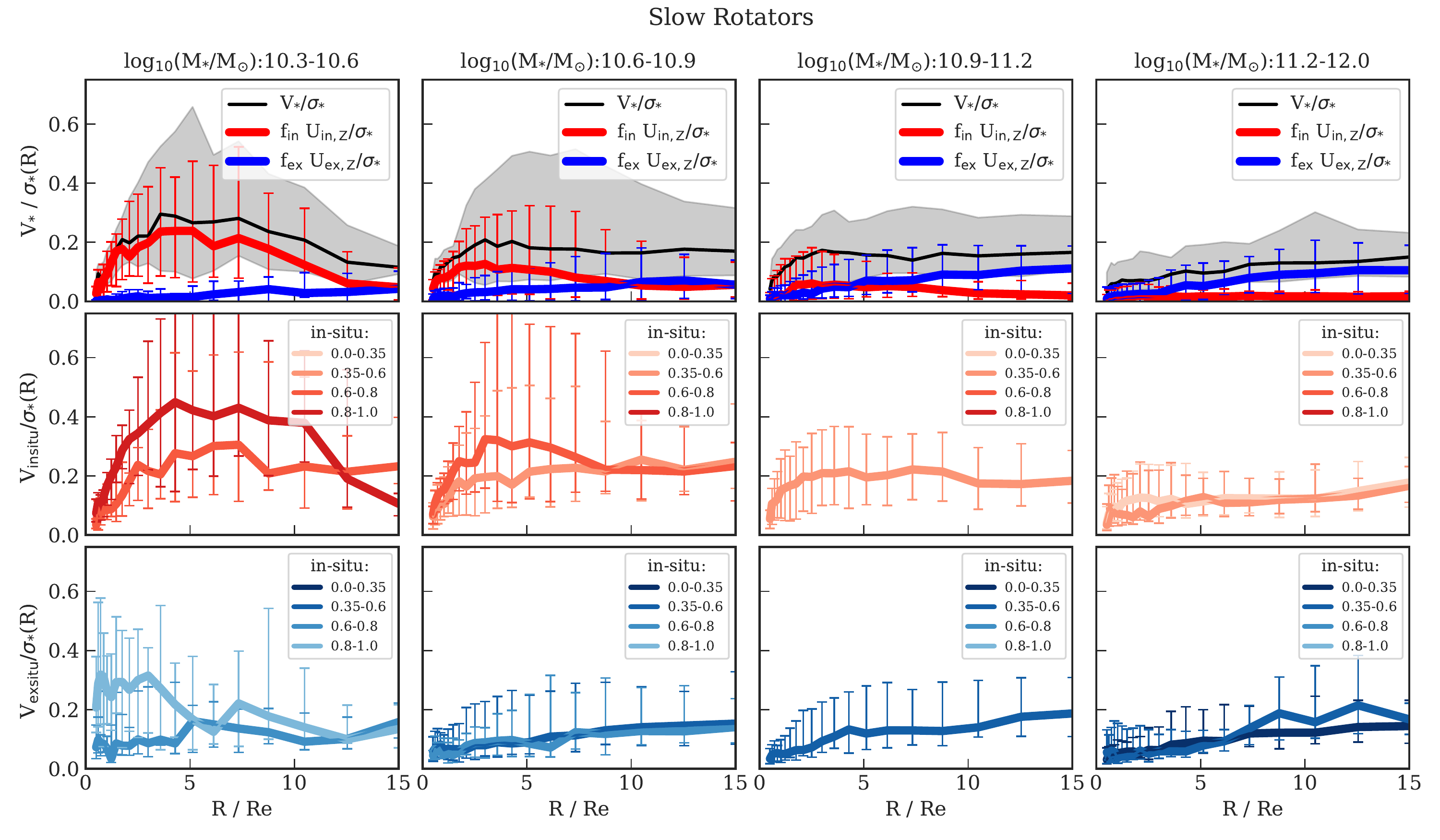}
    \caption{$V/\sigma(R)$ profiles for the TNG FRs (\textbf{top panels}) and SRs (\textbf{bottom panels}) divided in mass bins. In the two sets of panels, the \textbf{first rows} show the median $V_{*}/\sigma_{*}$ profiles of all the stars in black. The red and blue curves are the median $f_{i}(R) U_{i,Z}(R)/\sigma_{*}(R)$ profiles for the in-situ and ex-situ stars, respectively. The sum of these median profiles approximate the median $V_{*}/\sigma_{*}(R)$ profiles up to a median absolute difference of $0.02$. The \textbf{middle} and \textbf{bottom} rows show the median $V_i/\sigma_{*}(R) = \big{|}\sqrt{U_{i,X}^2(R) + U_{i,Z}^2(R)}\big{|}/\sigma_{*}(R)$ profiles of the in-situ and ex-situ components, respectively. Here the galaxies have been divided according to their total in-situ fraction, as shown in the legend. 
    Error-bars and shaded regions trace the quartiles of the distributions. The in-situ stars are characterized by a "peaked-and-outwardly-decreasing" $V_{\rm insitu}/\sigma_{*}(R)$ profiles, while the ex-situ stars typically add a constant level of (subdominant) rotational support.}
    \label{fig:Vsigma_profiles}
\end{figure*}

In this section we study how galaxy rotational support depends on accretion history. We do not distinguish galaxies using the accretion classes themselves as defined in the last section, which can not be readily derived from observations, but we rather divide them in stellar mass bins that best match the different classes.

\subsection{$\mathrm{V/\sigma}(R)$ profiles and in-situ fractions}\label{sec:Vsigma_profiles_insitu}

We start by quantifying the galaxy rotational support using the edge-on projected $V_{*}/\sigma_{*}(R)$ profiles of the all the stars derived as in Sect.~\ref{sec:measuring_rotational_support}. 
Figure~\ref{fig:Vsigma_profiles} shows median  $V/\sigma_{*}(R)$ profiles in stellar mass bins for FRs and SRs in two separate arrays of figures. 
The top panels of Fig.~\ref{fig:Vsigma_profiles} show the median $V_{*}/\sigma_{*}(R)$ for the total stellar component. For the FRs, the median profiles  typically have a "peaked-and-outwardly-decreasing" shape while the SRs show on average a mild increase of rotational support at large radii. However, in each mass bin, galaxies come with a variety of profile shapes, from approximately constant with radius to steeply declining for the FRs, and more or less increasing with radius for the SRs. This is reflected in the large scatter around the median. 
As for the $\lambda(R)$ profiles \citep{2020A&A...641A..60P}, the shapes of the median $V_{*}/\sigma_{*}(R)$ profiles weakly depend on stellar mass. For the most massive galaxies the peak in the $V_{*}/\sigma_{*}(R)$ profiles of the FRs is nearly absent, and the median halo rotation ($R\sim8\re{}$) is also suppressed compared to lower mass galaxies (see \citetalias{2020A&A...641A..60P}).

In Sect.~\ref{sec:measuring_rotational_support_insit_exsitu} we showed that the total $V_{*}/\sigma_{*}(R)$ profiles can be well approximated by the weighted sum of the rotational support of the in-situ and ex-situ stars (Eq.~\eqref{eq:Vsigma_decomp_inex}). This is done by neglecting the contribution to the angular momentum from minor axis rotation, which is small in most cases. As already observed in \citetalias{2020A&A...641A..60P}, only a small fraction of the selected TNG ETGs displays large kinematic twist.
The red and blue curves in the top panels of the two arrays of figures in Fig.~\ref{fig:Vsigma_profiles} show the median in-situ and ex-situ contribution $f_{i}(R) U_{i,Z}(R) / \sigma_{*}(R)$, with $i\in\{{\rm insitu, exsitu}\}$. The median in each radial bin takes into account of the few systems with counter-rotating in-situ and ex-situ  components, namely with $U_{i,Z}(R)$ having opposite sign (Sect~\ref{sec:measuring_rotational_support_insit_exsitu}. The sum of these median profiles approximate the total median $V_{*}/\sigma_{*}(R)$ profile in each mass bin very well, within an absolute median difference of 0.02 (the absolute maximum difference is 0.12). 

The contribution from the in-situ stars to $V_{*}/\sigma_{*}(R)$ is on average a peaked profile, even in the case of the low mass SRs. By comparison the ex-situ stars have flat 
$\fex{}(R)U_{{\rm exsitu},Z}(R)/\sigma_{*}(R)$ profiles with radius and contribute little to the total rotation, except for the high mass ($M_{*}>10^{11.2}\MSUN{}$) FRs and for the intermediate-to-high mass ($M_{*}>10^{10.9}\MSUN{}$) SRs, so that in most of the sample galaxy rotation is driven by the in-situ stars.

We see a clear dependence on stellar mass of contribution of the in-situ and ex-situ components to the total profiles, due to the strong mass dependence of the weights $f_i(R)$ (Fig.~\ref{fig:f_profiles}) more than to a variation with mass of the rotational support of the individual components (see below). At progressively higher masses, the contribution from the in-situ stars to the total rotational support decreases, while that from the accreted component mildly increases, resulting in a median total $V_{*}/\sigma_{*}(R)$ that is almost independent of stellar mass for FRs with $M_{*}\lesssim10^{11.2}\MSUN{}$ and for SRs with $M_{*}\lesssim10^{10.9}\MSUN{}$. At larger stellar masses the rotation of the in-situ stars is almost erased and the in-situ contribution $f_{\rm insitu}(R)U_{{\rm insitu},Z}/\sigma_{*}(R)$ becomes comparable to that of the ex-situ component.

The middle and bottom rows of both arrays in Fig.~\ref{fig:Vsigma_profiles} show the median $V_{\rm insitu}/\sigma_{*}(R)$ and $V_{\rm exsitu}/\sigma_{*}(R)$ profiles, now defined as the ratio of Eqs.~\eqref{eq:V_2D} and \eqref{eq:sigma_2D} so that $V_{i}/\sigma_{*}(R)$ is always a positive quantity.
In each mass bin we divide galaxies according to their total in-situ mass fraction (bins as in Fig.~\ref{fig:insituVSmass_classes}). Galaxies with similar stellar mass have, on average, more peaked $V_{\rm insitu}/\sigma_{*}(R)$ profiles if their in-situ fraction is higher, so that the peak value of the $V_{\rm insitu}/\sigma_{*}(R)$ profiles is approximately proportional to the in-situ mass fraction. The maximum rotational support of the in-situ stars themselves only weakly depends on stellar mass, again except for high mass FRs and intermediate-to-high mass SRs.

The median $V_{\rm exsitu}/\sigma_{*}(R)$ are generally quite flat and independent of the in-situ (or, equivalently, on the ex-situ) mass fraction. While for the SRs the $V_{\rm exsitu}/\sigma_{*}(R)$ profiles seem also insensitive to stellar mass, in the FRs we find an overall increase in the mean rotation of the ex-situ stars with stellar mass. On the other hand the scatter on the $V_{\rm exsitu}/\sigma_{*}(R)$ profiles is much larger than the median variations among stellar mass bins, reflecting the variety of merger histories of the FRs at all stellar masses (see also Fig.~\ref{fig:fraction_FRandSR_accretion_classes} and Sect.~\ref{sec:accretion_classes_mergerhistory}).
We verified that the scatter around the median in-situ and ex-situ profiles does not reduce for different binning choices of the in-situ mass fraction.

In the FRs the typical peaked-and-outwardly-decreasing shape of the total $V_{*}/\sigma_{*}(R)$ profiles is due to the in-situ stars, while the ex-situ component sets a constant level of rotational support at large radii. Galaxies with negligible accreted mass also display "peaked-and-outwardly-decreasing" $V_{*}/\sigma_{*}(R)$ profiles (see also Fig.~\ref{fig:Vsigma_profiles_mergers}B). We have checked, although we do not show for brevity, that this is true for both central and satellite galaxies. 
An increasing fraction of ex-situ stars (or a decreasing in-situ fraction) is accompanied with progressive weakening of the peak and a flattening of the  $V_{\rm insitu}/\sigma_{*}(R)$ profiles at large radii. At the same time the rotational support of the ex-situ stars, $V_{\rm exsitu}/\sigma_{*}(R)$, is insensitive to variations in the ex-situ fraction. These results imply that in the TNG FRs the decrease of rotational support in the stellar halo is not simply explainable as the weighted sum of a rotating disk-like in-situ component and a dispersion dominated spheroidal ex-situ component. The kinematic transition into the spheroidal dispersion-dominated halo in the TNG FRs is driven mainly by the in-situ stars. In addition, Fig.~\ref{fig:Vsigma_profiles} indirectly shows that (major) mergers dynamically suppress rotation also in the in-situ component thereby determining the flattening of the $V_{\rm insitu}/\sigma_{*}(R)$ profiles and an overall decreased rotation from the centers $1-2\re{}$ to the outskirts ($\gtrsim8\re{}$). 

This conclusion is also supported by the comparison of the $\mathrm{V_{insitu}/\sigma_{tot}}(R)$ profiles between FRs and SRs in Fig.~\ref{fig:Vsigma_profiles}. At fixed stellar mass, the SRs have on average larger fractions of major mergers and a more recent, more gas-poor accretion history (Sect.~\ref{sec:history}). Therefore at fixed stellar mass and in-situ fraction, SR galaxies have distinctly flatter rotation profiles.

\subsection{$\mathrm{V/\sigma}(R)$ profiles and merger mass ratio}
\label{sec:Vsigma_merger_mass_ratio}

\begin{figure}[ht]
    \centering
    
    \xincludegraphics[width=1.\linewidth,label=\textbf{A}]{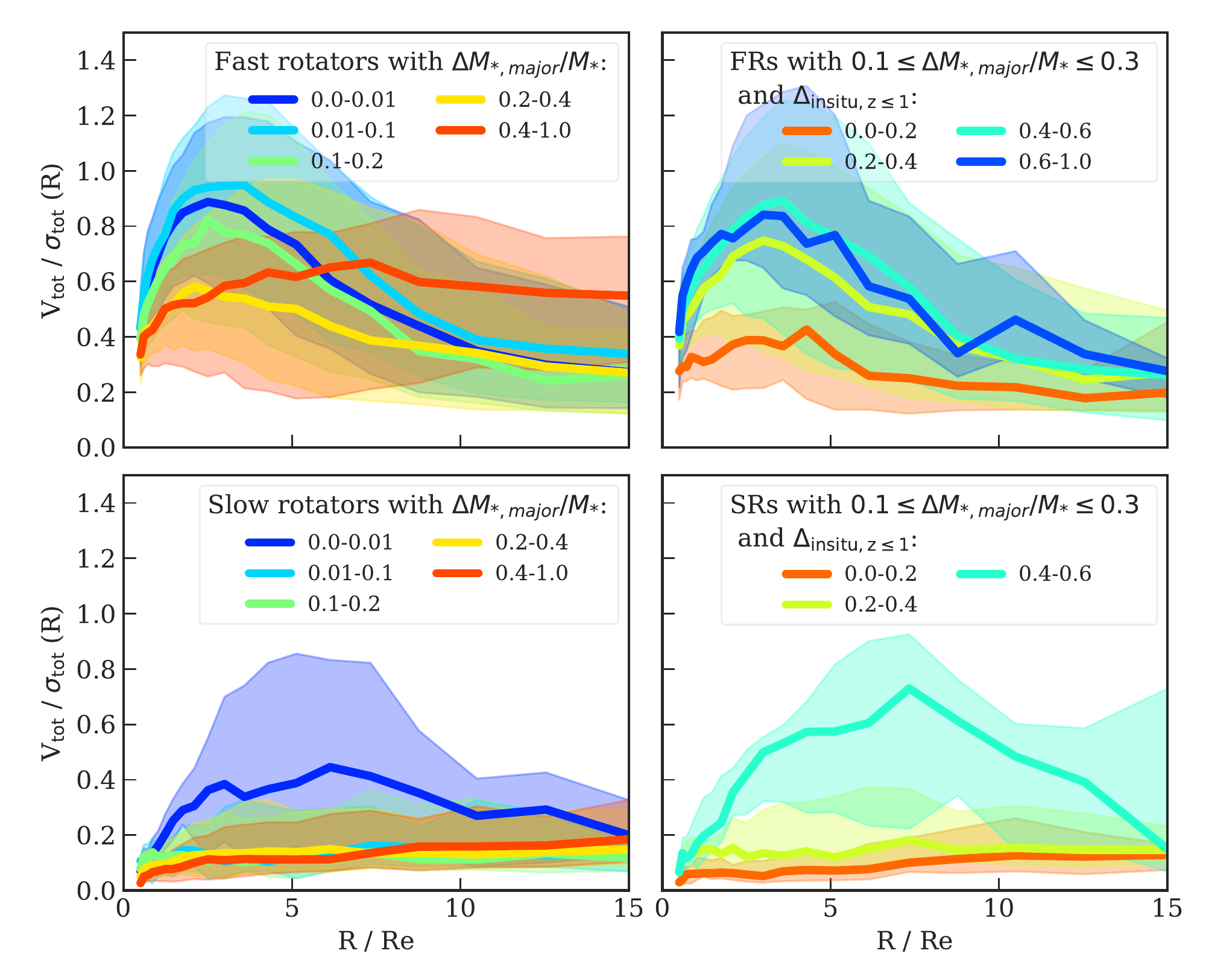}
    \xincludegraphics[width=1.\linewidth,label=\textbf{B}]{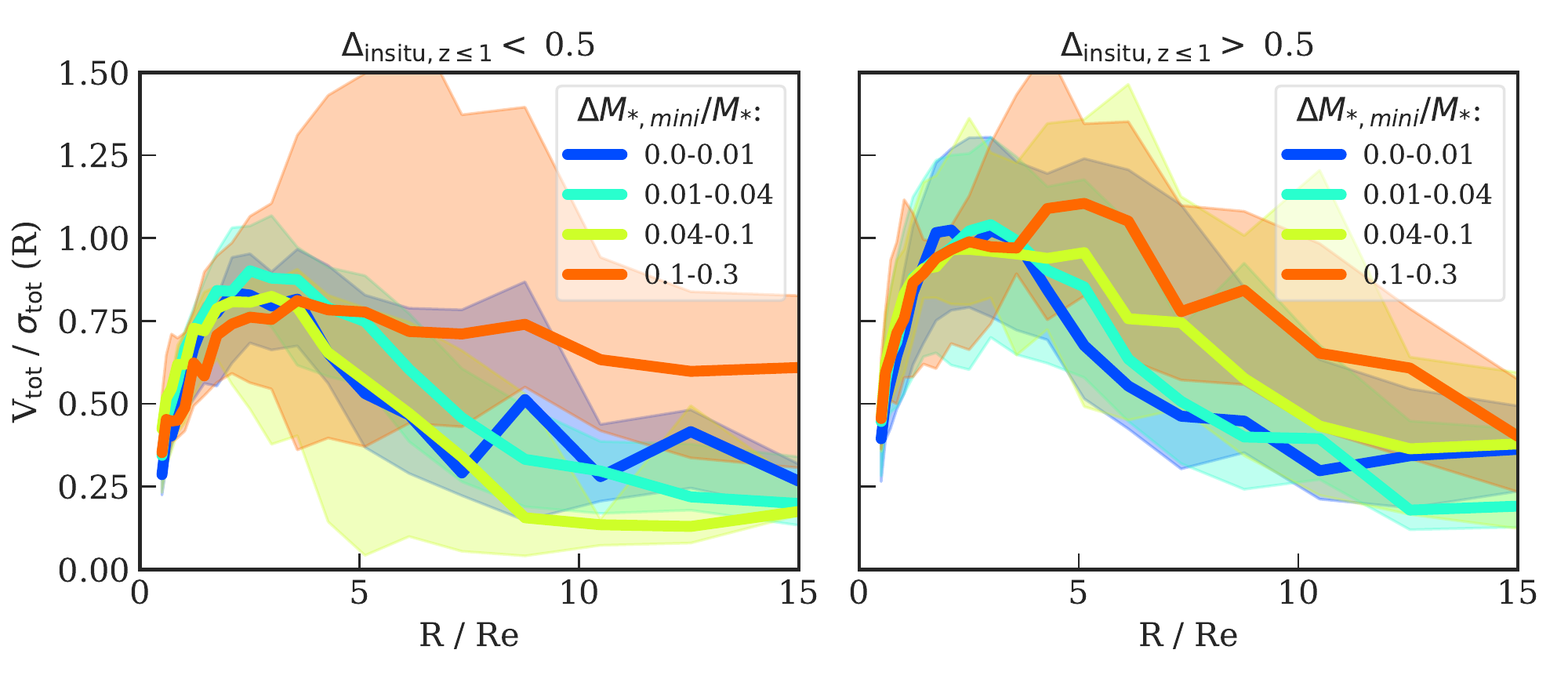}
    \caption{The effect of mergers on the $V_{*}/\sigma_{*}(R)$ profiles. \textbf{A}: Major mergers. Median profiles for the FRs (\textbf{top left}) and the SRs (\textbf{bottom left}) in bins of accreted mass fraction from major mergers $\Delta M_{\rm *,major}/M_{*}$. The \textbf{right panels} show the median $V_{*}/\sigma_{*}(R)$ for different $\Delta_{\rm insitu, z\leq1}$ fractions and within a narrow range of $\Delta M_{\rm *,major}/M_{*}$ fractions. Major mergers erase the peak in rotation and flatten the profiles. The presence of cold gas preserves or builds up the centrally rotating component.
    \textbf{B}: Mini mergers. Median profiles for ETGs with negligible fraction of accreted mass from major, minor mergers, or stripped from surviving galaxies in bins of accreted mass fraction from mini mergers $\Delta M_{\rm *,mini}/M_{*}$. The \textbf{left panel} shows galaxies with $\Delta_{\rm insitu, z\leq1}$ lower than the median $\Delta_{\rm insitu, z=0} = 0.61$ for class 1 galaxies (i.e. with lower recent star formation), the \textbf{right panel} those with $\Delta_{\rm insitu, z\leq1}$ higher than the median. Mini mergers are not massive enough to modify the peak in rotation but, on average and for low mass systems, tend to increase the rotational support at large radii.
    }
    \label{fig:Vsigma_profiles_mergers}
\end{figure}

In this section we study how mergers affect the shapes of the ETG $V_{*}/\sigma_{*}(R)$ profiles. We start with the effect of major mergers.

In Fig.~\ref{fig:accretion_classes_merger} we showed that the fraction of accreted mass from major mergers $\Delta M_{\rm *,major}/M_{*}$ depends on mass, with the lowest mass galaxies having negligible $\Delta M_{\rm *,major}/M_{*}$ while for higher mass galaxies ($M_{*}>10^{10.75}\MSUN{}$) the ex-situ fraction from major mergers spans a wide range, from 0 to 60\% and up to 80\% for a few objects above $10^{11.5}\MSUN{}$.
Figure~\ref{fig:Vsigma_profiles_mergers}A shows the effect of major mergers on the galaxy rotational support. The left panels of Fig.~\ref{fig:Vsigma_profiles_mergers}A show how the $V_{*}/\sigma_{*}(R)$ profiles vary in bins of $\Delta M_{\rm *,major}/M_{*}$ in FRs (top panels) and SRs (bottom panels). Increasing $\Delta M_{\rm *,major}/M_{*}$ clearly erases the peak in rotation and flattens the profiles. In extreme cases, where the FRs have $\Delta M_{\rm *,major}/M_{*}\gtrsim0.4$, the $\mathrm{V_{tot}/\sigma_{tot}(R)}$ profiles are flat from the central regions to the halos. In SRs, major mergers prove to be much more effective in flattening the profiles, as for these galaxies mergers are more gas poor and more recent (see Sect.~\ref{sec:accretion_classes_mergerhistory}). 

The left panels of Fig.~\ref{fig:Vsigma_profiles_mergers}A show together FRs and SRs of stellar masses from $10^{10.3}$ to $10^{12}\MSUN{}$, which assemble at different epochs and with different gas fractions. Each of these variables play a role in determining the final shape of the $V_{*}/\sigma_{*}(R)$ profiles. The right panels of Fig.~\ref{fig:Vsigma_profiles_mergers}A show, for example, the role of recent cold gas accretion (parametrized here by $\Delta_{\rm insitu, z\leq1}$) for galaxies with similar fraction of stellar mass accreted from major mergers (here we show $0.1\leq\Delta M_{\rm *,major}/M_{*}\leq0.3$). The presence of gas (in situ or accreted in the mergers) is essential to preserve/build up the central rotating disk-like structure. The comparison between the median profiles of FRs and SRs with similar $\Delta M_{\rm *,major}/M_{*}$ and $\Delta_{\rm insitu, z\leq1}$ hints at other variables determining the shape of the profiles like, for example, the timing of the merger. A galaxy might or not have its rotating component destroyed if its accreted mass comes from more recent merger, so that the galaxy does not have time to rebuild the disk.

To isolate the effect of mini mergers, we select galaxies that accreted a negligible fraction of mass from major, minor mergers, or stripped from surviving galaxies (less than $1\%$ of $M_{*}$ from each). Most of these galaxies have low stellar masses, $M_{*}<10^{10.5} \MSUN{}$, but some reach up to $M_{*}<10^{10.9} \MSUN{}$. The selected galaxies are mostly FRs ($\sim97\%$) and have accreted mass fractions from mini mergers $\Delta M_{\rm *,mini}/M_{*}$ ranging from 0 to 30\%. The median number of mini mergers per galaxy is 30. Figure~\ref{fig:f_profiles} shows that the accreted star from mini mergers for galaxies with $M_{*}<10^{10.9} \MSUN{}$ are mainly deposited in the outskirts. 

Figure~\ref{fig:Vsigma_profiles_mergers}B shows the median $V_{*}/\sigma_{*}(R)$ profiles for this group of galaxies for different $\Delta M_{\rm *,mini}/M_{*}$ fractions. We distinguished between galaxies with low fraction of recently formed in-situ stars $\Delta_{\rm insitu, z\leq1}$ (left panel) and galaxies with high $\Delta_{\rm insitu, z\leq1}$ (right panel). In this way we take into account of variations in the $V_{*}/\sigma_{*}(R)$ profiles due to the recent accretion of gas.
We find that, independent of $\Delta_{\rm insitu, z\leq1}$, galaxies with the lowest $\Delta M_{\rm *,mini}/M_{*}$ fractions have "peaked-and-outwardly-decreasing" profiles. Higher $\Delta M_{\rm *,mini}/M_{*}$ fractions progressively increase the rotational support of the outermost regions while roughly conserving the height of peak. These trends persist when considering only central galaxies (not shown here for brevity). The recent gas accretion parametrized by $\Delta_{\rm insitu, z\leq1}$ has the overall effect of increasing the peak of rotation from a median $\sim 0.75$ to a median $\sim1$.

Finally, we studied (although do not show here for brevity) the effect of minor mergers. We selected a sample of galaxies within a narrow range of accreted stellar mass from major mergers ($\Delta M_{*,\rm major}/M_{*}\in[0.1,0.2]$, so we have the minor mergers) to include more massive systems and at the same time limit the impact of major mergers. Minor mergers have a somewhat similar effect as the major mergers. If their contribution is high enough they can weaken the peak of rotation, so that galaxies with $\Delta M_{\rm *,minor}/M_{*}\gtrsim10\%$ have significantly lower maximum in $V_{*}/\sigma_{*}(R)$ profiles (peak value$\sim0.5$) with respect to galaxies with $\Delta M_{\rm *,minor}/M_{*}<5\%$ (peak value $\sim0.9$). On the other hand we find that the rotational support of the halo regions ($R\sim8\re{}$) is independent of $\Delta M_{\rm *,minor}/M_{*}$. 

\subsection{Dependence of the $\mathrm{V/\sigma}(R)$ profiles on other parameters}\label{sec:Vsigma_residuals}

Section~\ref{sec:Vsigma_merger_mass_ratio} demonstrated that the shape of the $V_{*}/\sigma_{*}(R)$  profiles is influenced by several physical processes that are at work in the galaxy assembly history. Accretion parameters correlate more or less tightly with stellar mass, and hence with each other \citep[Figs.~\ref{fig:trendswithmass} and \ref{fig:accretion_classes_merger}, see also][]{2017MNRAS.467.3083R}. For example, galaxies with the lowest stellar mass have on average the highest in-situ mass fraction and the highest mean accreted gas fraction.
Assuming that the total stellar mass and the in-situ mass fraction (or, equivalently, the ex-situ mass fraction) are the two fundamental variables (Sec.~\ref{sec:Vsigma_profiles_insitu}), we studied whether there are additional residual correlations with other accretion history properties. 

We have inspected (albeit do not show for brevity) the $V_{\rm insitu}/\sigma_{*}(R)$ profiles in bins of stellar mass and in-situ fractions for variations of mean gas fraction, in-situ stellar mass formed after $z=1$, stellar mass accreted via major mergers, redshift of the last major merger, and number of major mergers: we find no secondary dependencies.
The lack of such correlations indicate that even though the shape of the $\mathrm{V_{insitu}/\sigma_{tot}}(R)$ profiles are sensitive to different accretion parameters which describe the details of the galaxy formation histories, their effect is already taken into account by the total stellar mass and the total in-situ mass fraction.
We repeated this study for the total $V_{*}/\sigma_{*}(R)$ profiles and found almost identical results.

\subsection{Rotational support and the local accreted fraction}\label{sec:local_Vsigma_fex}

The $V_{*}/\sigma_{*}(R)$ profile shapes at different stellar masses are primarily regulated by the in-situ mass fraction (Sect.~\ref{sec:Vsigma_profiles_insitu}). The typical profile of a low mass galaxy with negligible accreted fraction is a peaked profile that decreases at large radii. At higher stellar masses and ex-situ fractions, mergers modify the rotational support of the in-situ components and add the contribution of the ex-situ stars, so that the outskirts of galaxies ($R>2-3\re{}$) show a large variety of stellar halo rotational support in qualitative agreement with extended kinematic studies of observed ETGs \citep[][see also \citetalias{2020A&A...641A..60P}]{2018A&A...618A..94P}. However Fig.~\ref{fig:Vsigma_profiles} shows that, at fixed stellar mass and in-situ mass fraction, galaxies exhibit a large scatter around the median profiles. In Sect.~\ref{sec:Vsigma_residuals} we showed that this scatter is not due to the additional accretion parameters investigated.

A possible explanation for the large scatter in the profiles comes from Figs.~\ref{fig:insituVSmass_classes} and \ref{fig:f_profiles}: galaxies with similar masses and similar total in-situ fractions can have different spatial distribution of in-situ and accreted stars, for example if they belong to different accretion classes. This results in a scatter around the median $\fex{}(R)$ profiles.
In this section we investigate whether the stellar halo rotational support has a clearer dependence on the local fraction of accreted stars (i.e. measured both at the same radius) rather than the total accreted mass fraction. 

\begin{figure}
    \centering
    \includegraphics[width=1\linewidth]{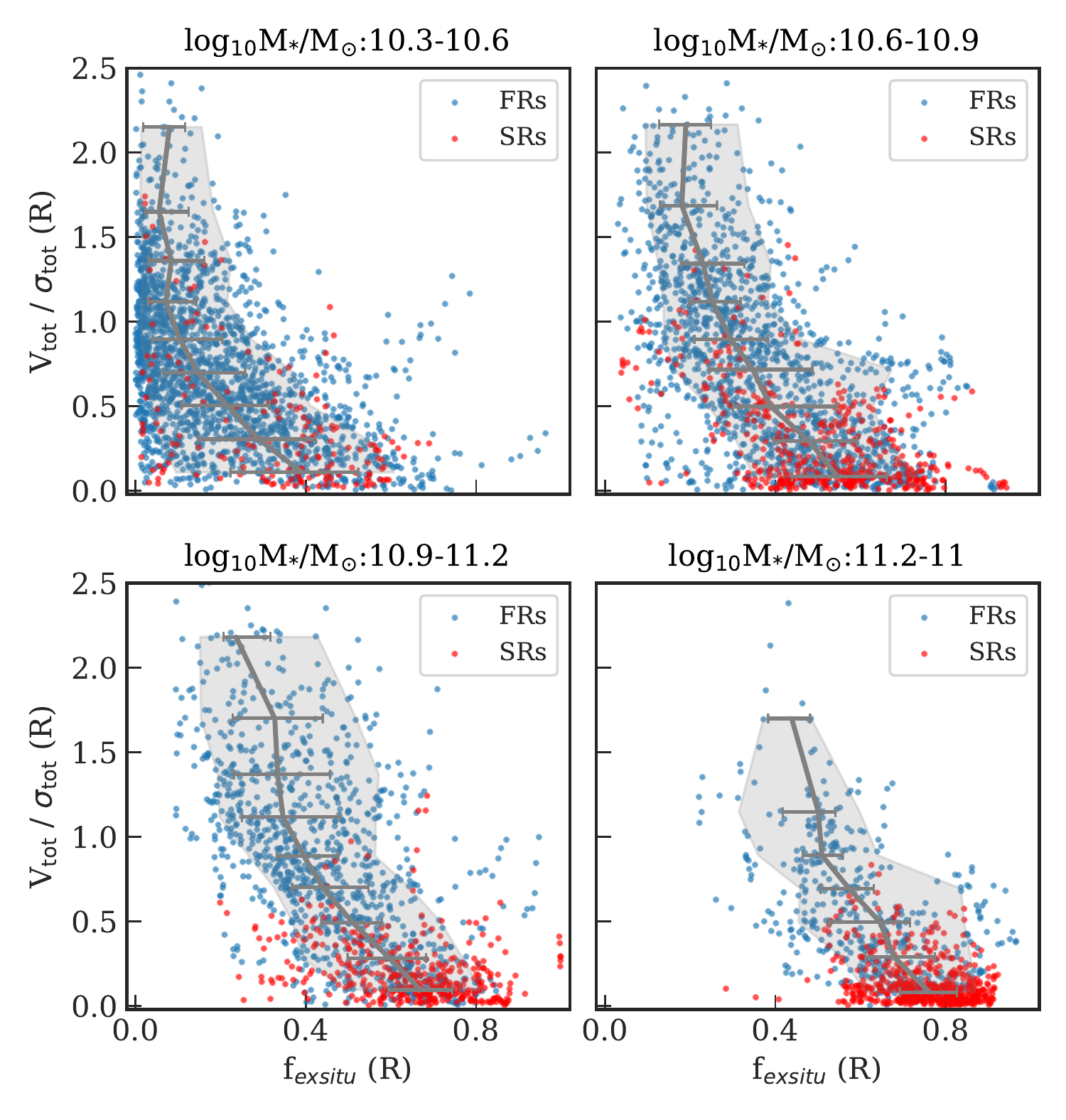}
    \caption{Local correlation between rotational support and accreted fraction in the stellar halos. Each data point correspond to a local measurement at radius $R\in[3-8\re{}]$ within each galaxy. Blue circles show measurements within FRs, red circles within SRs. The solid lines show the median, the error-bars the quartiles, and the shaded regions the 10-90 percentiles of the $V_{*}/\sigma_{*}(\fex{})$ distributions. Galaxies are divided in stellar mass bins as labelled. At high local ex-situ mass fractions the stellar halo rotational support $V_{*}/\sigma_{*}$ decreases.}
    \label{fig:Vsigma_local}
\end{figure}

We proceed as follows. The outer regions ($3-8\re{}$) of each TNG100 galaxy are divided into seven elliptical shells with semi-major axis $R$ and flattening following the ellipticity profile of the galaxy. In each shell we measure the rotational support of the stellar particles $V_{*}/\sigma_{*}(R)$ and the local ex-situ mass fraction $\fex{}$. The minimum radius of $3\re{}$ is motivated by the requirement that $R$ should be large enough so that the $V_{*}/\sigma_{*}(R)$ profiles of the TNG FRs have reached their peak rotation, which occurs at a median radius  $\sim2.7\re{}$.

Figure \ref{fig:Vsigma_local} shows the rotational support $V_{*}/\sigma_{*}(R)$ as a function of  $\fex{}(R)$ measured in the same radial bins. Each data point in the diagram corresponds to a local measurement within one galaxy, meaning that each galaxy is represented by seven data points. 
Galaxies are divided into stellar mass bins, FRs and SRs are shown with different colors.

Figure \ref{fig:Vsigma_local} reveals that there is a strong anti-correlation between these $V_{*}/\sigma_{*}$ and $\fex{}$: wherever the local ex-situ contribution $\fex{}$ is high the rotational support is low. As $\fex{}$ decreases $V_{*}/\sigma_{*}$ tends to increase but at low $\fex{}\lesssim0.2$, $V_{*}/\sigma_{*}(R)$ is independent of $\fex{}$. These cases correspond mainly to low-mass in-situ dominated galaxies, in which the kinematic transition to the dispersion-dominated stellar halo is caused by the in-situ stars alone. 

The strong dependence of $V/\sigma$ on the local $\fex{}$, together with the variations of $\fex{}$ among different galaxies at the same radius, partially explains the large scatter in $V/\sigma(R)$ profiles. The scatter on the median $\mathrm{V_{tot}/\sigma_{tot}}(\fex{})$ profiles is reduced by a factor $\sim 1.4$ compared to the scatter on the median $V_{*}/\sigma_{*}(R)$. 

SRs sit on the the same anti-correlation traced by the FRs and populate the tail of the FR stellar halo parameter distributions at high \fex{}.
This continuity of local properties between the two classes can be reconciled with the systematic differences in median $V_{*}/\sigma_{*}(R)$ profiles of FRs and SRs with similar mass and in-situ fraction seen in Fig.~\ref{fig:Vsigma_profiles}, by considering their different radial distribution of in-situ versus ex-situ stars as implied in Figs.~\ref{fig:f_profiles} and \ref{fig:fraction_FRandSR_accretion_classes}. 
Finally we remark we observe that if we assume that a similar $V_{*}/\sigma_{*} - \fex{}$ correlation holds in real galaxies, then one could in principle use the observationally accessible $\mathrm{V/\sigma}(R)$ profiles in the stellar halos and stellar masses to estimate the local ex-situ contribution $\fex{}(R)$ in observed galaxies within approximately $\pm0.16$. This range is derived from the average width of the $\fex{}$ distribution at the 10-90 percentiles for different $V_{*}/\sigma_{*}$ and stellar mass bins. The median and quartiles of the local $V_{*}/\sigma_{*}$ distribution as a function of the local $\fex{}$ for ETGs with $M_{*}>10^{10.6}\MSUN{}$ is reported in Table~\ref{tab:local_fex_T_Vs}.

\section{Intrinsic shapes and accretion history}\label{sec:Intrinsic_shapes}

In \citetalias{2020A&A...641A..60P} we determined the intrinsic shapes of our sample of simulated ETGs and found that kinematics and intrinsic shapes are closely related, such that the kinematic transitions into the stellar halos are accompanied by changes in the galaxy intrinsic shapes. The FRs that decrease in rotational support at large radii tend also to become more spherical in the outskirts, while a fraction with high stellar halo rotation show constant axis ratio $q(r)$ profiles out to large radii. The majority of low mass FRs have stellar halos consistent with oblate shapes, but at higher stellar masses the fraction of FRs with triaxial stellar halos increases. By comparison, the SRs display milder structural changes with radius, exhibiting rather constant $q(r)$ and triaxiality $T(r)$ profiles. In this section we study the dependence of galaxy intrinsic shapes on accretion.

\subsection{Intrinsic shape profiles and in-situ fractions}\label{sec:shape_profiles_insitu}

\begin{figure}
    \centering
    \includegraphics[width=\linewidth]{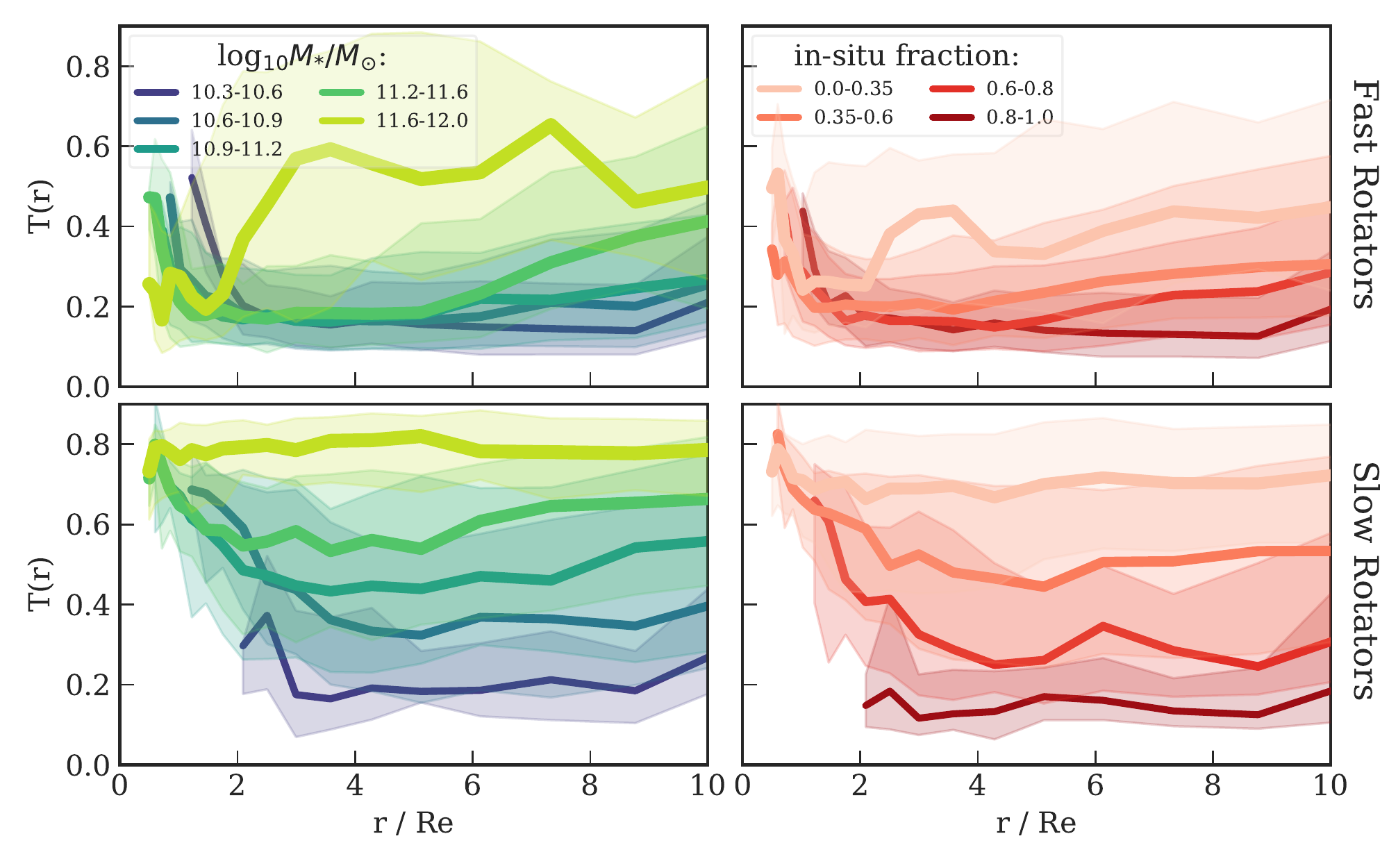}
    \includegraphics[width=\linewidth]{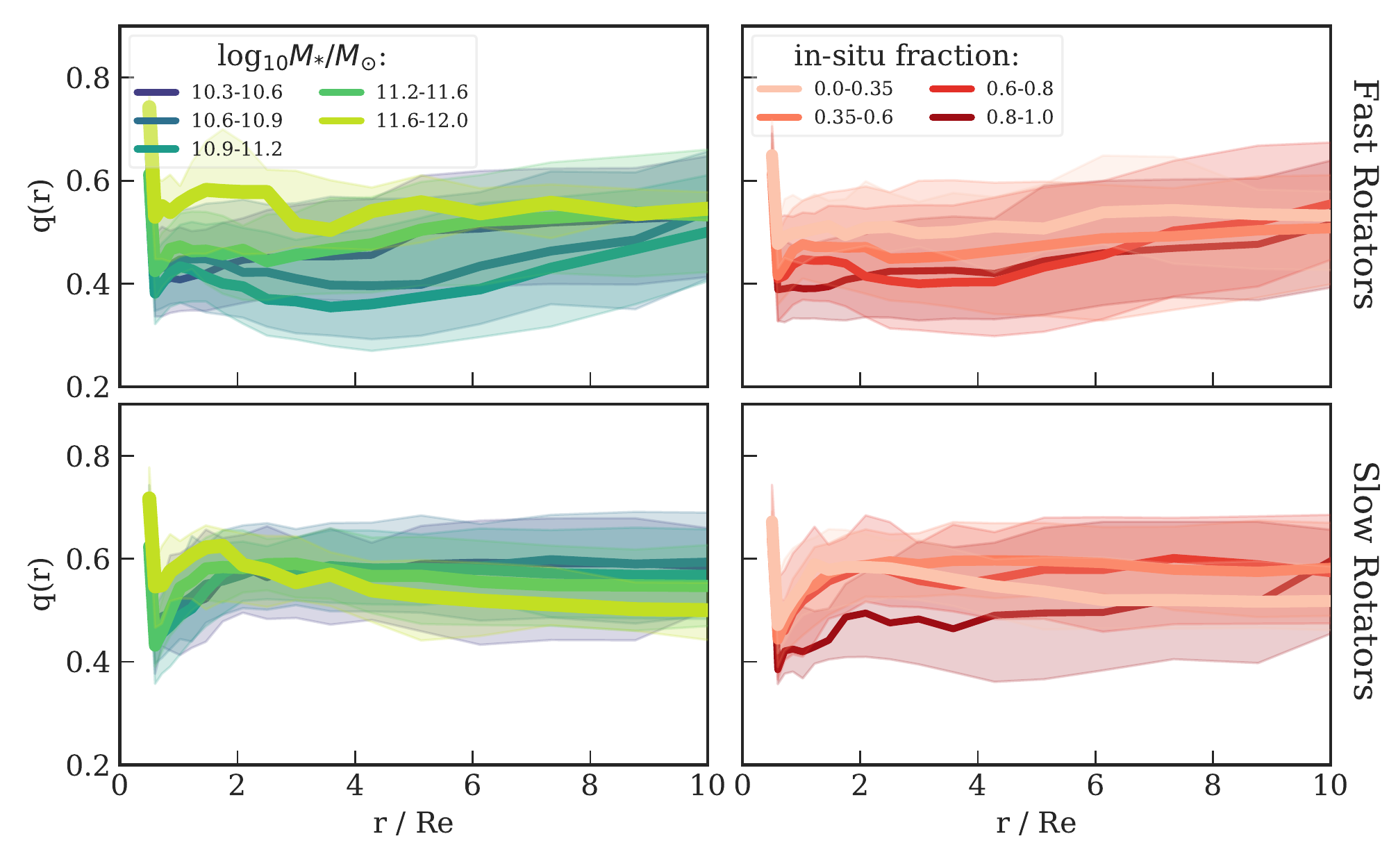}
    \caption{Median stellar intrinsic shape profiles for FRs  and SRs as labelled on the right margins. The \textbf{top} panels show the triaxiality $T(r)$, the \textbf{bottom} panels show the axis ratio $q(r)$. Galaxies on the \textbf{left} are divided in stellar mass bins, galaxies on the \textbf{right} are divided in bins of total in-situ mass fraction. The solid lines show the median profiles, the shaded regions trace the quartiles of the distributions. Higher stellar masses and accreted fractions correspond to higher triaxiality and less flattened shapes.}
    \label{fig:shape_profiles_insitu}
\end{figure}
 
Figure \ref{fig:shape_profiles_insitu} shows median $q(r)$ and $T(r)$ profiles for FRs and SRs divided into bins of stellar mass and in-situ fraction. 
The median profiles are built using the profiles of the single galaxies at the radii where resolution effects are small ($r\geq2\rsoft{}$ for $q(r)$ and $r\geq9\rsoft{}$ for $T(r)$, see Sect.~\ref{sec:measuring_IntrinsicShape}). 

In \citetalias{2020A&A...641A..60P} we found that triaxiality increases with radius in FRs and that in both FRs and SRs the stellar halo triaxiality ($T(8\re{})$) increases with stellar mass. These trends are visible also in Fig.~\ref{fig:shape_profiles_insitu}.
Since $M_{*}$ tightly correlates with the total fraction of in-situ stars (Fig.~\ref{fig:trendswithmass}), similar behaviors are found as a function of this variable. 

FR galaxies with higher $M_{*}$ and in-situ fractions reach the largest values of $T$ in the stellar halo. In the SRs the dependence of $T(r)$ on these parameters is much more striking, such that at similar stellar mass and total in-situ mass fraction triaxiality is overall systematically higher than in FRs. 

From the median $q(r)$ profiles, we see that FRs tend to become slightly rounder at increasing $M_{*}$ and decreasing total in-situ fraction. FRs with higher $M_{*}$/lower in-situ fraction have rather constant median $q(r)$ profiles while those with lower $M_{*}$/higher in-situ fractions show more pronounced dips in $q$ at radii corresponding to the positions of the peaks in the median $V_{*}/\sigma_{*} (R)$ profiles of Fig.~\ref{fig:Vsigma_profiles}. 
The $q(r)$ profiles in SRs are almost independent of these parameters, although SRs with $M_{*}>10^{11.2}\MSUN{}$ or with in-situ fractions $<0.35$ have significantly flatter intrinsic shapes at large radii (see also Sect.~\ref{sec:local_shape_fex}), similarly to the few SRs with in-situ fractions $>0.8$. 

Overall galaxy intrinsic shapes depend on both stellar mass and in-situ fraction but we find (although we do not show here) that at fixed stellar mass the galaxies with highest triaxiality and lowest flattening are those with low in-situ mass fraction (see the scatter in in-situ fraction values in  Fig.~\ref{fig:trendswithmass}).
The results of Fig.~\ref{fig:shape_profiles_insitu} indicate that the stellar component that is born in-situ tends to preserve the near-oblate flattened disk-like structure of the star forming gas, at least in the central regions. Therefore FRs with higher in-situ fractions and lower stellar masses show larger variations of the median $q(r)$ profiles with radius, with the outskirts being rounder than the centers. Instead galaxies with higher accreted fractions and stellar masses have milder variations of flattening with radius but have overall increased triaxiality at large (FRs) or all (SRs) radii.
The collisionless dynamics of gas-poor mergers, which enrich the galaxies with ex-situ stars, is the primary mechanism responsible for changing the intrinsic shapes from near-oblate to spheroidal triaxial, as we demonstrate in  Sect.~\ref{sec:shape_profiles_mergers}.

\subsection{Intrinsic shape profiles and mergers}\label{sec:shape_profiles_mergers}

\begin{figure}
    \centering
    \xincludegraphics[width=1.\linewidth,label=\textbf{A}]{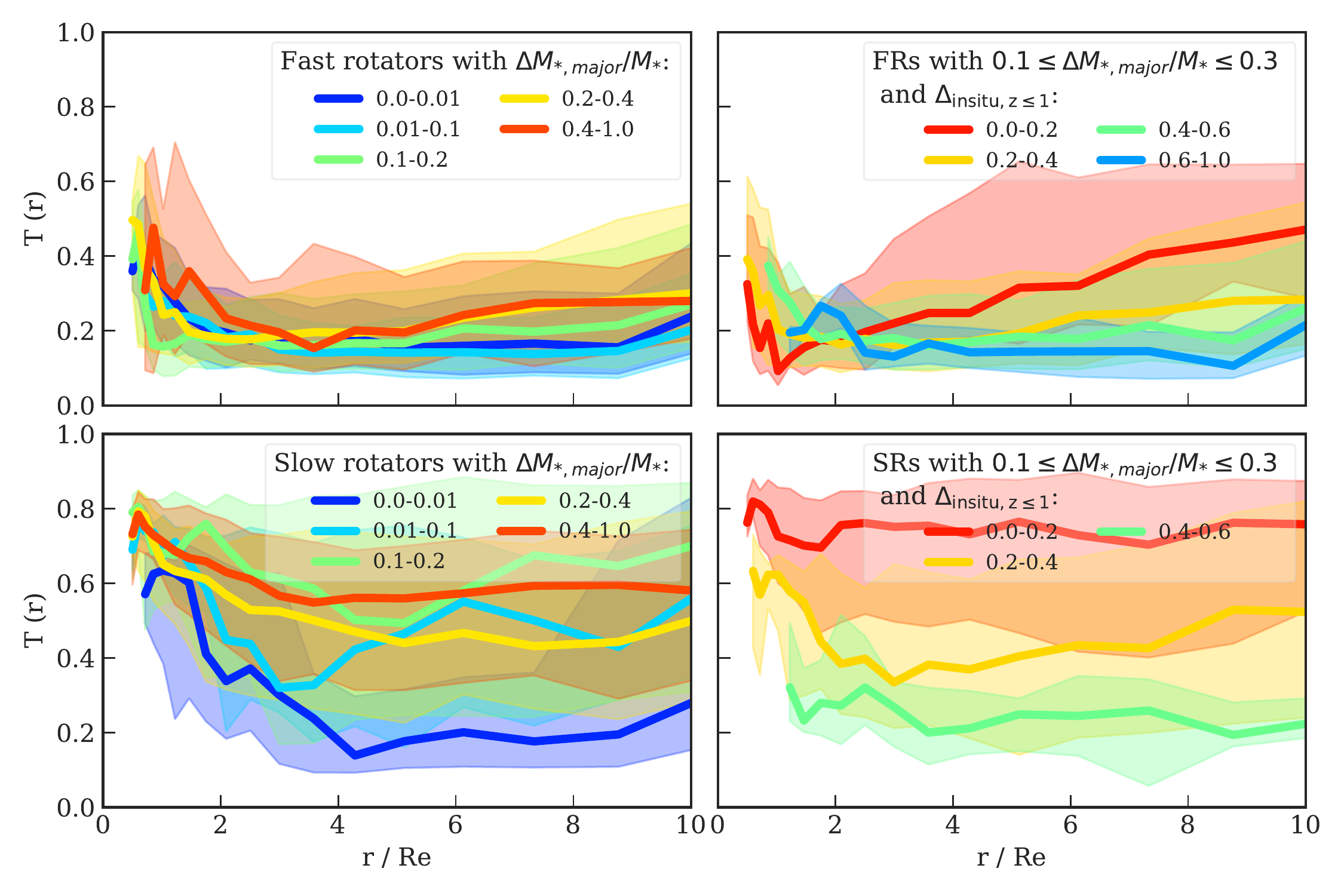}
    \xincludegraphics[width=1.\linewidth,label=\textbf{B}]{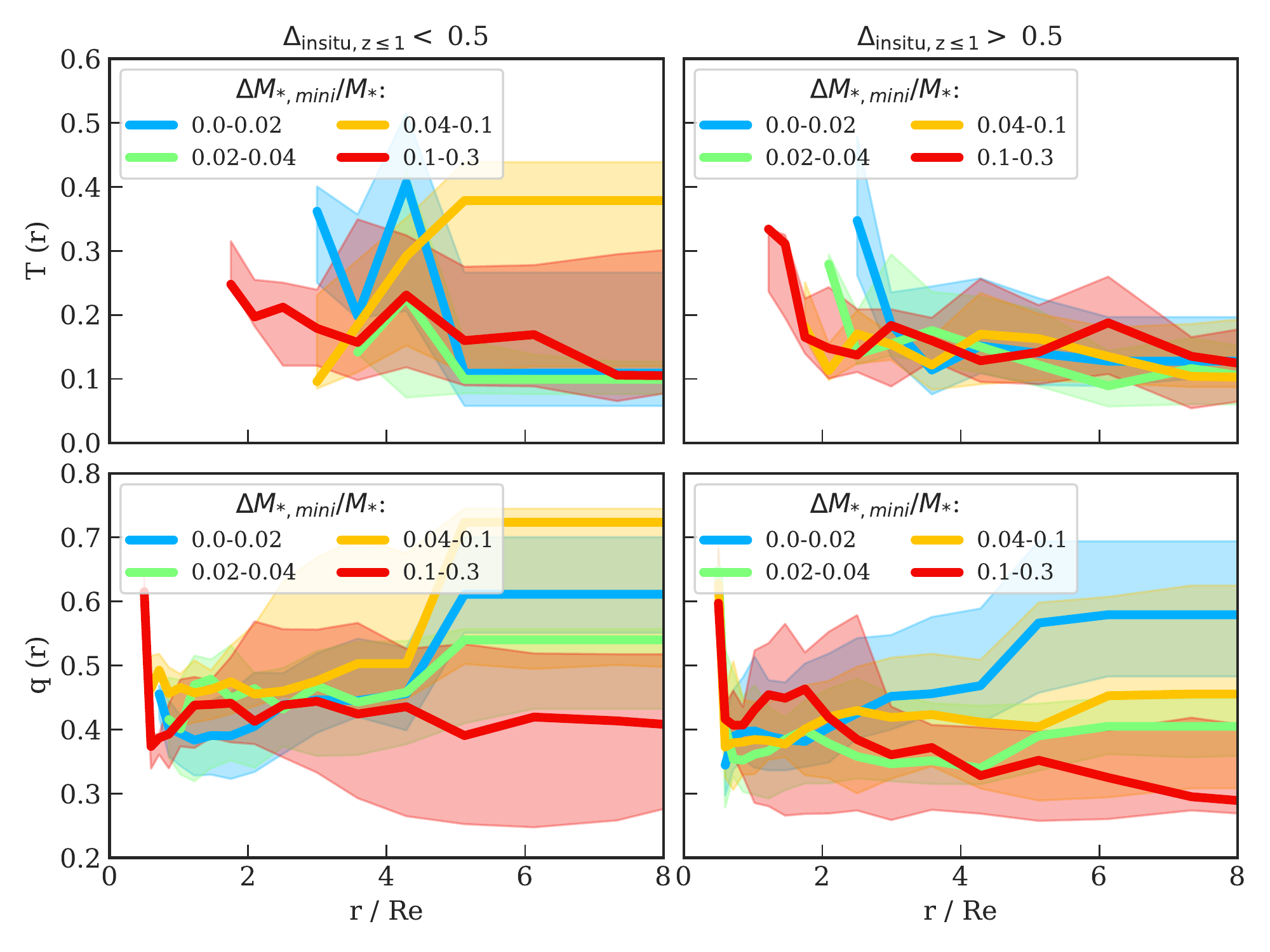} 
    \caption{The effect of mergers  on  the intrinsic shape profiles. \textbf{A}: Major mergers in FRs (\textbf{top}) and SRs (\textbf{bottom}). The solid lines show the median triaxiality profiles in bins of accreted mass fraction from major mergers $\Delta M_{\rm *,major}/M_{*}$ (\textbf{left panel}) and in bins of $\Delta_{\rm insitu, z\leq1}$ for $0.1\leq \Delta M_{\rm *,major}/M_{*}\leq0.3$ (\textbf{right panel}). The shaded regions show the quartiles of the distributions. Major mergers increase the triaxiality at large radii, especially when the fraction of recently accreted gas is low. \textbf{B}: The effect of mini mergers on the intrinsic shape profiles $T(r)$ (\textbf{top}) and $q(r)$ (\textbf{bottom}) for ETGs with negligible accreted mass from major mergers, minor mergers, and stripped from surviving galaxies. The solid lines show median profiles in bins of accreted mass fraction from mini mergers $\Delta M_{\rm *,mini}/M_{*}$; the shaded regions show the quartiles of the distributions. The \textbf{left panel} shows galaxies with $\Delta_{\rm insitu, z\leq1}<0.5$ (i.e. with lower recent star formation), the \textbf{right panel} those with $\Delta_{\rm insitu, z\leq1}>0.5$. Mini mergers do not influence significantly the galaxy triaxiality but tend to increase the stellar halo flattening.}
    \label{fig:shape_mergers}
\end{figure}

In this section we analyze how mergers affect the intrinsic shapes of galaxies. 
The effects of major mergers on the triaxiality profiles is shown in Fig.~\ref{fig:shape_mergers}A. The left panels show median profiles in bins of $\Delta M_{\rm *,major}/M_{*}$ for our sample of ETGs divided in FRs and SRs. While for the SRs the effect of $\Delta M_{\rm *,major}/M_{*}$ in increasing the median $T(r)$ is rather clear, for the FRs the picture is complicated by the higher fractions of accreted gas (Fig.~\ref{fig:trendswithmass}) which, on the other hand, tend to decrease $T(r)$. The right panels of Fig.~\ref{fig:shape_mergers}A show that, at fixed fraction of accreted mass from major mergers $\Delta M_{\rm *,major}/M_{*}$, the galaxies with lowest recent in-situ star formation $\Delta_{\rm insitu, z\leq1}$ (i.e. those that did not recently accrete gas), whose recent mergers were then gas-poor, are the FRs with increasing $T(r)$ profiles and the SRs with the highest median $T(r)$.
Major mergers also affect the $q(r)$ profiles (not shown here) in that galaxies with higher $\Delta M_{\rm *,major}/M_{*}$ and lower $\Delta_{\rm insitu, z\leq1}$ tend to be on average more spherical, although the effect is not as pronounced as on the $T(r)$ profiles (see Sect.~\ref{sec:local_shape_fex}).

Figure~\ref{fig:shape_mergers}B shows the effect of mini mergers on the intrinsic shape profiles. Analogously to Sect.~\ref{sec:Vsigma_merger_mass_ratio} we isolate the effect of mini mergers by considering galaxies with negligible accreted mass from other accretion channels (i.e. with $\Delta M_{\rm *,major}/M_{*}$, $\Delta M_{\rm *,minor}/M_{*}$, and $\Delta M_{\rm *,stripped}/M_{*}$ less than $1\%$ each). The selected galaxies, mostly FRs with $M_{*}< 10^{10.5}\MSUN{}$, have their accreted mass accumulated mainly at large radii (Fig.~\ref{fig:f_profiles}). We group the galaxies in bins of accreted mass fraction from mini mergers ($\Delta M_{\rm *,mini}$) and use $\Delta_{insitu, z\leq1}=0.5$ to distinguish between galaxies with recent gas accretion and in-situ star formation (right panels in Fig.~\ref{fig:shape_mergers}B) and those with more quiescent recent star formation history (left panels). 

We see that mini mergers do not clearly affect the triaxiality profiles, the majority of the selected galaxies are consistent with being near-oblate ($T\leq 0.3$) at all radii. On the other hand we find that, independent of $\Delta_{insitu, z\leq1}$, galaxies with negligible $\Delta M_{\rm *,mini}/M_{*}$ have low $q\sim 0.4$ at $r\sim1\re{}$ which increases to $q\sim 0.6$ at $r>4\re{}$. Increasing contributions of mini mergers tend to flatten the stellar halos along with increasing their rotational support (Fig.~\ref{fig:Vsigma_profiles_mergers}B).
Finally, galaxies with higher recent star formation (i.e. recent cold gas accretion) tend to have on average more flattened shapes, both at the centers and at large radii.
Figure~\ref{fig:shape_mergers}B shows results consistent those in Fig.~\ref{fig:Vsigma_profiles_mergers}B. Galaxies dominated by in-situ stars at all radii with negligible accreted mass ($\Delta M_{\rm *,mini}/M_{*}<0.02$ in Fig.~\ref{fig:shape_mergers}B) display characteristic 
"peaked-and-outwardly-decreasing" $V_{*}/\sigma_{*}(R)$ profile shapes and growing $q(r)$ profiles in the stellar halo. The kinematic transition is accompanied by a change in the dynamical structure of these galaxies, both completely driven by the in-situ stars. We have checked, although we do not show for brevity, that these results also hold if we consider central galaxies alone.

Figure~\ref{fig:shape_mergers}B indirectly shows that high mass ratio mergers are essential for galaxies to reach high values of triaxiality. 
We found (but not showed here for brevity) that minor mergers, studied in a sample of galaxies selected within a narrow range of accreted mass fractions from major mergers $\Delta M_{\rm *,major}/M_{*}\in[0.1,0.2]$ (see also Sect.~\ref{sec:Vsigma_merger_mass_ratio}), can have the effect of increasing the galaxy triaxiality at large radii if their contribution is large enough ($>10\%$ of the total stellar mass). Contrary to mini mergers, higher fractions of $\Delta M_{\rm *,minor}/M_{*}$ lead to less flattened shapes ($q\gtrsim0.5$).

The results described in this section demonstrate that the galaxy accretion history models their intrinsic shapes in different ways according to the contribution of the various accretion channels. On the other hand, the fact that the gas fraction, the recent formation of in-situ stars, and the fraction of accreted mass from major mergers are all tightly correlated with stellar mass (Fig.~\ref{fig:accretion_classes_merger}), reduces the degree of freedom of these parameters. 
We conducted a study on the intrinsic shape profiles similar to that for the $V_{*}/\sigma_{*}(R)$ profiles in Sect.~\ref{sec:Vsigma_residuals} to test their dependence on accretion parameters other than the total in-situ mass fraction. As for the $V_{*}/\sigma_{*}(R)$ profiles, we find no additional dependence once the total stellar mass and the total in-situ mass fraction have been taken into account. 


\subsection{Intrinsic shapes and the local accreted fractions}\label{sec:local_shape_fex}

\begin{figure}
    \centering
    \includegraphics[width=1\linewidth]{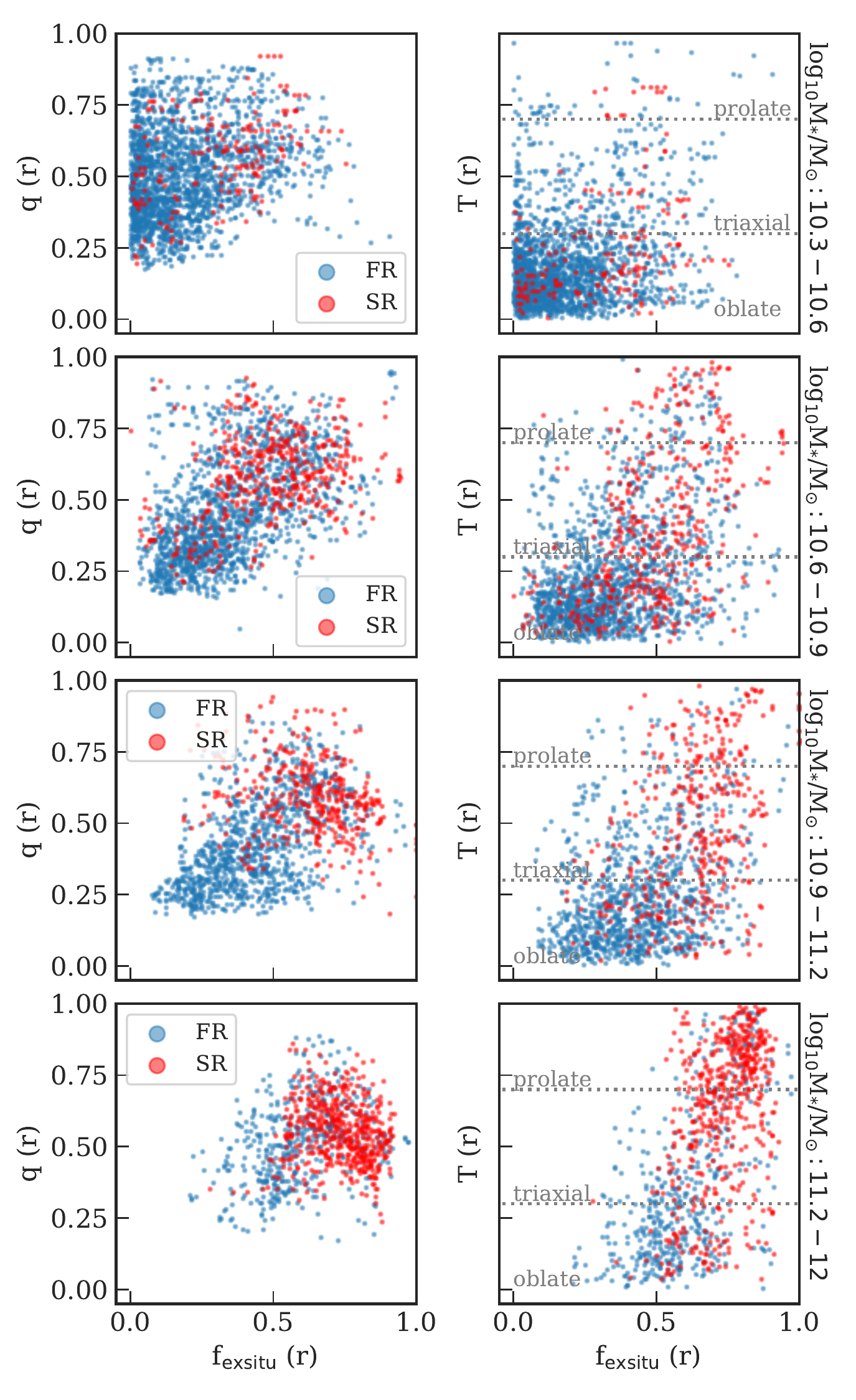}
    \caption{Local correlation between stellar halo intrinsic shape and fraction of accreted stars in stellar mass bins. Each galaxy is represented by $\sim6$ data points, measured in ellipsoidal shell with semi-major axis $r\in[3.5-8\re{}]$. The correlations are clear for galaxies more massive that $M_{*}>10^{10.6}\MSUN{}$. Low $\fex{}$ is consistent with near-oblate shapes with low axis ratio $q$. At higher $\fex{}$ the stellar halo triaxiality parameter increases while the $q(\fex)$ correlation reaches a maximum of $q\sim0.6$ for $\fex{}\sim0.7$, beyond which the $q$ decreases. }
    \label{fig:shape_fex_local}
\end{figure}

\begin{figure}
    \centering
    \includegraphics[width=1\linewidth]{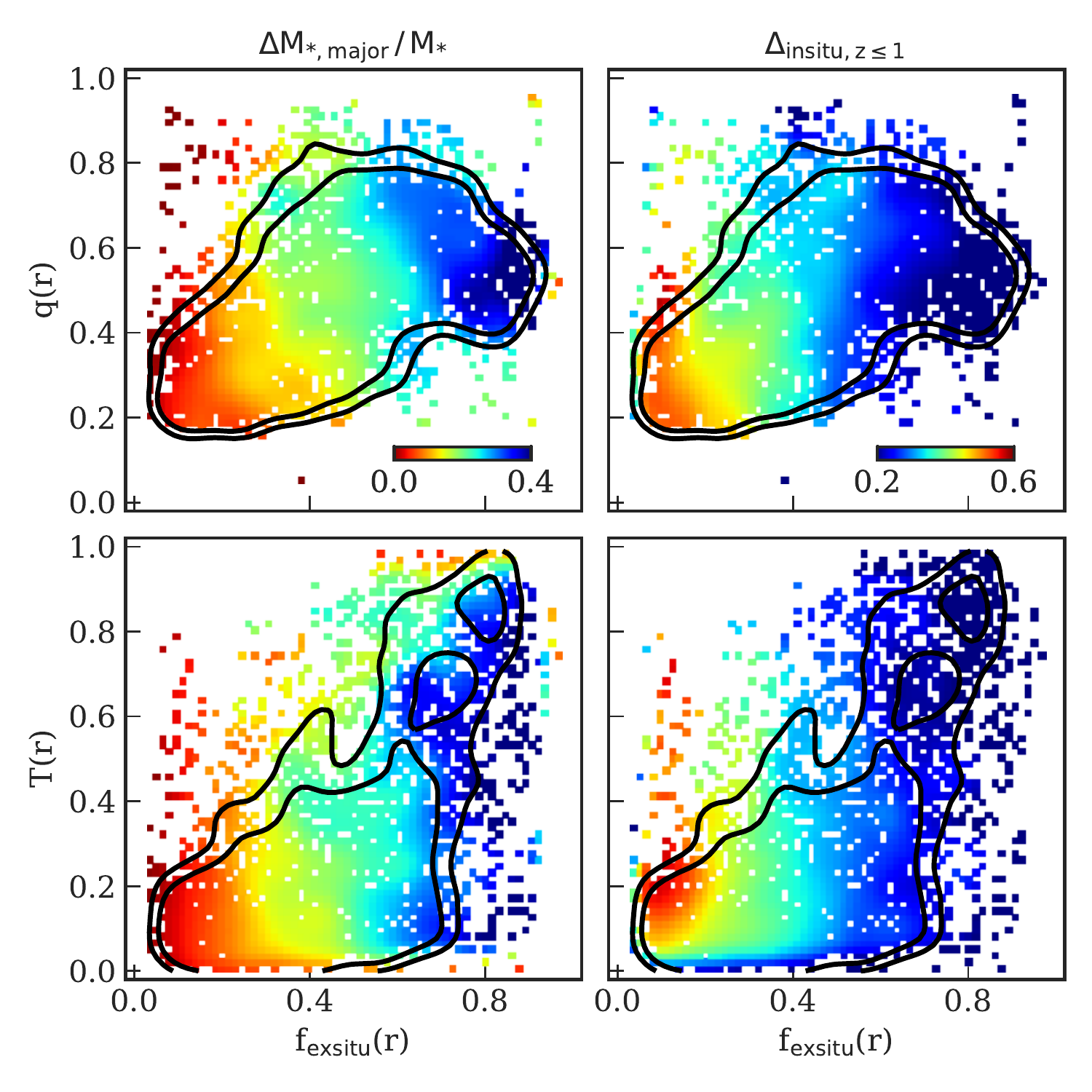}
    \caption{Local relation between galaxy intrinsic shapes and the local fraction of accreted stars for galaxies with $M_{*}\geq10^{10.6}\MSUN{}$. The \textbf{top} panels show $q(\fex{})$, the \textbf{bottom} panels $T(\fex{})$. In the \textbf{left} panels galaxies are color-coded according to the fraction of accreted mass from major mergers $\Delta M_{\rm *,major}/M_{*}$, in the \textbf{right} panels colors indicate the fraction of in-situ stars produced since $z=1$, $\Delta_{\rm in-situ, z\leq1}$. The local ex-situ fraction parametrizes the different galaxy accretion histories: the break in $q(\fex{})$ is driven by recent dry major mergers in massive systems; the most triaxial galaxies are systems with the lowest fractions of recently accreted cold gas, while low $T$ stellar halos are formed in more gas rich systems.}
    \label{fig:shape_fex_local_mergers}
\end{figure}

Here we explore the relation between local galaxy intrinsic shapes and accreted stellar mass fraction. This is done by measuring the mass fraction of accreted stars within  three-dimensional ellipsoidal shells that approximate the isodensity surfaces.
The outskirts of galaxies between $3.5\leq r/\re{}\leq8$ are divided into six spherical radial bins.
Each of these spherical bins is deformed into a homeoid with fixed semi-major axis length following the iterative procedure described in Sect.~\ref{sec:measuring_IntrinsicShape}. The ratios of the homeoid semi-axes give the local intrinsic shapes; within each homeoid we derive the total three-dimensional stellar mass density $\rho_{\rm *}(r)$, and three-dimensional density of the ex-situ component $\rho_{\rm exsitu}(r)$. From the ratio $\fex{}(r) = \rho_{\rm exsitu}(r)/ \rho_{\rm tot}(r)$ we obtain the local three dimensional ex-situ fraction. 

Figure \ref{fig:shape_fex_local} shows the intrinsic shape parameters $q(r)$ and $T(r)$ as functions of the local ex-situ fraction $\fex{}(r)$. Galaxies are plotted in stellar mass bins, SRs and FRs are shown with different colors. 
We find that the halo intrinsic shapes and $\fex$ are correlated. 
Their relation is more complicated than the simple linear correlation between rotational support and $\fex{}(R)$ observed in Fig.~\ref{fig:Vsigma_local}. We find that stellar halos are near-oblate where $\fex{}<0.4$ but at larger local ex-situ fractions they show increasingly larger triaxiality. 
The axis ratio $q(r)$ strongly correlates with $\fex{}(r)$ up to a break at $\fex{}\sim0.7$, beyond which the $q(\fex{})$ is an anti-correlation. 
The break of the $q(\fex{})$ relation explains the modest effects of the total in-situ fraction on the median $q(r)$ profiles for galaxies with large total ex-situ fractions, such as the massive FRs and the intermediate to high mass SRs, as well the change in trend seen in the median profiles of the SRs with $M_{*}>10^{11.2}\MSUN{}$ in Fig.~\ref{fig:shape_profiles_insitu}. 
We note that class 1 galaxies, mainly present in the lowest stellar mass bin at low $\fex{}$, do not follow the $q(\fex{})$ correlation, as changes in flattening are driven by the in-situ stars only (see Sect.~\ref{sec:shape_profiles_mergers}). For $M_{*}\geq10^{10.6}\MSUN{}$ galaxies the $q(\fex{})$ and $T(\fex{})$ relations are clear. The median and quartiles of the distributions of $q$ and $T$ as a function of the local $\fex{}$ for ETGs with $M_{*}>10^{10.6}\MSUN{}$ are reported in Table~\ref{tab:local_fex_T_Vs}.

The $q(\fex{})$ and $T(\fex{})$ relations are followed by both FRs and SRs in each stellar mass bin, but with relative fractions varying with stellar mass.
FRs and SRs show again a continuity of stellar halo properties across the two classes, with the SRs significantly overlapping with the FRs but crowding at the high $\fex{}$ extreme.
Therefore the systematic differences between median shape profiles of FRs and SRs with similar stellar mass and in-situ fractions observed in Fig.~\ref{fig:shape_profiles_insitu} are due to the different radial distribution of in-situ versus ex-situ stars in the two families, which is probably at the root of the FR-SR bimodality. At large radii, where the local $\fex{}$ fractions are higher, stellar halos display a more continuous sequence of properties.

Figure~\ref{fig:shape_fex_local_mergers} shows the local $q(\fex{})$ and $T(\fex{})$ correlations for all the galaxies with $M_{*}\geq10^{10.6}\MSUN{}$, colored according to the total accreted mass fraction from major mergers and the fraction of in-situ stars produced since $z=1$. 
Since both $\Delta M_{\rm *,major}/M_{*}$ and $\Delta_{\rm insitu, z\leq1}$ correlate with stellar mass (Fig.~\ref{fig:accretion_classes_merger}), as well as with the total ex-situ fraction (Fig.~\ref{fig:insituVSmass_classes}), gradients of these global properties with the local $\fex{}$ fraction result. These gradients indicate that galaxies reaching high $\fex{}$ have accreted larger fractions of their mass from major mergers and did not produce many in-situ stars recently, implying that their recent merger history is gas-poor.
Figure~\ref{fig:shape_fex_local_mergers} shows that the most triaxial galaxies (i.e. those with $T>0.4-0.5$) have a wide range of $\Delta M_{\rm *,major}/M_{*}$ fractions, but they all had a recent dry merger history ($\Delta_{\rm insitu, z\leq1}\lesssim0.3$). On the other hand, galaxies with high $\fex{}>0.4$ that have near-oblate intrinsic shapes accreted less mass from major mergers and have accreted only slightly more gas than their near-prolate counterparts. 
Finally Fig.~\ref{fig:shape_fex_local_mergers} reveals that the knee in the $q(\fex{})$ relation at $\fex{}\sim0.7$ is driven by galaxies with the highest accreted mass fractions from major mergers with the lowest $\Delta_{\rm insitu, z\leq1}$. 


\section{Stars and dark matter halo relation}\label{sec:dmVSstars}

\begin{figure*}
    \centering
    \includegraphics[width=\linewidth]{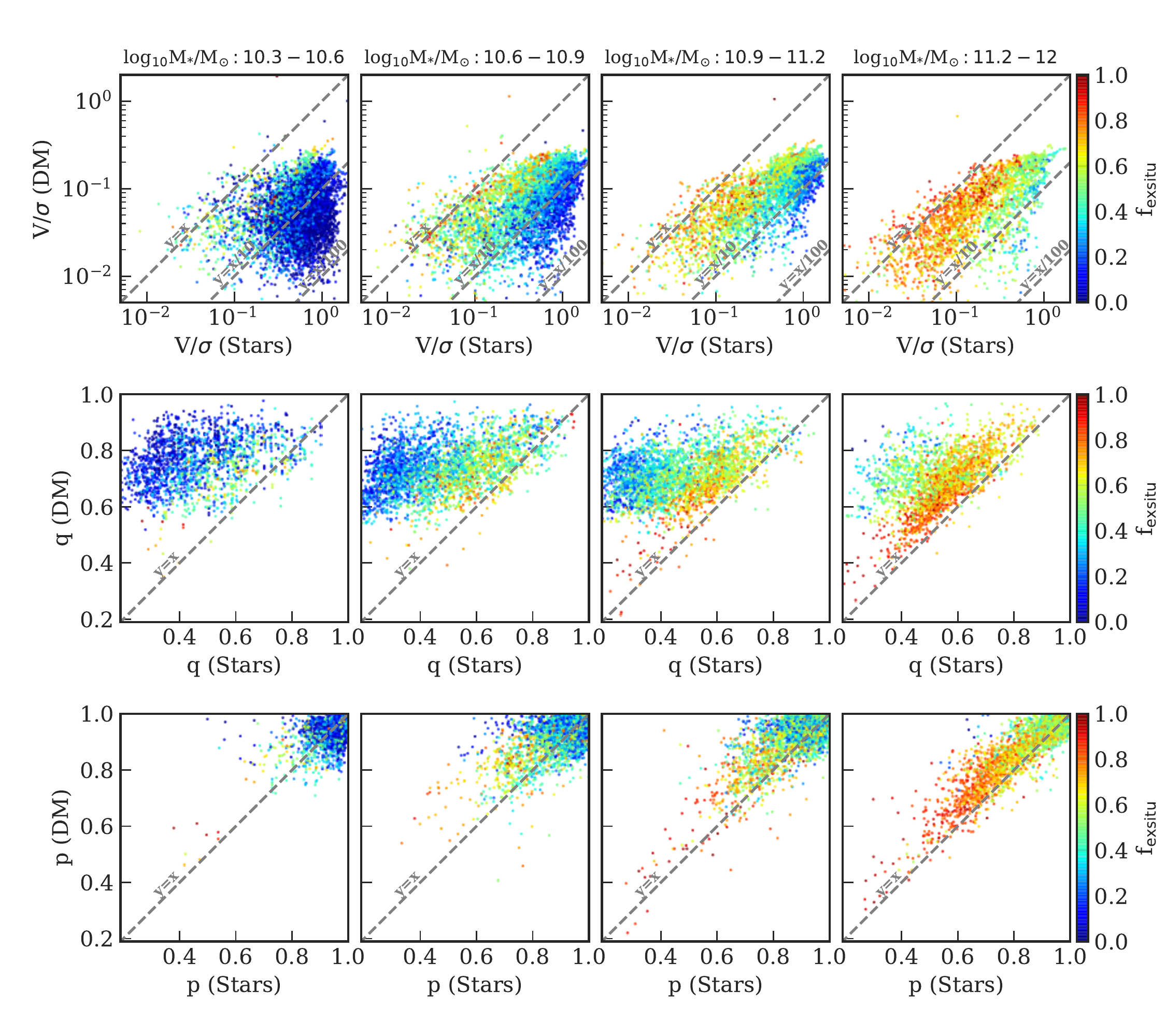}
    \caption{Local stellar versus dark matter rotational support ($V/\sigma$, \textbf{top panels}) and intrinsic shape (axis ratios $q$ and $p$, respectively in the \textbf{middle} and \textbf{bottom panels}) for all galaxies in 12 radial shells with $r\in[1\re{}-8\re{}]$. Local measurements are color coded according to the local ex-situ fraction \fex. Galaxies are divided in stellar mass bins as labelled on top the respective columns. The dashed diagonal lines show the relations $y = x$, $y = x/10$, and $y=x/100$. Rotation and shapes of stars and dark matter halo become similar with increasing $\fex{}$.}
    \label{fig:local_shape_and_rotation_starsVSdm}
\end{figure*}

In this part of the paper we investigate the relation between the rotation and structural properties of the stellar and the dark matter halos.
Because the local rotational support and the intrinsic shape of the stellar halos are linked to the accretion history (parametrized in Figs.~\ref{fig:Vsigma_local} and \ref{fig:shape_fex_local} by $\fex{}$) and because stars and dark matter are accreted jointly, we expect that accretion has a similar influence also on the dark matter component.

\subsection{Relation between dark matter and stellar halo parameters and local ex-situ fraction}\label{sec:local_relation_dmVSstars}

The intrinsic shape of the dark matter halo is derived following the same iterative procedure used in Sect. \ref{sec:measuring_IntrinsicShape} for the stellar particles. For both components we use shells with the same fixed semi-major axes lengths $r$ and $r+\Delta r$, but we allow the direction of the principal axes to vary, as well as the axis ratios $p$ and $q$. 
The rotational support of the stellar and the dark matter components is derived in spherical shells of radii $r$ and $r+\Delta r$. In each radial bin we use the stellar and the dark matter particles separately to calculate the $V/\sigma(r)$ of the two components. In the same spherical shell we derive the local fraction of ex-situ stars $\fex(r)$ from the ratio of the stellar mass densities within the shell $\rho_{\rm exsitu}(r) / \rho_{\rm stars}(r)$. 

Figure~\ref{fig:local_shape_and_rotation_starsVSdm} shows the relation between rotational support $V/\sigma$ and axis ratios  $q$ and $p$ of the stellar and the dark matter components. Each galaxy has been divided in 12 radial bins (ellipsoidal or spherical, as described above) in the range $r\in[1\re{},8\re{}]$, thus each galaxy is represented by $\sim12$ data points\footnote{As already discussed in Sect.~\ref{sec:measuring_IntrinsicShape}, we can reliably measure intrinsic shapes out to 8\re{} for 96\% of the selected ETGs. } in Fig.~\ref{fig:local_shape_and_rotation_starsVSdm}.

We see that in general the stellar component rotates faster that the dark matter and it is more flattened (i.e., $q(\rm stars)<q(\rm DM)$). In particular, while the stars have axis ratio $q(\rm stars)$ that can vary from 0.2 to 0.8, the dark matter flattening is contained in a much narrower range of values ($0.6\lesssim q(\rm DM) \lesssim 0.9$, stretching to lower values only in very few cases). In high mass galaxies ($M_{*}>10^{10.9}\MSUN{}$) the dark matter is more elongated than the stars ($p(\rm stars) > p(\rm DM)$) but at lower masses and at low $\fex{}$ it is as oblate as the stellar component ($p(\rm DM) \sim p(\rm stars)>0.8$).
The stars show progressively lower rotational support with increasing stellar mass and move from the lower right corner of the diagrams, around the $y=x/10$ line, towards the 1:1 line. Their intrinsic shapes show a similar behavior, becoming rounder (i.e., with higher $q$), while both stars and dark matter become more triaxial (i.e., with lower $p$). 

Each data point in Fig.~\ref{fig:local_shape_and_rotation_starsVSdm} is color coded by the local ex-situ fraction $\fex$ derived at the same radius $r$ at which the rotational support and the intrinsic shapes are measured. The colors show that the stellar and the dark matter components tend to have similar amount of rotational support and similar intrinsic shapes where the fraction of the ex-situ stars is high ($\fex{}\gtrsim0.5$). 

\begin{figure}
    \centering
    \xincludegraphics[width=1.\linewidth,label=\textbf{A}]{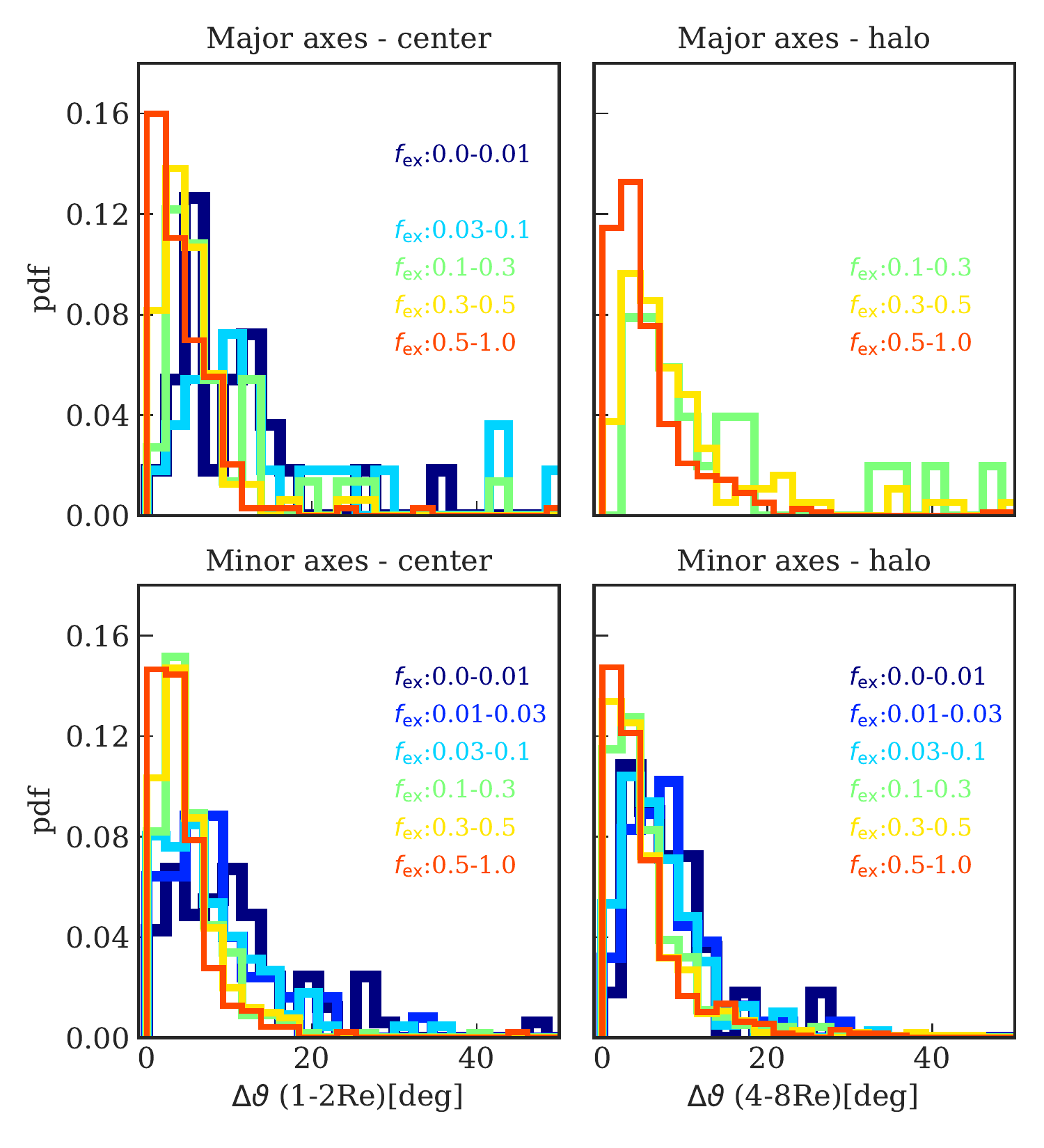}
    \xincludegraphics[width=1.\linewidth,label=\textbf{B}]{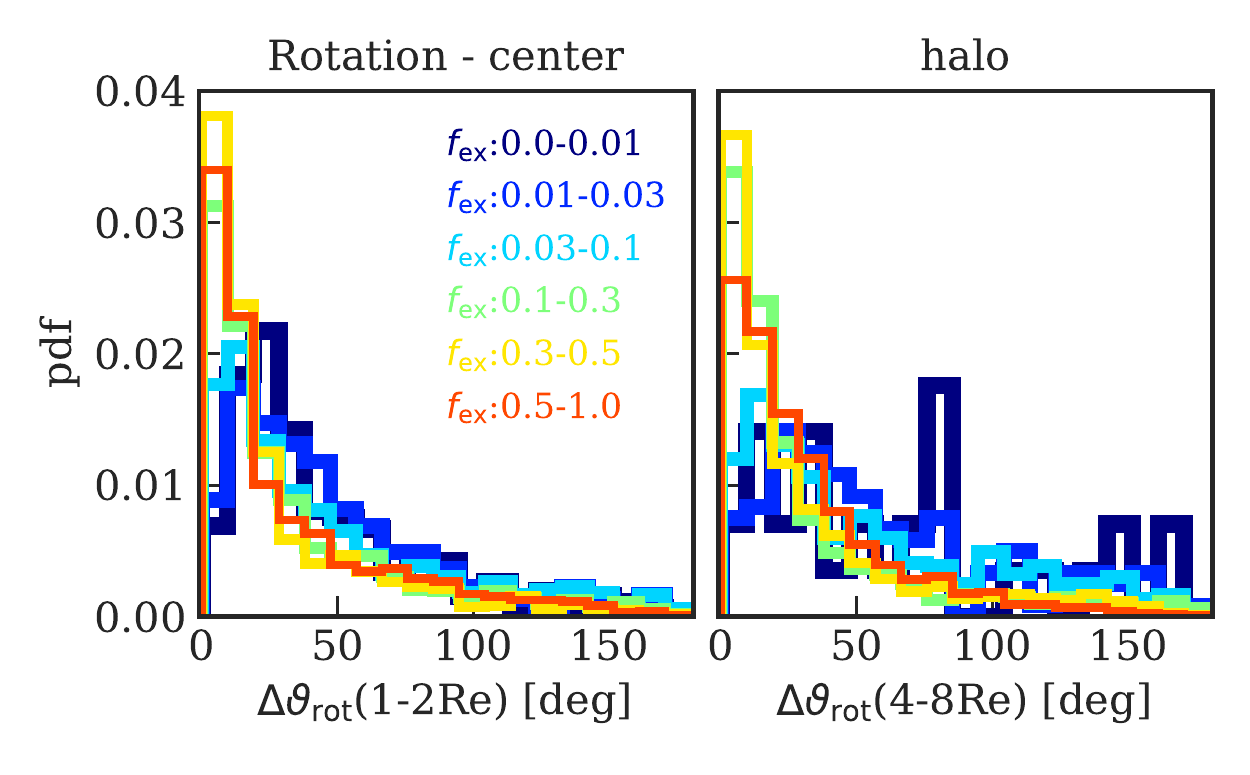}
     \caption{\textbf{A:} Distribution of misalignments of the principal axes of the stellar and dark matter components for different local ex-situ fractions $\fex$ as shown in the legend. \textbf{Top panels}: misalignments of major axes for the non-oblate systems; \textbf{bottom panels}: misalignments of minor axes for the non-prolate systems. \textbf{B:} Distribution of misalignments in the direction of rotation of the stellar and dark matter components for different local ex-situ fractions $\fex$. In each figure, we distinguish between central regions (1-2\re{}) and outskirts (4-8\re{}) and we show only distributions containing at least 20 galaxies. Stellar and dark matter components are increasingly aligned with increasing $\fex$.}
    \label{fig:stars_dm_misalignment}
\end{figure}

\subsection{Principal axes and direction of rotation of dark matter and stellar components}\label{sec:misalignment_dmVSstars}

The similarity between stellar and dark matter components at growing $\fex{}$ extends also to their directions of rotation and directions of the principal axes. 
Misalignments between the major axes and between the minor axes $\Delta\vartheta_j$ of the stellar and dark matter components are derived from the dot product of the eigenvectors $\hat{e}_j$ of their inertia tensor at each radius $r$:
\begin{equation}
    \Delta\theta_j = \cos^{-1}(\hat{e}_{{\rm stars},j} \cdot \hat{e}_{{\rm DM},j}),
\end{equation}
where $j\in\left[{\rm major\,axis, minor\,axis}\right]$.

The misalignment angle between the direction of rotation of the stars and of the dark matter, $\Delta\theta_{\rm rot}(r)$, is measured in each spherical shell of radius $r$ from the direction of the velocity vectors including their sign, so that $\Delta\theta_{\rm rot}\in[0,180]$ degrees. For each galaxy we derive $\Delta\theta_j(r)$, $\Delta\theta_{\rm rot}(r)$, and $\fex(r)$ in radial bins of radius $r$. Figure~\ref{fig:stars_dm_misalignment} shows the distributions of $\Delta\theta_j(r)$ and  $\Delta\theta_{\rm rot}$ in bins of $\fex$:  the left panels show the distributions from shells  with radii $1\re{}\leq r\leq 2\re{}$, while the right panels those from shells with radii $4\re{}\leq r\leq 8\re{}$.  \footnote{In the distributions of the principal axes misalignments, we excluded $\Delta\theta_{\rm major}(r)$ and $\Delta\theta_{\rm minor}(r)$ measurements from ellipsoids with intrinsic shapes respectively close to oblate ($p(\rm stars)>0.9$ or $p(\rm dm)>0.9$, $77\%$ of the ellipsoids) and to prolate ($q/p(\rm stars)>0.9$ or $q/p(\rm dm)>0.9$, $19\%$ of the ellipsoids), for which the direction of the principal axes is more uncertain (see Sect.~\ref{sec:measuring_IntrinsicShape}). Thus the distribution of $\Delta\theta_{\rm major}(r)$ applies for the remaining triaxial to prolate shapes, while the 
$\Delta\theta_{\rm minor}(r)$ for the triaxial to oblate shapes. The consequence is that, for example, the distribution of $\Delta\theta_{\rm major}(r)$ for galaxies with low $\fex$ is not well sampled, as most galaxies are near-oblate (see Fig.~\ref{fig:shape_fex_local}). Therefore we should assess misalignments by considering together the $\Delta\theta_{\rm major}(r)$ and $\Delta\theta_{\rm minor}(r)$ distributions.}

Stellar and dark matter halo components generally have well aligned principal axes, within $\sim20$ deg, both in the centers and in the outskirts.
Their alignment improves with higher ex-situ stellar contribution in both regions. Together Figs.~\ref{fig:local_shape_and_rotation_starsVSdm} and \ref{fig:stars_dm_misalignment} reveal that wherever the local fraction of ex-situ stars is large enough ($\fex\gtrsim50\%$) stellar and dark matter components tend to have similar intrinsic shapes and better aligned principal axes. 
The right panels in Fig.~\ref{fig:stars_dm_misalignment} show that also the directions of rotation of the stellar and dark matter components tend to better align for high $\fex$. 
Therefore, with growing \fex{}, stellar and dark matter halo acquire similar degree of rotational support (Fig.~\ref{fig:local_shape_and_rotation_starsVSdm}).

\subsection{Mergers and coupling between stellar and dark matter components}

\begin{figure}
    \centering
    \xincludegraphics[width=0.9\linewidth,label=\textbf{A}]{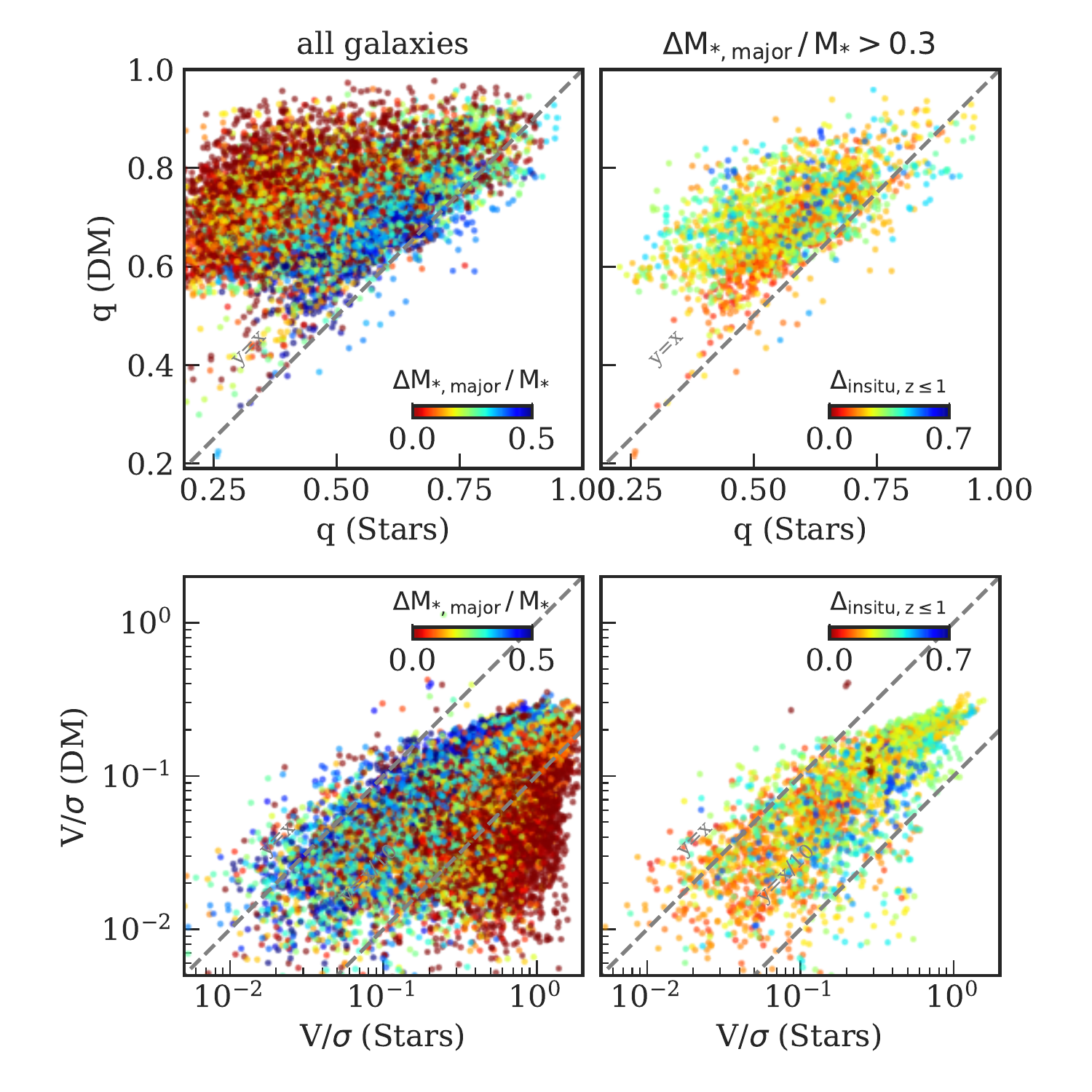}
    \xincludegraphics[width=0.9\linewidth,label=\textbf{B}]{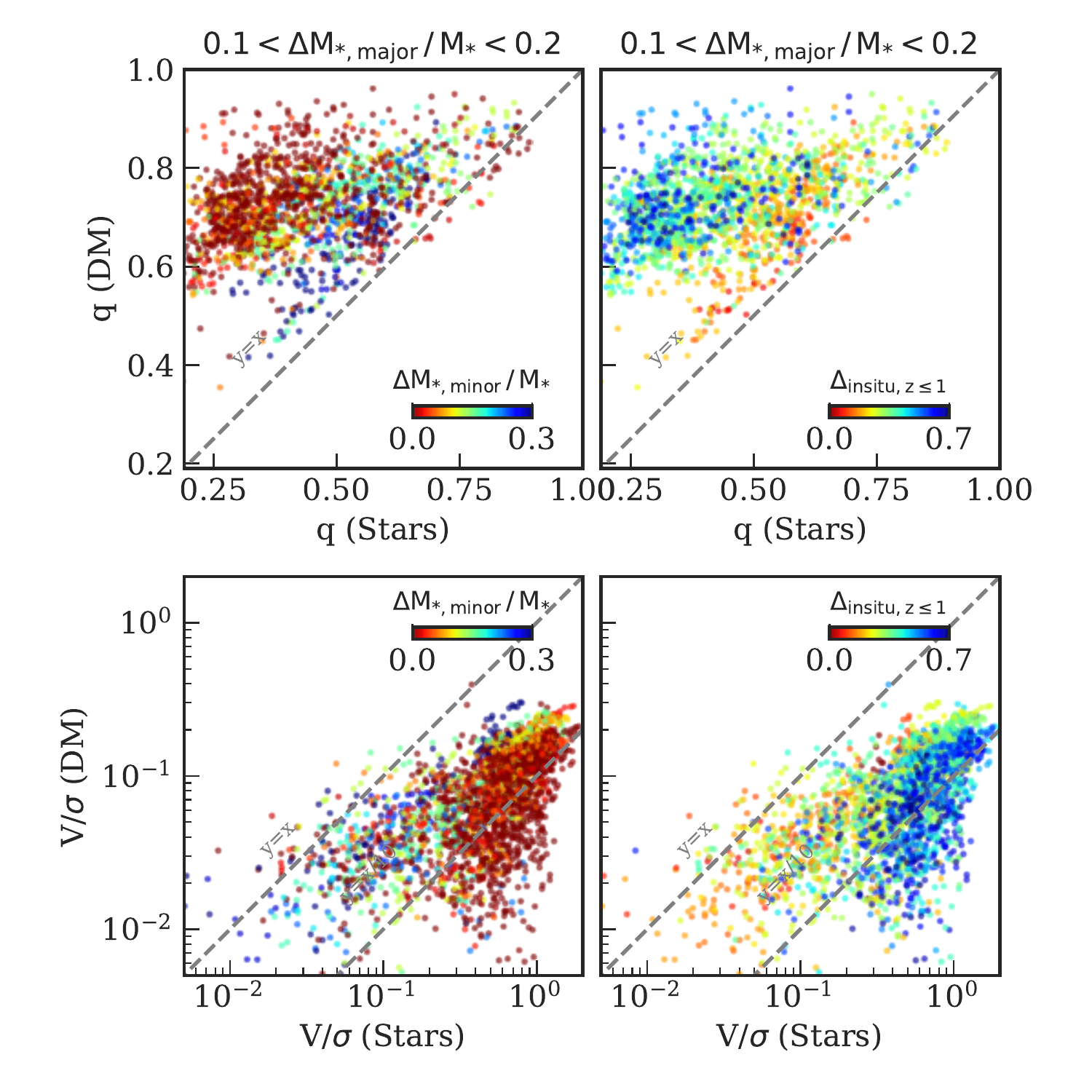}
    \caption{Coupling between stellar and dark matter components versus mergers. \textbf{A:} Major mergers. The \textbf{left} panels show all the TNG ETGs, color coded by their fraction of stellar mass accreted by major mergers. On the \textbf{right} we consider only galaxies with at least $30\%$ of their stellar mass accreted from major mergers and color code them according to the fraction of recently produced in-situ stars $\Delta_{\rm insitu,z\leq1}$. Major mergers cause the coupling between stellar and dark matter components, especially in galaxies that did not accrete cold gas recently. \textbf{B}: Minor mergers. Here we consider galaxies with stellar mass fraction accreted from major mergers in the narrow range $0.1<\Delta M_{\rm *,major} / M_{*}<0.2$. In the \textbf{left} panels galaxies are color coded by the fraction of stellar mass accreted from minor mergers $\Delta M_{\rm *,minor} / M_{*}$ and on the \textbf{right} by the fraction of recently produced in-situ stars $\Delta_{\rm insitu,z\leq1}$. Dry minor mergers also help to set up the similarity between stellar and dark matter halo structure.}
    \label{fig:Coupling_mergers}
\end{figure}

Next we explore the dependence of the coupling between stellar and dark matter component on the merger history. 

The first suspects are the major mergers. Major mergers are known to efficiently mix together stellar and dark matter particles through violent relaxation \citep[e.g.][]{2012MNRAS.425.3119H}.  Figure~\ref{fig:Coupling_mergers}A shows again the local relation between intrinsic shapes and rotational support of the dark matter and stellar components. All galaxies, each represented by $\sim12$ data points as in Fig.~\ref{fig:local_shape_and_rotation_starsVSdm}, are now color-coded according to the total fraction of stellar mass from major mergers $\Delta M_{\rm *,major}/M_{*}$. From Fig.~\ref{fig:Coupling_mergers}A it is clear that in galaxies with higher $\Delta M_{\rm *,major}/M_{*}$ stars and dark matter have more similar structure. In particular, if we isolate the galaxies with the highest $\Delta M_{\rm *,major}/M_{*}$ fractions, we see that those with the lowest fraction of new in-situ stars $\Delta_{\rm insitu,z\leq1}$ are closest to the 1:1 line. Since $\Delta_{\rm insitu,z\leq1}$, $\Delta M_{\rm *,major}/M_{*}$, and $z_{\rm last}$ tightly depend on stellar mass (Fig.~\ref{fig:accretion_classes_merger}), the galaxies with high $\Delta M_{\rm *,major}/M_{*}$ and low $\Delta_{\rm insitu,z\leq1}$ are massive recent dry major merger remnants. 
The stars of low mass galaxies with low ex-situ fractions (see the leftmost panels in Fig.~\ref{fig:local_shape_and_rotation_starsVSdm}) are not coupled with the dark matter component. Most of these systems are in-situ dominated and their accreted mass is contributed mainly by mini mergers (Fig.~\ref{fig:f_profiles}), which are not massive enough to influence the dark matter halo.
However minor mergers can play a role in producing the stars/dark matter coupling. Figure~\ref{fig:Coupling_mergers}B isolates galaxies within a narrow range of $\Delta M_{\rm *,major}/M_{*}\in[0.1,0.2]$. This selection includes higher mass systems but, at the same time, limits the contribution from major mergers.
Fig.~\ref{fig:Coupling_mergers}B shows a correlation with the fraction of mass from minor mergers $\Delta M_{\rm *,minor}/M_{*}$. The most coupled galaxies are, on average, those with the largest $\Delta M_{\rm *,minor}/M_{*}$ fractions and those with lowest recent in-situ star formation. On the other hand, we find that the position of this sample of galaxies relative to the 1:1 line does not correlate with the fraction of mass accreted from mini mergers $\Delta M_{\rm *,mini}/M_{*}$, neither with that from major mergers within the interval $0.1\leq\Delta M_{\rm *,major}/M_{*}\leq0.2$. 
A coupling between star and dark matter from minor mergers might originate from the simultaneous stripping of stellar and dark matter particles, which are added coherently to the host.

\section{ETGs with cores made of stars from a high-redshift compact progenitor}
\label{sec:discussion_RNcores}

Massive ETGs are thought to form early, at $z>2$, beginning with intense bursts of star formation producing compact, massive ($M_{*}\gtrsim10^{11}M_{\odot}$), disk-like galaxies, which quickly evolve into passive red galaxies (\citealt{2005ApJ...626..680D, 2008ApJ...687L..61B, 2008ApJ...677L...5V, 2009ApJ...695..101D}, as also found in cosmological simulations, e.g.  \citealt{2015MNRAS.449..361W} using Illustris). 
The evolution of these objects into present day ETGs is dominated by a strong size growth driven by mergers and renewed star formation \citep[e.g.][]{1993ApJ...416..415H, 2006ApJ...650...18T, 2009ApJ...699L.178N, 2010ApJ...709.1018V, 2014ApJ...788...28V, 2017MNRAS.465..722F}.

\begin{figure}
\begin{center}
    \includegraphics[width=\linewidth]{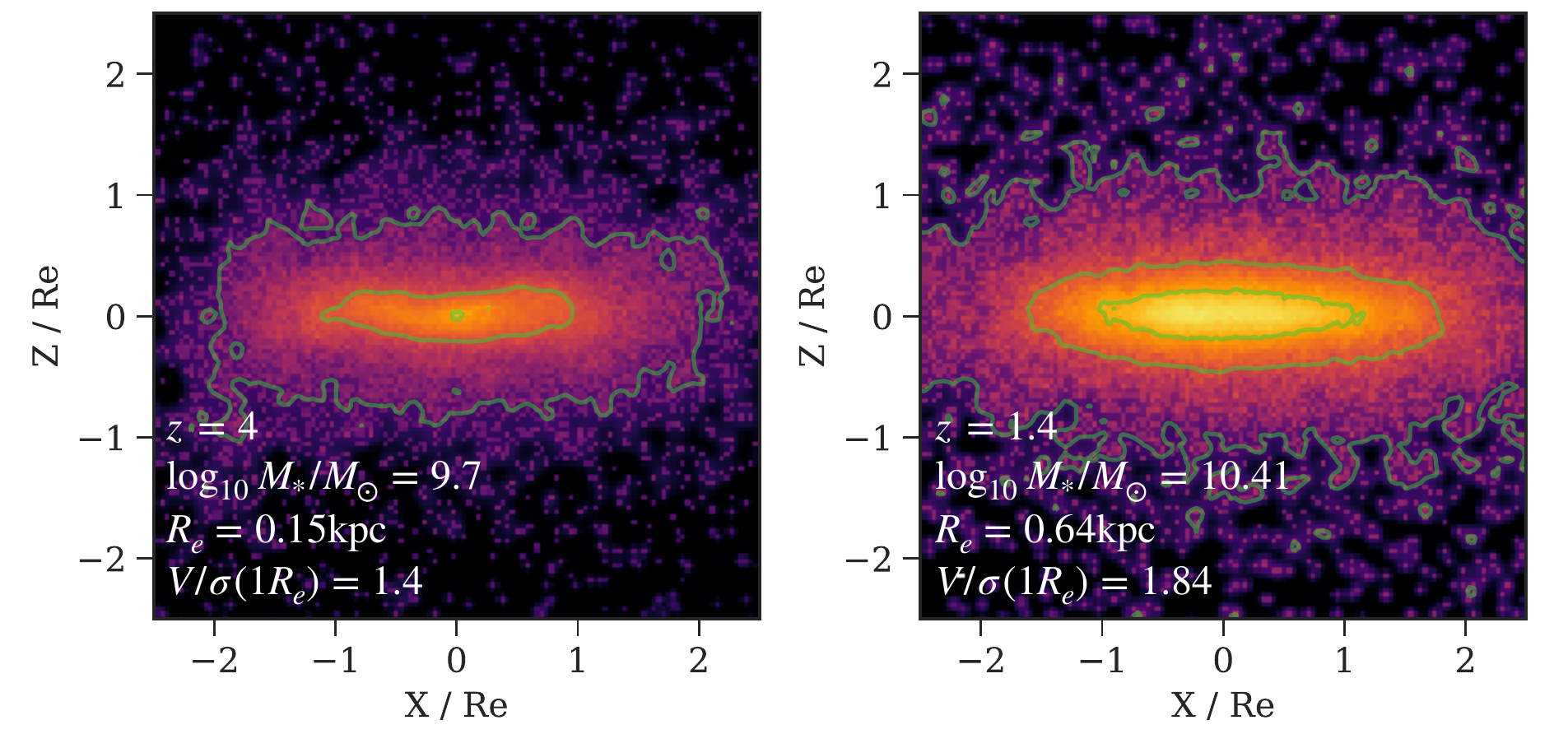}
    
    \vspace{0,5cm}
    
    \includegraphics[width=\linewidth]{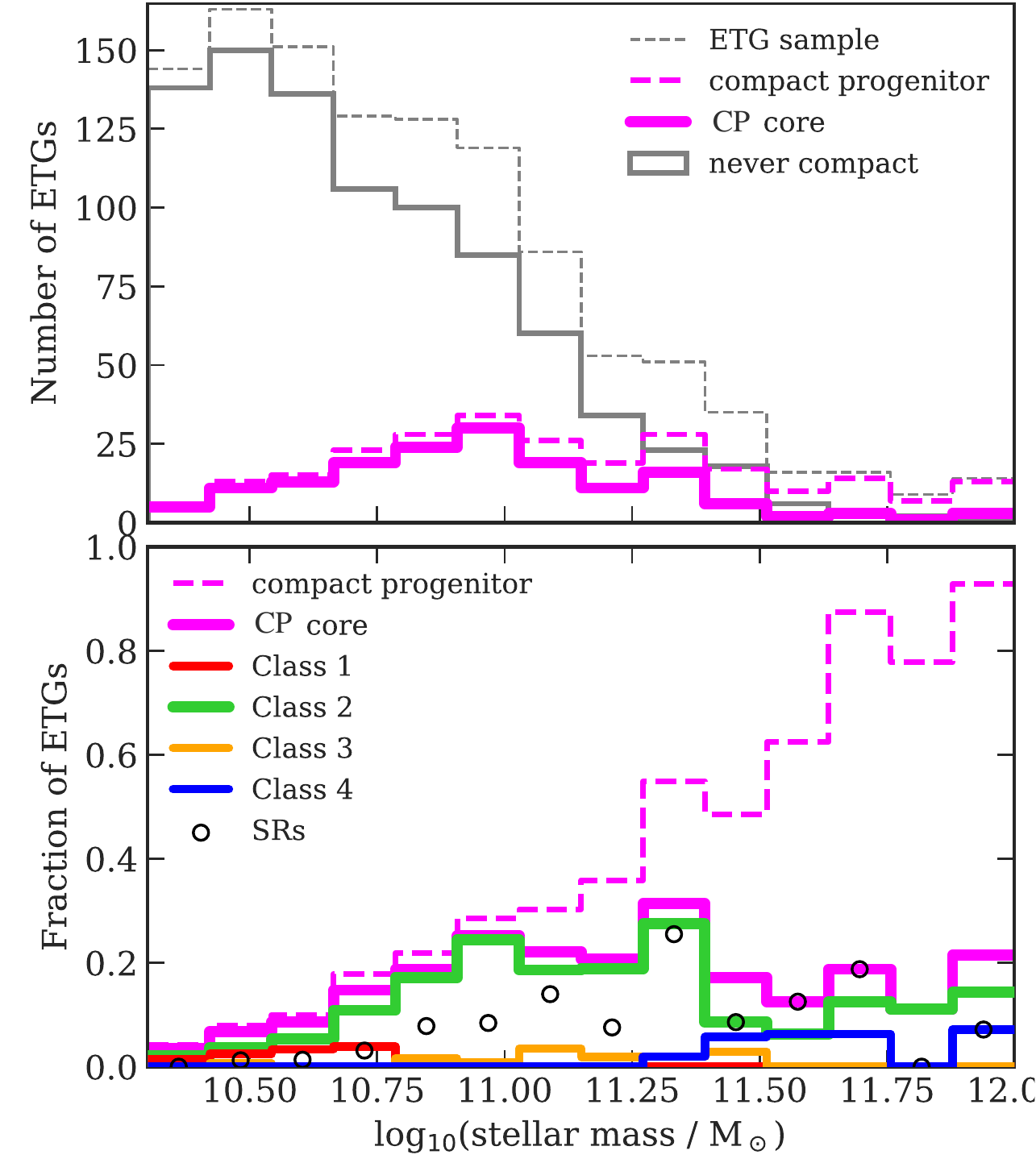}
    \vspace{0cm}
    \caption{\textbf{Top}: Two examples of high-redshift compact progenitors (CPs) in TNG100, progenitors of a FR (left) and of a SR (right). Both CPs are shown edge-on and have rather disky shapes. Physical parameters are reported on the figures. 
    \textbf{Center}: Mass functions of the entire sample of TNG ETGs (grey dashed), of the ETGs descending from a compact progenitor (magenta dashed), and of the ETGs with "surviving compact progenitor cores" (CP-cores, magenta solid line). The distribution of galaxies whose main progenitors were never compact is also shown (grey solid line). \textbf{Bottom}: Fraction of galaxies, in mass bins, with compact progenitors, with surviving CP-cores, total and for SRs, and with surviving CP-cores split by accretion class. The fraction of galaxies with compact progenitors is a strong function of stellar mass but the fraction of objects with surviving CP-cores is nearly constant at $\sim20\%$ at $M_{*}\gtrsim10^{10.7}\MSUN{}$, and decreases at lower masses.  }
    \label{fig:GalaxiesCompactProgenitor}
\end{center}
\end{figure}

A fraction of 
these compact red early galaxies formed at high $z$ may survive intact to present times as relic galaxies \citep{2018A&A...619A.137B, 2020ApJ...893....4S}, but these are extremely rare objects and, to date, only a handful have been spectroscopically confirmed \citep[][]{2014ApJ...780L..20T, 2017MNRAS.467.1929F,2020arXiv201105347S}. However, the compact red progenitor galaxies are also expected to be found in the cores of present-day massive ETGs \citep[e.g][]{2009MNRAS.398..898H,2009ApJ...697.1290B}. The study of such systems at $z=0$ is crucial for understanding the properties of the primordial in-situ component of ETGs, giving insights into the conditions for the formation of these objects and on the mechanisms that regulated star formation at high $z$. 

Simulations such as IllustrisTNG can provide understanding of which present-day ETGs are likely to contain cores dominated by stars from a compact progenitor (CP) galaxy
and how these could be disentangled from the accreted component. In Sect.~\ref{sec:accretion_classes} we showed that ex-situ stars can be distributed very differently in ETGs according to their merger history and that the most massive galaxies in our sample are dominated by ex-situ stars up to their very central regions (Fig.~\ref{fig:definition_classes}). In this section we identify simulated ETGs that still contain a core made of CP stars and characterize their structural and accretion properties.

We first identified the galaxies whose progenitors are compact galaxies by studying the compactness of each galaxy as function of time. The compactness can be quantified through the observationally motivated parameter \citep[e.g.][]{2013ApJ...765..104B}
\begin{equation}
    \Sigma_{1.5} = M_*/r_e^{1.5},
\end{equation}
where $r_e$ is the spherical stellar half mass radius.
For each galaxy, we tracked $\Sigma_{1.5}$ back in redshift along the main progenitor branch of its merger tree. Following \citet{2016MNRAS.456.1030W}, we consider any galaxy exceeding the threshold 
\begin{equation}
    \log_{10}\Sigma_{1.5} \mathrm{kpc^{1.5}/M_{\odot}}= M_*/r_e^{1.5}>10.5
    \label{eq:compactness}
\end{equation}
as compact.

As shown by the mass functions in the central panel of Fig.~\ref{fig:GalaxiesCompactProgenitor}, not all ETGs in our TNG sample descend from a CP; i.e., only a fraction of the galaxies overcomes the threshold in Eq.~\eqref{eq:compactness} at some point in their history. The fraction of ETGs with a CP is a strong function of stellar mass: in the range $10^{10.5}\MSUN{}\leq M_*\leq10^{11}\MSUN{}$ only 18\% of the ETGs were compact in the past, while at $M_*>10^{11.5}\MSUN{}$ this fraction rises to 80\%. We find that the compact phase of a galaxy can occur at different epochs, in the range $1\lesssim z \lesssim 4$. \citet{2016MNRAS.456.1030W} found very similar results using the original Illustris simulation. In particular, they observed a trend between compactness of the progenitor, the present-day stellar mass, and the galaxy formation time. This correlation is still present for the TNG100 ETGs and can, to some extent, explain the strong mass dependence of the fraction of ETGs with CPs: low mass galaxies tend to form later, when the universe is less dense and less gas-rich, and the conditions are less favorable to the formation of compact objects. 

Once the sample of ETGs with a CP is identified, we use the unique indices for each stellar particle to determine the mass fraction of compact progenitor stars in present day ETGs, within a radius of $2$ kpc from the center. We refer to galaxies with central CP mass fraction higher than 50\% as ETGs with "surviving compact progenitor core" or CP-core ETGs. Figure~\ref{fig:GalaxiesCompactProgenitor} shows the stellar mass distribution of these objects. We find that the fraction of CP-core ETGs grows with stellar mass in the range $10.3\leq \log_{10}M_*/\MSUN{}\lesssim10.7$, together with the fraction of ETGs with CPs. At higher masses, while the latter continues to grow, the fraction of surviving CP-cores remains constant at $\sim 20\%$ almost independent of stellar mass. This indicates that at low masses, whenever a galaxy descends from a CP, the primordial in-situ population typically survives in its central regions. At stellar masses above $10^{10.7}\MSUN{}$, even though the fraction of CPs increases with mass, the chances of CP-core survival decrease, as the CP-cores are more likely destroyed by major merger events (see Fig.~\ref{fig:accretion_classes_merger}). This is also shown by the distribution of CP-cores divided by accretion class in Fig.~\ref{fig:GalaxiesCompactProgenitor}: CP-core ETGs belong almost exclusively to class 2 and include both FRs and SRs.\footnote{We note the presence of a few class 4 galaxies among CP-core ETGs. For these galaxies, the central CP mass fraction is not larger than 65\%, and already their CPs contain a large population of ex-situ stars, up to 73\%.}

\begin{figure}
    \centering
    \includegraphics[width=\linewidth]{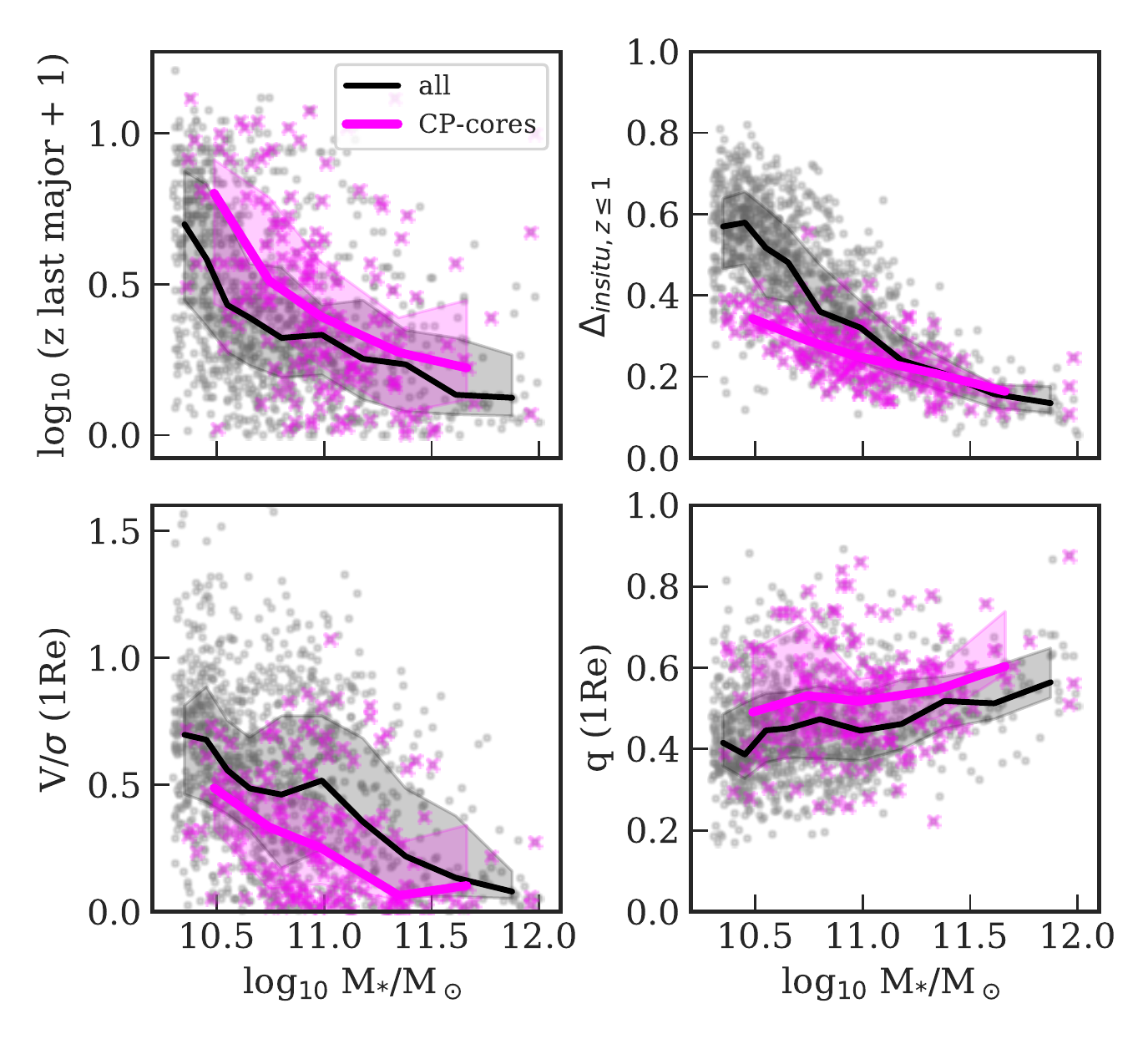}
    \caption{The sample of galaxies with surviving compact progenitor (CP) cores (magenta) compared to the whole sample of ETGs (grey). Panels show the redshift of the last major merger, fraction of recently formed in-situ stars $\Delta_{\rm insitu, z\leq1}$, rotational support $V/\sigma$, and minor-to-major axis ratio $q$ at 1 \re{} as functions of stellar mass. Solid lines show the median profiles and shaded regions enclose the quartiles of the distributions. Galaxies with CP-cores had fewer recent major mergers, formed fewer in-situ stars after redshift $z=1$, and are characterized by lower rotational support and rounder shapes than all ETGs of similar mass.}
    \label{fig:PropertiesGalaxiesCompactProgenitor}
\end{figure}

Figure~\ref{fig:PropertiesGalaxiesCompactProgenitor} compares the properties of the CP-core ETGs to those of the whole sample of TNG100 ETGs. While the total in-situ mass fractions and the mean accreted gas fractions (not shown here for brevity) are similar among galaxies of similar mass, the fraction of recently formed in-situ stars is systematically lower in CP-core ETGs, especially at $M_*<10^{11}\MSUN{}$. In other words, the recent accretion of gas and subsequent formation of new in-situ stars contributes to reducing the central mass fraction of primordial CP stars. 
Similarly mild differences are also found in the  recent accretion history of CP-cores: even though their accreted stellar mass fractions from major mergers are comparable to ETGs with similar mass (not shown here), these accretion events happened at earlier times (Fig.~\ref{fig:PropertiesGalaxiesCompactProgenitor}). 
Therefore, CP-core ETGs formed their in-situ population earlier than normal ETGs and their subsequent evolution is relatively quieter in terms of recent gas accretion, star formation, and accretion from late mergers.
The lower rate of recent gas accretion in CP-core galaxies is related to lower rotational support in their central regions as well as somewhat more spherical shapes (bottom panels of Fig.~\ref{fig:PropertiesGalaxiesCompactProgenitor}. 

We note that many of the early progenitors of the CP-core ETGs have rather disky shapes and are dominated by rotational support; the two examples shown in the top panels of Fig.~\ref{fig:GalaxiesCompactProgenitor} have $V/\sigma(1\re{})>1$. On the other hand, in Fig.~\ref{fig:PropertiesGalaxiesCompactProgenitor} almost none of the CP-core ETGs overcomes $V/\sigma(1\re{})>1$. This is in agreement with kinematic results from relic galaxies \citep[e.g.][]{2017MNRAS.467.1929F}, which are all characterized by high rotation velocity ($\sim 200-300$ km/s), while the CP-core galaxy NGC3311 has negligible rotation \citep{2018A&A...609A..78B,2020arXiv201211609B}. The investigation of the merger histories of these systems and of how mergers modify the dynamical properties of the CP-cores will be the subject of a future study.

Based on the results from the TNG100 simulation, selecting for slowly-rotating, rounder galaxies in observational studies can increase the probability of finding a CP-core. For example, a limit on $V/\sigma(1\re{})<0.1$ or, alternatively, a limit on $q(1\re{})>0.6$, doubles the fraction of CP-cores in a sample with $M_*>10^{10.7}\MSUN{}$. 

\section{Discussion}\label{sec:discussion}

\subsection{Rotational support and intrinsic shapes versus accretion history}\label{sec:discussion_stars}

Sections \ref{sec:Vsigma_profiles} and \ref{sec:Intrinsic_shapes} were dedicated to the study of how the rotational support and intrinsic shape profiles of the ETGs in TNG100 depend on their accretion history.
In the selected sample of ETGs, we find that low mass in-situ dominated galaxies with negligible accreted fractions are characterized by "peaked-and-outwardly-decreasing" $V_{*}/\sigma_*(R)$ profiles (Fig.~\ref{fig:Vsigma_profiles}), near-oblate stellar shapes, and increasing axis ratio $q(r)$ profiles at large radii that follow the decrease in rotation (e.g. Fig.~\ref{fig:shape_mergers}B). In galaxies with larger masses the in-situ components have similarly shaped rotation profiles, but the galaxy structure is altered by the accreted stars and gas from merger events in a way that is strongly dependent on stellar mass. In this work we mostly focused on the effect of different mass ratio mergers and of the recently accreted cold gas (parametrized by $\Delta_{\rm insitu, z\leq1}$).

We find that the merger history of TNG100 ETGs is dominated by major and mini mergers (see Sect.~\ref{sec:accretion_classes_fprofiles}).
Major mergers, especially the gas-poor ones, are effective in erasing the peak of rotation and flattening the $V_{*}/\sigma_*(R)$ profiles (Fig.~\ref{fig:Vsigma_profiles_mergers}A), while at the same time they lead to more spherical shapes and increased triaxiality (Fig.~\ref{fig:shape_mergers}A). Minor mergers are less effective than major mergers but they can also contribute to more spherical-triaxial stellar halo shapes and suppressed rotation. 
In low mass galaxies without major and minor mergers, mini mergers have the somewhat unexpected effect of increasing the median galaxy rotational support at large radii (Fig.~\ref{fig:Vsigma_profiles_mergers}B) and, at the same time, decreasing their median axis ratio $q$ while keeping the shapes near-oblate (Fig.~\ref{fig:shape_mergers}B), as if they contributed to a size increase of the host disk. This effect would only be possible if the accretion of these low mass satellites occurred in the plane of the in-situ rotating disk-like component on rather circular orbits \citep{2017MNRAS.464.2882A}. \citet{2019MNRAS.487..318K} found that mini mergers can indeed increase the size of the galaxy disk provided that the impact occurs in the disk plane. Whether mini mergers are generally accreted from preferential directions is an interesting subject for a future study.

The dependence of the galaxy structure and rotational support on their merger history generates a correlation between the $V_{*}/\sigma_*(R)$, $q(r)$, and $T(r)$ profile shapes and the total stellar mass, and therefore fraction of in-situ (or accreted) stars (Figs.~\ref{fig:Vsigma_profiles} and \ref{fig:shape_profiles_insitu}), and between the local rotational support and intrinsic shapes and the local ex-situ mass fraction $\fex$ (Figs.~\ref{fig:Vsigma_local} and \ref{fig:shape_fex_local}). At increasing contribution of ex-situ stars, the stellar halo $V_{*}/\sigma_*$ decreases and shapes become rounder and more triaxial. Low mass, in-situ dominated galaxies show poor correlations between rotational support and shapes and the local $\fex{}$, as they are completely determined by the in-situ star. 

\subsection{Kinematic transition radius}\label{sec:discussion_RTkin}

\begin{figure}
    \centering
    \includegraphics[width=1\linewidth]{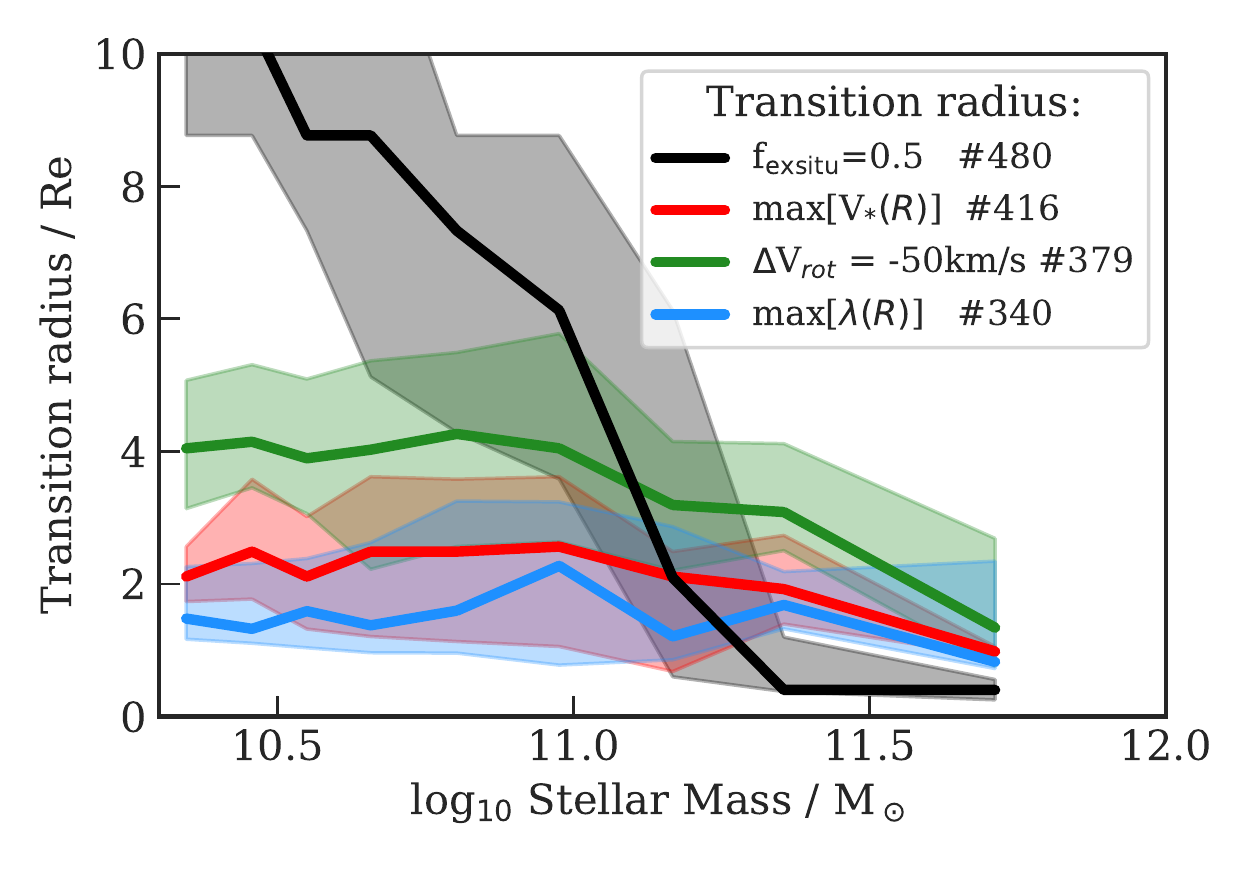}
    \caption{Median TNG100 kinematic transition radii (red, green, and blue solid curves) as a function of galaxy stellar mass, compared with the median transition radius $R_{\rm T, exsitu}$ between in-situ and ex-situ dominated regions (black line). All the curves are obtained from FR galaxies of class 2 (see Fig.~\ref{fig:definition_classes}) with well defined $R_{\rm exsitu}$ between in-situ dominated centers and ex-situ dominated outskirts. The number of galaxies used to produce each curve is reported in the legend.}
    \label{fig:transradius_mass}
\end{figure}

Previous studies have proposed kinematic transitions as signatures of the two-phase formation scenario, with the drop in rotational support marking the transition between in-situ to ex-situ dominated regions \citep[e.g.][]{2016MNRAS.457..147F, 2018A&A...618A..94P}. In Sects.~\ref{sec:Vsigma_profiles_insitu} we demonstrated that, for the TNG ETGs, the decrease in rotation is mainly driven by the in-situ stars. In particular we observe that kinematic transitions typically occur in the in-situ dominated parts of galaxies, where the fraction of accreted stars is everywhere irrelevant ($\fex{}(R)<0.1$).
Hence a kinematic transition radius $R_{\rm T,kin}$ defined by either by the drop in rotation \citep{2018A&A...618A..94P} or by the location of the peak in $\lambda$ parameter \citep{2020MNRAS.493.3778S} does not generally trace a particular value of $\fex{}$ in the TNG galaxies. This means that the kinematic transition radius in TNG galaxies does not  generally correspond to the transition radius $R_{\rm T,exsitu}$ between the in-situ and the ex-situ dominated regions where $\fex{}$ = 0.5. 

This is shown in Fig.~\ref{fig:transradius_mass}. We considered FRs in accretion class 2 for which we can define a transition radius $R_{\rm T, exsitu}$ from in-situ dominated centers to ex-situ dominated outskirts. This is similar, although not identical, to the definition of $R_{\rm T, exsitu}$ in \citet{2016MNRAS.458.2371R} for the original Illustris simulation. The median $R_{\rm T, exsitu}$ as a function of stellar mass is shown in Fig.~\ref{fig:transradius_mass} with a black curve (see also Table~\ref{tab:transradius_mass}). The other curves represent the median kinematic transition radii $R_{\rm T,kin}$ for the same group of galaxies derived using different definitions:
\begin{itemize}
    \item by taking the position of the maximum of $V_{*}(R)$ as defined in Eq.~\ref{eq:V_2D} (red curve)
    \item by considering the mean radius between the position of the peak in the rotational velocity profile $V_{\rm rot}(R)$ and the radius at which $V_{\rm rot}(R)$ decreases by 50 km/s (as in \citealt{2018A&A...618A..94P},
    $V_{\rm rot}(R)$ derived as in \citetalias{2020A&A...641A..60P}, green curve)
    \item by taking the peak in the differential $\lambda(R)$ profile (as in \citealt{2020MNRAS.493.3778S},
    $\lambda(R)$ derived as in \citetalias{2020A&A...641A..60P}, blue curve).
\end{itemize}
Figure~\ref{fig:transradius_mass} shows that the kinematic transition radius in TNG is not related to the transition radius between in-situ and ex-situ dominated regions. While the median $R_{\rm T, exsitu}(M_{*})$ decreases quickly with increasing stellar mass, the median $R_{\rm T, kin}$ shows a weaker dependence on $M_{*}$. Lower mass systems have kinematic transition radii systematically smaller than $R_{\rm T, exsitu}$. For these galaxies the kinematic transition occurs in in-situ dominated regions (at $\fex{}<0.2$ for $M_{*}<10^{10.7}\MSUN{}$), while the ex-situ transition radius occurs at much larger radii. Since $\fex{}(R)$ depends on stellar mass (Fig.~\ref{fig:f_profiles}), in more massive systems the contribution of ex-situ stars to the rotational support is larger and sufficient to determine the anti-correlation $V_{*}/\sigma_{*}(\fex{})$. In these cases the kinematic transition accidentally occurs at radii closer to $R_{\rm T, exsitu}$, since also for these galaxies the "peaked-and-outwardly-decreasing" rotation profile shape is mainly determined by the rotation profile of in-situ stars (Fig.~\ref{fig:Vsigma_profiles}). This is demonstrated by the galaxies in the high mass end of Fig.~\ref{fig:transradius_mass} whose kinematic transition radius is much larger than $R_{\rm T, exsitu}$, showing again that the kinematic transition radius does not trace $R_{\rm T, exsitu}$ in the IllustrisTNG ETGs.

\citet{2020MNRAS.493.3778S} recently showed similar kinematic results using the Magneticum Pathfinder simulations but drew different conclusions. In their work they selected 492 galaxies with stellar masses larger than $10^{10.3}\MSUN{}$, although their cuts on the number of particles at $R>3\re{}$ excluded the majority of the galaxies with $M_{*}<10^{10.6}\MSUN{}$. A subset of this sample (19\%) have "peaked-and-outwardly-decreasing" differential $\lambda(R)$ profiles: this is made of galaxies that, on average, had the smallest fraction of accreted stars and did not undergo major mergers. \citet{2020MNRAS.493.3778S} argue that for these galaxies the accretion histories, dominated by mini and minor mergers, lead to the characteristic profile shape. This is motivated by the fact that the disruption of low mass satellites merging from random directions would enhance the dispersion in the halo without affecting the central embedded disk and create the decreasing $\lambda(R)$ profiles. On the other hand, their Fig. 11 reveals that the fraction of galaxies with this characteristic $\lambda(R)$ profile shape was much higher at $z=2$, when the galaxies are mostly made of in-situ stars. At lower redshifts, when galaxies progressively increase their stellar mass through both in-situ star formation and accretion, the fraction of "peaked-and-outwardly-decreasing" $\lambda(R)$ profiles drops in favour of other profile shapes. This suggests that the "peaked-and-outwardly-decreasing" $\lambda(R)$ profile is already set at high $z$ by the in-situ stars and is not determined the subsequent minor and mini mergers. The example galaxy shown in their Figs. 12 and 13 seems to confirm this conclusion.

Figure 15 in \citet{2020MNRAS.493.3778S} shows the relation between $R_{\rm T, kin}$ and $R_{\rm T, exsitu}$ for a fraction of their sample of galaxies with such “peaked-and-outwardly-decreasing" $\lambda(R)$ profiles. For these galaxies the figure shows a trend for galaxies with higher $\Delta M_{\rm *,major}/M_{*}$ to have kinematic transition radii larger that $R_{\rm T, exsitu}$, and for galaxies with lower $\Delta M_{\rm *,major}/M_{*}$ to have $R_{\rm T, kin} \sim R_{\rm T, exsitu}$.
Even though it is not straightforward without further data to estimate the stellar masses of the systems shown, if we assume for the Magneticum galaxies a similar merger history as for the TNG100 ETGs, the galaxies with higher $\Delta M_{\rm *,major}/M_{*}$ are also the more massive systems.
Therefore the apparent dependence on merger mass ratio in their Fig.~15 together with the selection against low-mass systems may actually be consistent with a stellar mass dependence similar to that observed for the TNG ETGs in Fig.~\ref{fig:transradius_mass}, in which the kinematic transition radius traces $R_{\rm T, exsitu}$ only for a special mass range, but not generally.

\subsection{Stars and dark matter halo relation}

In Sect.~\ref{sec:dmVSstars} we investigated the relation between the stellar and the dark matter components and found that the two are coupled where the fraction of accreted stars is highest; at $\fex{}\gtrsim0.5$ the stars have similar intrinsic shape and rotational support as the underlying dark matter component. Such high local ex-situ fractions occur in the most massive systems ($M_{*}>10^{11.2}\MSUN{}$) but also at sufficiently large radii in intermediate mass systems (Fig.~\ref{fig:f_profiles}) such as the class 2 galaxies which dominate at $M_{*}>10^{10.6}\MSUN{}$ (Fig.~\ref{fig:insituVSmass_classes}). 

The relation between the intrinsic shapes of the galaxies and their dark matter halos, as well as the alignment of their spins and of their principal axes, has been widely investigated in the literature with different sets of simulations.
The co-evolution of galaxies and inner dark matter halos ($r\sim0.1 r_{200}$) correlates their shapes and angular momenta in the central regions \citep[e.g.,][]{2004ApJ...611L..73K, 2013MNRAS.429.3316B, 2019MNRAS.484..476C}, while centers and outer dark matter halos are found to be more misaligned \citep[e.g.,][]{2005ApJ...627L..17B, 2011MNRAS.415.2607D, 2015MNRAS.453..469T}. 

In this work we show that if we follow the variations of galaxy properties with radius and analyse the relation between stellar and dark matter components locally, we find that the two components are correlated also at large radii \citep[see also][]{2014MNRAS.438.2701W}.
We find that gas poor accretion, quantified by \fex{}, efficiently couples the stars to the dark matter component. This might be due to either the efficient mixing of stars and dark matter in major mergers or to the coherent accretion of baryonic and dark matter at large radii in the case of minor mergers. Then the accreted stellar and dark matter components gather on roughly similar orbits, so the rotational support and shapes of both components would be similar. Thus where the accreted component dominates the mass, they would follow each other in the overall dynamics.
In-situ dominated regions instead rotate faster than the dark matter halo, have more flattened shapes, and show larger misalignments with the dark matter component at the same radius. 

Our results are in line with previous studies. For example, \cite{2015MNRAS.453..469T}, using the MassiveBlack-II cosmological hydrodynamic simulation, observed a tendency for galaxies with high triaxiality parameter (those with highest $\fex{}$ in TNG, see Fig.~\ref{fig:shape_fex_local}) to be more aligned with the dark matter halo, while oblate galaxies have higher misalignments.
\cite{2017MNRAS.466.1625Z}, using the original Illustris simulation, found that the stellar and dark matter total angular momenta are progressively better aligned at higher masses, where the mostly old stellar populations interact gravitationally with the dark matter and not hydro-dynamically with the gas. In lower mass halos, galaxies are typically younger and their stellar angular momentum aligns well with the gas, which, as opposed to the dark matter, is strongly influenced by feedback processes.
\cite{2019MNRAS.487.1607G} found that the galaxies in the EAGLE simulation with spin oriented perpendicularly relative to their filament, which are galaxies that are more massive and of earlier type, are the best aligned with their host halo, while galaxies with parallel spin show poor alignment. 
Lastly, for the most massive galaxies, which are those with the highest global ex-situ fractions, the slope of the spherically-averaged radial mass profiles of the stellar halos has been found to tend towards that of the underlying dark matter (slope of about -3,  \citealt{2014MNRAS.444..237P, 2018MNRAS.475..648P}).

The local coupling of stellar and dark matter halos at high local ex-situ fractions ($\fex>0.5$) implies that in massive enough galaxies, or at sufficiently large radii, we can approximate the intrinsic shapes and rotational support of the dark matter component with that measured from deep photometry of stellar halos and from kinematic tracers such as planetary nebulae or globular clusters.

\section{Summary}\label{sec:summary}

This paper is a follow up study of \citet{2020A&A...641A..60P}, where we characterized the stellar halo kinematic properties in a sample of early type galaxies (ETGs) selected from the IllustrisTNG cosmological simulations and connected them to variations in their intrinsic shapes. In the present paper we studied the dependence of the kinematics and structure of the simulated galaxies on their accretion history, using the same sample of ETGs from TNG100. Then we extended the investigation to the dark matter halos and studied how their rotation and shape properties are related with those of the stellar halos.

In Sect.~\ref{sec:history} we characterised the accretion history of the selected sample of ETGs by considering the total fraction of the accreted mass, the mass fraction contributed by different mass ratio mergers, the fraction of accreted cold gas, and the recent in-situ star formation. The tight correlation between stellar mass and these accretion parameter implies that a (i) galaxy’s stellar mass is a good indicator of its accretion history. The different merger histories of the TNG ETGs lead to different radial distribution of accreted stars within the galaxies so that (ii) galaxies can be divided in four accretion classes, each dominating in different intervals of stellar mass and total in-situ mass fraction (Fig.~\ref{fig:insituVSmass_classes}): low mass ETGs (up to $M_{*}\sim10^{10.6}\MSUN{}$) are in-situ dominated at all radii; intermediate ETGs have in-situ dominated central regions and ex-situ dominated outskirts, with a smaller group in which in-situ and ex-situ components interchangeably dominate at small and intermediate radii; and high mass galaxies ($M_{*}>10^{11.2}\MSUN{}$) are mostly ex-situ dominated from the center to the halo. (iii) Fast (FRs) and slow rotators (SRs) follow similar trends with stellar mass but, at fixed mass, SRs on average contain more accreted stars, they accreted less gas, and they had more recent major mergers (Figs.~\ref{fig:trendswithmass} and \ref{fig:fraction_FRandSR_accretion_classes}).

In Sect.~\ref{sec:Vsigma_profiles} we characterized the shape of the $V_{*}/\sigma_{*}(R)$ profiles as a function of stellar mass and total in-situ fraction (Fig.~\ref{fig:Vsigma_profiles}). For both FRs and SRs, the total median $V_{*}/\sigma_{*}(R)$ profiles in stellar mass bins can be approximated by the weighted sum of the rotational profiles of the in-situ and ex-situ components. We showed that (i) the $V_{\rm insitu}/\sigma_{*}(R)$ profiles of the in-situ stars typically have a "peaked-and-outwardly-decreasing" form, with peaks in rotation decreasing at higher stellar masses and accreted fractions, due to the effects of mergers; (ii) the ex-situ stars have median $V_{\rm exsitu}/\sigma_{*}(R)$ profiles that are largely constant with radius and independent of stellar mass (Fig.~\ref{fig:Vsigma_profiles}); and (iii) the median shapes of the overall $V_{*}/\sigma_{*}(R)$ profiles are essentially determined by the stellar mass and total in-situ fraction (or equivalently, total accreted mass fraction), whereas we found no residual correlations with other accretion parameters.  
    
Galaxy intrinsic shapes are also found to vary as a function of the in-situ fraction and stellar mass (Sect.~\ref{sec:Intrinsic_shapes}), such that (i) low mass galaxies with negligible fractions of accreted stars are consistent with near-oblate shapes and increasing axial ratio $q(r)$ in the halo, while (ii) galaxies with higher stellar mass and accreted fractions are less flattened and more triaxial (Fig.~\ref{fig:shape_profiles_insitu}).

In Sections~\ref{sec:Vsigma_merger_mass_ratio} and \ref{sec:shape_profiles_mergers} we found that mergers and recent gas accretion, whose importance strongly depends on stellar mass, change the galaxy structure by modifying both their rotational support and intrinsic shapes: (i) major mergers tend to erase the peak in rotation and flatten the $V_*/\sigma_*(R)$ profiles (Fig.~\ref{fig:Vsigma_profiles_mergers}A) while at the same time they increase the triaxiality and make the halo rounder; (ii) the recent accretion of cold gas acts in the opposite direction by preserving or building up the central rotation and making shapes closer to oblate (Fig.~\ref{fig:shape_mergers}A).  (iii) Minor mergers are less effective than major mergers but they can also contribute to more spherical-triaxial stellar halo shapes and suppressed rotation.
We found that in low mass galaxies with no major and minor mergers, mini mergers tend to increase the size of the host disks by building up an extended, flattened, rotating component (Figs.~\ref{fig:Vsigma_profiles_mergers}B and ~\ref{fig:shape_mergers}B).

Mergers and recent gas accretion play an important role also for the evolution of the stellar population mixture within galaxies. In Sect.~\ref{sec:discussion_RNcores} we found that in about $\sim20\%$ of high-mass ETGs the central regions are dominated by stars from a compact high-redshift progenitor galaxy. These systems had a quieter recent evolution with respect to similar mass galaxies in terms of late mergers and recent star formation. In particular, the lower rate of recent cold gas accretion in these galaxies is related to a higher probability of observing them among ETGs with slow rotation and rounder shapes.

Since the "peaked-and-outwardly-decreasing" $V_*/\sigma_*(R)$ profiles in the IllustrisTNG galaxies are characteristic of the low mass ($M_{*}<10^{10.6}\MSUN{}$) class 1 galaxies with little ex-situ stars and, more generally, of the in-situ component,  a kinematic transition radius defined either by the rotation peak or the drop in rotation does not trace a particular $\fex{}$ value (Fig.~\ref{fig:transradius_mass}). Only in sufficiently massive systems does the kinematic transition radius become (accidentally) similar to the transition radius between in-situ dominated and ex-situ dominated regions ($R_{\rm exsitu}$). Therefore our results cast some doubts on the use of the kinematic transition radius as tracer of $R_{\rm exsitu}$. 

However, if we exclude the low mass galaxies with $M_{*}<10^{10.6}\MSUN{}$) in which radial variations of rotation and flattening are driven by the in-situ stars alone, we find clear relations between the stellar halo structural parameters and the local fraction of ex-situ stars $\fex{}$. 
(i) The median stellar halo rotational support decreases strongly with $\fex{}$, with an average scatter of 0.16 in $\fex{}$ at constant  $V_*/\sigma_*(R)$ (Fig.~\ref{fig:Vsigma_local}). 
(ii) Low $\fex{}$ is consistent with near-oblate shapes with low $q$. At higher $\fex{}$ the stellar halo triaxiality increases, while the $q(\fex{})$ reaches a maximum of $q\sim0.6$ at $\fex{}\sim0.7$, beyond which it decreases again (Fig.~\ref{fig:shape_fex_local}).
(iii) SRs populate the same correlations in the stellar halo parameter distribution as the FRs, but at the high $\fex{}$ end. At large radii, where the local ex-situ fractions are high the two classes show a continuous sequence of stellar halo properties with a significant overlap.

If these correlations between the stellar halo rotational support, intrinsic shapes, and local $\fex$ found here also hold for real massive galaxies (with $M_{*}>10^{10.6}$), the measurement of these observationally accessible quantities could provide an estimate of the local ex-situ contribution. The sensitivity of these correlations to the adopted galaxy formation model in the simulations, which could, for example, influence the distribution of $\fex$ with radius (e.g. see Sect.~\ref{sec:accretion_classes}), should be tested with different sets of cosmological simulations.

In Sect.~\ref{sec:dmVSstars} we found that stellar and dark matter components are dynamically coupled in regions of high ex-situ fraction:
(i) For local ex-situ fractions $\fex{}\gtrsim0.5$ stellar and dark matter halos have similar intrinsic shapes and well-aligned principal axes, similar amount of rotational support, and a similar direction of rotation (Figs.~\ref{fig:local_shape_and_rotation_starsVSdm} and \ref{fig:stars_dm_misalignment}).
(ii) Galaxies with higher accreted fractions from (dry) major mergers are the systems with the closest coupling between stellar and dark matter components (Fig.~\ref{fig:Coupling_mergers}A), but we find that also minor mergers can contribute (Fig.~\ref{fig:Coupling_mergers}B).
The coupling of stars and dark matter may be due to the simultaneous mixing of stars and dark matter in major mergers (Fig.~\ref{fig:Coupling_mergers}A), or also the coherent accretion of stars and dark matter on similar orbits in smaller mass-ratio mergers (Fig.~\ref{fig:Coupling_mergers}B).

\section{Conclusions}\label{sec:conclusions}
Finally, we summarize our main conclusions from these results:

\vspace{6pt}
1) The stellar mass of an early type galaxy (ETG) in the TNG100 cosmological simulations is tightly correlated with its accretion history and with the distribution of the accreted stars. Low mass galaxies are dominated by the in situ component, intermediate mass systems typically have in situ-dominated cores and ex-situ dominated halos, and the highest mass systems are everywhere dominated by accreted, ex-situ stars.

\vspace{6pt}
2) The in-situ stars dominating in low mass ETGs typically have characteristic peaked-and-outwardly-decreasing rotation profiles and near-oblate shapes with axial ratio q(r) increasing in the halo. At higher stellar masses and accreted mass fractions, the rotation peak decreases but is still dominated by the in situ stars, and the galaxy halos become more triaxial. A kinematic transition radius defined by the position of either the peak or the drop in rotation does therefore not trace the transition between in-situ dominated and ex-situ dominated regions in a galaxy. 

\vspace{6pt}
3) Major mergers (with mass ratio $\mu>$1:4) dominate the accreted mass fraction in TNG100 ETGs over mini mergers ($\mu<$1:10) and finally minor mergers (1:4$>\mu>$1:10). Dry major mergers tend to decrease the peak in rotation and flatten the $V_{*}/\sigma_{*}(R)$ profiles, while at the same time they increase the triaxiality $T$ and make the halo rounder. Thus an approximate estimate of the local ex-situ mass fraction $\fex{}(R)$ can be obtained from the local rotational support $V_{*}/\sigma_{*}(R)$ and intrinsic shape parameters $q(r)$ and $T(r)$.

\vspace{6pt}
4) Fast rotators (FRs) and slow rotators (SRs) populate the same overall correlations between stellar halo rotational support, intrinsic shape, and local ex situ fraction $\fex{}$, but SRs are concentrated at the high $\fex{}$ end. Low-mass ETGs are mostly FRs with near-oblate shapes while only a few are SRs. At higher masses, FRs acquire accretion-dominated halos with slow rotation and triaxial shapes, while at the highest stellar masses most ETGs are SRs.

\vspace{6pt}
5) ETGs with a high-redshift compact progenitor galaxy surviving in their cores are systems which had a quieter recent evolution  compared to similar mass galaxies, in terms of mergers and recent star formation. This prevented the central regions from being diluted both by accreted and newly formed in-situ stars, leading to lower average rotation.

\vspace{6pt}
6) In regions where accreted stars dominate (i.e. where the local $\fex{}>0.5$), stellar and
dark matter halos have similar intrinsic shapes and rotational support, and similar principal and rotation axes. Therefore in sufficiently massive ETGs or at sufficiently large radii the local intrinsic shape and rotational support of the dark matter halos can be approximated with those obtained from photometry and extended kinematics of their stellar halos.

\begin{acknowledgements}
  \\
  We thank the anonymous referee for the helpful comments and valuable suggestions on this manuscript. C.P. is grateful to F. Hofmann for his support.
\end{acknowledgements}

\bibliographystyle{aa}
\bibliography{auto}

\begin{appendix}
\section{Tables}
\label{appendix:tables}
In this appendix, we provide tables reporting relevant physical relations derived throughout this paper. Tables~\ref{tab:definition_class1}, \ref{tab:definition_class2}, \ref{tab:definition_class3}, and \ref{tab:definition_class4} contain the median stellar mass density and cumulative stellar mass fraction profiles for the in-situ and the ex-situ components in each of the four accretion classes, as shown in Fig.~\ref{fig:definition_classes}. Table~\ref{tab:local_fex_T_Vs} reports the median local relations between ex-situ mass fraction $\fex{}$, rotational support $V_*/\sigma_*$, and intrinsic shapes (quantified by the triaxiality parameter $T$ and the minor-to-major axis ratio $q$),  as in Figs.~\ref{fig:Vsigma_local} and \ref{fig:shape_fex_local}. These relations are derived using all the ETGs more massive than $10^{10.6}M_{\odot}$, without dividing into stellar mass bins. Table~\ref{tab:transradius_mass} contains the median transition radius $R_{\rm T,exsitu}$ between the in-situ and ex-situ dominated regions in class 2 ETGs (as in Fig.~\ref{fig:transradius_mass}, black curve).

\begin{table*}
\caption{Stellar mass density and cumulative stellar mass fraction profiles for the in-situ and ex-situ components of ETGs in accretion class 1. We report the median, first (Q1), and third quartiles (Q3) of the distribution. }             
\label{tab:definition_class1}  
\centering    
\begin{tabular}{ c| c c c| c c c| c c c| c c c }
\hline\hline      
\textbf{Class 1} & \multicolumn{6}{c|}{In-situ component}& \multicolumn{6}{c}{Ex-situ component}\\
\hline
R/$\re{}$ &  \multicolumn{3}{c|}{$\log_{10}\Sigma / ({\rm M}_{\odot} {\rm kpc}^{-2} )$} &  \multicolumn{3}{c|}{Cum. mass fract.} & \multicolumn{3}{c|}{$\log_{10}\Sigma / ({\rm M}_{\odot} {\rm kpc}^{-2} )$} &  \multicolumn{3}{c}{Cum. mass fract.} \\
\cline{2-13}
   & Q1 & median & Q3 & Q1 & median & Q3 & Q1 & median & Q3 & Q1 & median & Q3\\
 \hline
0.1	&	9.55	&	9.64	&	9.75	&	0.01	&	0.020	&	0.02	&		&	7.52	&	8.17	&	0.00	&	0.00	&	0.00	\\
0.2	&	9.52	&	9.61	&	9.71	&	0.03	&	0.040	&	0.06	&	6.79	&	7.57	&	8.14	&	0.00	&	0.00	&	0.00	\\
0.3	&	9.45	&	9.56	&	9.68	&	0.05	&	0.080	&	0.1	&	6.78	&	7.53	&	8.06	&	0.00	&	0.00	&	0.00	\\
0.4	&	9.37	&	9.48	&	9.63	&	0.09	&	0.130	&	0.16	&	6.85	&	7.5	&	8	&	0.00	&	0.00	&	0.00	\\
0.5	&	9.26	&	9.39	&	9.55	&	0.14	&	0.180	&	0.22	&	6.81	&	7.44	&	7.95	&	0.00	&	0.00	&	0.01	\\
0.6	&	9.13	&	9.28	&	9.43	&	0.22	&	0.250	&	0.28	&	6.81	&	7.45	&	7.94	&	0.00	&	0.00	&	0.01	\\
0.7	&	9.02	&	9.19	&	9.36	&	0.26	&	0.300	&	0.32	&	6.73	&	7.38	&	7.87	&	0.00	&	0.00	&	0.01	\\
0.8	&	8.93	&	9.11	&	9.27	&	0.31	&	0.340	&	0.37	&	6.69	&	7.33	&	7.84	&	0.00	&	0.00	&	0.02	\\
0.9	&	8.84	&	9.02	&	9.17	&	0.36	&	0.380	&	0.41	&	6.67	&	7.28	&	7.8	&	0.00	&	0.01	&	0.02	\\
1	&	8.74	&	8.93	&	9.09	&	0.4	&	0.420	&	0.44	&	6.65	&	7.23	&	7.76	&	0.00	&	0.01	&	0.03	\\
2	&	8.07	&	8.24	&	8.40	&	0.61	&	0.660	&	0.7	&	6.38	&	6.87	&	7.32	&	0.00	&	0.02	&	0.05	\\
3	&	7.60	&	7.73	&	7.86	&	0.69	&	0.770	&	0.83	&	6.1	&	6.59	&	6.97	&	0.01	&	0.02	&	0.07	\\
4	&	7.21	&	7.37	&	7.50	&	0.74	&	0.830	&	0.89	&	5.85	&	6.33	&	6.7	&	0.01	&	0.03	&	0.07	\\
6	&	6.52	&	6.73	&	6.93	&	0.81	&	0.880	&	0.94	&	5.43	&	5.89	&	6.26	&	0.01	&	0.04	&	0.09	\\
8	&	5.95	&	6.19	&	6.48	&	0.84	&	0.910	&	0.96	&	5.09	&	5.51	&	5.89	&	0.01	&	0.04	&	0.10	\\
10	&	5.50	&	5.77	&	6.10	&	0.86	&	0.930	&	0.97	&	4.81	&	5.19	&	5.62	&	0.01	&	0.04	&	0.10	\\
12	&	5.14	&	5.44	&	5.76	&	0.86	&	0.930	&	0.97	&	4.6	&	4.95	&	5.39	&	0.02	&	0.05	&	0.10	\\
14	&	4.85	&	5.19	&	5.50	&	0.87	&	0.930	&	0.97	&	4.43	&	4.79	&	5.18	&	0.02	&	0.05	&	0.10	\\
\hline  
\end{tabular}
\end{table*}

\begin{table*}
\caption{As in Table~\ref{tab:definition_class1}, but for ETGs in accretion class 2.}             
\label{tab:definition_class2}  
\centering    
\begin{tabular}{ c| c c c| c c c| c c c| c c c }
\hline\hline      
\textbf{Class 2} & \multicolumn{6}{c|}{In-situ component}& \multicolumn{6}{c}{Ex-situ component}\\
\hline
R/$\re{}$ &  \multicolumn{3}{c|}{$\log_{10}\Sigma / ({\rm M}_{\odot} {\rm kpc}^{-2} )$} &  \multicolumn{3}{c|}{Cum. mass fract.} & \multicolumn{3}{c|}{$\log_{10}\Sigma / ({\rm M}_{\odot} {\rm kpc}^{-2} )$} &  \multicolumn{3}{c}{Cum. mass fract.} \\
\cline{2-13}
   & Q1 & median & Q3 & Q1 & median & Q3 & Q1 & median & Q3 & Q1 & median & Q3\\
 \hline
0.1	&	9.60	&	9.70	&	9.81	&	0.02	&	0.02	&	0.03	&	7.99	&	8.64	&	9.16	&	0.00	&	0.00	&	0.01	\\
0.2	&	9.48	&	9.62	&	9.74	&	0.04	&	0.06	&	0.08	&	7.96	&	8.59	&	9.08	&	0.00	&	0.01	&	0.02	\\
0.3	&	9.27	&	9.49	&	9.65	&	0.08	&	0.10	&	0.14	&	7.91	&	8.50	&	8.93	&	0.00	&	0.01	&	0.04	\\
0.4	&	9.03	&	9.30	&	9.53	&	0.12	&	0.15	&	0.19	&	7.86	&	8.39	&	8.74	&	0.00	&	0.02	&	0.06	\\
0.5	&	8.79	&	9.10	&	9.40	&	0.16	&	0.20	&	0.24	&	7.81	&	8.28	&	8.60	&	0.00	&	0.03	&	0.08	\\
0.6	&	8.60	&	8.91	&	9.26	&	0.21	&	0.24	&	0.28	&	7.75	&	8.17	&	8.47	&	0.01	&	0.04	&	0.10	\\
0.7	&	8.47	&	8.78	&	9.15	&	0.24	&	0.28	&	0.32	&	7.70	&	8.08	&	8.37	&	0.01	&	0.05	&	0.12	\\
0.8	&	8.36	&	8.66	&	9.06	&	0.27	&	0.31	&	0.35	&	7.66	&	8.00	&	8.28	&	0.01	&	0.06	&	0.13	\\
0.9	&	8.24	&	8.55	&	8.97	&	0.29	&	0.35	&	0.38	&	7.61	&	7.91	&	8.19	&	0.01	&	0.07	&	0.14	\\
1	&	8.14	&	8.45	&	8.88	&	0.31	&	0.38	&	0.42	&	7.55	&	7.84	&	8.11	&	0.02	&	0.08	&	0.15	\\
2	&	7.46	&	7.85	&	8.14	&	0.41	&	0.51	&	0.62	&	7.13	&	7.37	&	7.55	&	0.04	&	0.13	&	0.23	\\
3	&	7.03	&	7.42	&	7.67	&	0.47	&	0.57	&	0.71	&	6.80	&	7.02	&	7.20	&	0.07	&	0.16	&	0.27	\\
4	&	6.70	&	7.06	&	7.30	&	0.5	&	0.62	&	0.76	&	6.56	&	6.76	&	6.94	&	0.08	&	0.19	&	0.31	\\
6	&	6.07	&	6.38	&	6.63	&	0.53	&	0.66	&	0.8	&	6.07	&	6.31	&	6.52	&	0.10	&	0.22	&	0.35	\\
8	&	5.55	&	5.85	&	6.16	&	0.54	&	0.68	&	0.82	&	5.71	&	5.97	&	6.21	&	0.12	&	0.24	&	0.37	\\
10	&	5.12	&	5.43	&	5.74	&	0.55	&	0.69	&	0.83	&	5.39	&	5.68	&	5.95	&	0.12	&	0.25	&	0.38	\\
12	&	4.76	&	5.10	&	5.40	&	0.55	&	0.70	&	0.83	&	5.12	&	5.44	&	5.73	&	0.13	&	0.26	&	0.39	\\
14	&	4.48	&	4.84	&	5.15	&	0.56	&	0.70	&	0.83	&	4.91	&	5.24	&	5.55	&	0.13	&	0.26	&	0.40	\\

\hline  
\end{tabular}
\end{table*}

\begin{table*}
\caption{As in Table~\ref{tab:definition_class1}, but for ETGs in accretion class 3.}             
\label{tab:definition_class3}  
\centering    
\begin{tabular}{ c| c c c| c c c| c c c| c c c }
\hline\hline      
\textbf{Class 3} & \multicolumn{6}{c|}{In-situ component}& \multicolumn{6}{c}{Ex-situ component}\\
\hline
R/$\re{}$ &  \multicolumn{3}{c|}{$\log_{10}\Sigma / ({\rm M}_{\odot} {\rm kpc}^{-2} )$} &  \multicolumn{3}{c|}{Cum. mass fract.} & \multicolumn{3}{c|}{$\log_{10}\Sigma / ({\rm M}_{\odot} {\rm kpc}^{-2} )$} &  \multicolumn{3}{c}{Cum. mass fract.} \\
\cline{2-13}
   & Q1 & median & Q3 & Q1 & median & Q3 & Q1 & median & Q3 & Q1 & median & Q3\\
 \hline
0.1	&	9.27	&	9.46	&	9.62	&	0.010	&	0.01	&	0.020	&	9.43	&	9.67	&	9.9	&	0.01	&	0.02	&	0.04	\\
0.2	&	9.15	&	9.31	&	9.47	&	0.030	&	0.04	&	0.060	&	9.21	&	9.4	&	9.6	&	0.04	&	0.06	&	0.08	\\
0.3	&	8.91	&	9.12	&	9.3	&	0.050	&	0.07	&	0.090	&	8.93	&	9.13	&	9.28	&	0.07	&	0.09	&	0.12	\\
0.4	&	8.68	&	8.90	&	9.11	&	0.070	&	0.1	&	0.130	&	8.65	&	8.86	&	9.07	&	0.09	&	0.13	&	0.15	\\
0.5	&	8.49	&	8.70	&	8.91	&	0.090	&	0.13	&	0.160	&	8.46	&	8.66	&	8.84	&	0.11	&	0.15	&	0.18	\\
0.6	&	8.33	&	8.53	&	8.75	&	0.120	&	0.16	&	0.190	&	8.29	&	8.46	&	8.65	&	0.13	&	0.18	&	0.21	\\
0.7	&	8.23	&	8.41	&	8.61	&	0.140	&	0.18	&	0.210	&	8.17	&	8.34	&	8.54	&	0.15	&	0.19	&	0.23	\\
0.8	&	8.14	&	8.32	&	8.51	&	0.150	&	0.19	&	0.240	&	8.06	&	8.23	&	8.41	&	0.16	&	0.2	&	0.25	\\
0.9	&	8.05	&	8.23	&	8.42	&	0.170	&	0.21	&	0.260	&	7.96	&	8.12	&	8.29	&	0.18	&	0.22	&	0.26	\\
1	&	7.97	&	8.15	&	8.35	&	0.180	&	0.22	&	0.280	&	7.88	&	8.02	&	8.21	&	0.19	&	0.23	&	0.28	\\
2	&	7.45	&	7.65	&	7.86	&	0.280	&	0.34	&	0.400	&	7.32	&	7.46	&	7.63	&	0.26	&	0.3	&	0.36	\\
3	&	7.05	&	7.29	&	7.51	&	0.340	&	0.4	&	0.460	&	6.94	&	7.08	&	7.23	&	0.3	&	0.35	&	0.41	\\
4	&	6.69	&	7.01	&	7.2	&	0.370	&	0.44	&	0.510	&	6.64	&	6.79	&	6.95	&	0.32	&	0.37	&	0.44	\\
6	&	6.04	&	6.39	&	6.61	&	0.410	&	0.49	&	0.560	&	6.08	&	6.28	&	6.43	&	0.36	&	0.41	&	0.48	\\
8	&	5.45	&	5.83	&	6.16	&	0.440	&	0.5	&	0.590	&	5.59	&	5.85	&	6.09	&	0.38	&	0.43	&	0.5	\\
10	&	4.96	&	5.34	&	5.72	&	0.440	&	0.51	&	0.590	&	5.17	&	5.49	&	5.81	&	0.38	&	0.45	&	0.52	\\
12	&	4.53	&	4.92	&	5.34	&	0.440	&	0.51	&	0.590	&	4.89	&	5.22	&	5.54	&	0.39	&	0.46	&	0.52	\\
14	&	4.20	&	4.60	&	5.03	&	0.440	&	0.51	&	0.590	&	4.64	&	4.99	&	5.29	&	0.39	&	0.46	&	0.53	\\

\hline  
\end{tabular}
\end{table*}

\begin{table*}
\caption{As in Table~\ref{tab:definition_class1}, but for ETGs in accretion class 4.}             
\label{tab:definition_class4}  
\centering    
\begin{tabular}{ c| c c c| c c c| c c c| c c c }
\hline\hline      
\textbf{Class 4} & \multicolumn{6}{c|}{In-situ component}& \multicolumn{6}{c}{Ex-situ component}\\
\hline
R/$\re{}$ &  \multicolumn{3}{c|}{$\log_{10}\Sigma / ({\rm M}_{\odot} {\rm kpc}^{-2} )$} &  \multicolumn{3}{c|}{Cum. mass fract.} & \multicolumn{3}{c|}{$\log_{10}\Sigma / ({\rm M}_{\odot} {\rm kpc}^{-2} )$} &  \multicolumn{3}{c}{Cum. mass fract.} \\
\cline{2-13}
   & Q1 & median & Q3 & Q1 & median & Q3 & Q1 & median & Q3 & Q1 & median & Q3\\
 \hline
0.1	&	9.02	&	9.24	&	9.47	&	0.010	&	0.01	&	0.020	&	9.68	&	9.81	&	9.95	&	0.03	&	0.05	&	0.06	\\
0.2	&	8.65	&	8.88	&	9.22	&	0.020	&	0.03	&	0.050	&	9.27	&	9.42	&	9.61	&	0.09	&	0.1	&	0.12	\\
0.3	&	8.30	&	8.60	&	8.92	&	0.030	&	0.05	&	0.060	&	8.9	&	9.06	&	9.22	&	0.14	&	0.15	&	0.17	\\
0.4	&	8.07	&	8.37	&	8.65	&	0.040	&	0.06	&	0.080	&	8.65	&	8.8	&	8.95	&	0.17	&	0.19	&	0.21	\\
0.5	&	7.92	&	8.18	&	8.45	&	0.050	&	0.07	&	0.100	&	8.43	&	8.59	&	8.75	&	0.2	&	0.22	&	0.24	\\
0.6	&	7.76	&	8.05	&	8.28	&	0.070	&	0.08	&	0.120	&	8.26	&	8.41	&	8.57	&	0.23	&	0.25	&	0.28	\\
0.7	&	7.64	&	7.94	&	8.17	&	0.070	&	0.09	&	0.130	&	8.14	&	8.3	&	8.45	&	0.25	&	0.28	&	0.3	\\
0.8	&	7.55	&	7.84	&	8.08	&	0.080	&	0.11	&	0.140	&	8.03	&	8.19	&	8.35	&	0.27	&	0.3	&	0.33	\\
0.9	&	7.47	&	7.74	&	7.99	&	0.090	&	0.12	&	0.150	&	7.93	&	8.09	&	8.26	&	0.29	&	0.32	&	0.35	\\
1	&	7.38	&	7.66	&	7.91	&	0.100	&	0.13	&	0.160	&	7.85	&	7.99	&	8.16	&	0.31	&	0.34	&	0.37	\\
2	&	6.78	&	7.08	&	7.34	&	0.150	&	0.19	&	0.230	&	7.25	&	7.39	&	7.56	&	0.42	&	0.47	&	0.5	\\
3	&	6.36	&	6.64	&	6.97	&	0.180	&	0.22	&	0.270	&	6.86	&	7	&	7.18	&	0.48	&	0.53	&	0.59	\\
4	&	6.05	&	6.35	&	6.64	&	0.190	&	0.24	&	0.300	&	6.58	&	6.74	&	6.89	&	0.52	&	0.57	&	0.63	\\
6	&	5.50	&	5.82	&	6.03	&	0.200	&	0.25	&	0.320	&	6.12	&	6.27	&	6.4	&	0.57	&	0.63	&	0.68	\\
8	&	5.09	&	5.37	&	5.59	&	0.210	&	0.26	&	0.330	&	5.72	&	5.89	&	6.04	&	0.6	&	0.65	&	0.71	\\
10	&	4.71	&	4.99	&	5.22	&	0.220	&	0.27	&	0.340	&	5.39	&	5.55	&	5.78	&	0.62	&	0.66	&	0.72	\\
12	&	4.38	&	4.66	&	4.93	&	0.220	&	0.27	&	0.340	&	5.12	&	5.28	&	5.54	&	0.63	&	0.67	&	0.73	\\
14	&	4.13	&	4.42	&	4.7	&	0.220	&	0.27	&	0.340	&	4.93	&	5.1	&	5.33	&	0.63	&	0.68	&	0.74	\\

\hline  
\end{tabular}
\end{table*}

\begin{table*}
\caption{Local relations between the stellar halo structural parameters (rotational support $V_{\rm tot}/\sigma_{\rm tot}$, minor-to-major axis ratio $q$, and triaxiality parameter $T$) and the fraction of ex-situ stars $\fex{}$. For each value of $\fex{}$, we provide the median, first (Q1), and third quartiles (Q3) of the distribution of the structural parameters in ETGs with $M_{*}>10^{10.6}\MSUN{}$ (Figs.~\ref{fig:Vsigma_local} and \ref{fig:shape_fex_local}).}             
\label{tab:local_fex_T_Vs} 
\centering    
\begin{tabular}{ c| c c c| c c c| c c c}
\hline\hline   
$\fex{}$ &  \multicolumn{3}{c|}{$V_{\rm tot}/\sigma_{\rm tot}$} &  \multicolumn{3}{c|}{$q$} & \multicolumn{3}{c|}{$T$}\\
\cline{2-10}
   & Q1 & median & Q3 & Q1 & median & Q3 & Q1 & median & Q3\\
 \hline
0.1	&	0.85	&	1.24	&	1.64	&	0.25	&	0.3	&	0.39	&	0.08	&	0.14	&	0.22	\\
0.2	&	0.82	&	1.17	&	1.49	&	0.27	&	0.32	&	0.41	&	0.08	&	0.14	&	0.23	\\
0.3	&	0.59	&	0.93	&	1.22	&	0.32	&	0.39	&	0.5	&	0.09	&	0.17	&	0.27	\\
0.4	&	0.31	&	0.65	&	0.95	&	0.36	&	0.47	&	0.6	&	0.11	&	0.22	&	0.39	\\
0.5	&	0.21	&	0.46	&	0.8	&	0.41	&	0.52	&	0.64	&	0.13	&	0.24	&	0.41	\\
0.6	&	0.12	&	0.29	&	0.51	&	0.49	&	0.58	&	0.67	&	0.19	&	0.36	&	0.58	\\
0.7	&	0.07	&	0.13	&	0.27	&	0.53	&	0.59	&	0.66	&	0.27	&	0.52	&	0.71	\\
0.8	&	0.05	&	0.11	&	0.23	&	0.47	&	0.54	&	0.6	&	0.49	&	0.71	&	0.85	\\
0.9	&	0.04	&	0.1	&	0.21	&	0.45	&	0.52	&	0.58	&	0.53	&	0.73	&	0.86	\\
\hline  
\end{tabular}
\end{table*}

\begin{table*}
\caption{Median transition radius $R_{\rm T,exsitu}$ between in-situ and ex-situ dominated regions in class 2 ETGs as a function of stellar mass (Fig.~\ref{fig:transradius_mass}). Q1 and Q3 are the first and third quartiles of the distribution of $R_{\rm T,exsitu}$.}             
\label{tab:transradius_mass}
\centering    
\begin{tabular}{ c| c c c}
\hline\hline   
$\log_{10}M_{*}/\MSUN{}$ &  \multicolumn{3}{c|}{$R_{\rm T,exsitu}$} \\
\cline{2-4}
   & Q1 & median & Q3\\
 \hline
10.35	&	8.77	&	10.49	&	14.9	\\
10.4	&	8.77	&	10.49	&	13.8	\\
10.5	&	8.1	&	9.68	&	12.54	\\
10.6	&	6.29	&	8.77	&	12.54	\\
10.8	&	4.3	&	7.36	&	8.84	\\
11	&	3.2	&	5.61	&	8.43	\\
11.2	&	0.56	&	1.8	&	5.27	\\
11.4	&	0.36	&	0.4	&	1.12	\\
11.7	&	0.25	&	0.4	&	0.58	\\

\hline  
\end{tabular}
\end{table*}

\end{appendix}

\end{document}